\documentclass[pra,aps,reprint,superscriptaddress,onecolumn,longbibliography]{revtex4-2}
\usepackage{amsmath}
\usepackage{amssymb}
\usepackage{graphicx}
\usepackage{dcolumn}
\usepackage{bm}
\usepackage{hyperref}
\usepackage[svgnames]{xcolor}
\usepackage{comment}
\usepackage{braket}
\usepackage{xfrac}
\usepackage{hyperref}
\hypersetup{
    pdftitle = {Motion-Insensitive Time-Optimal Control of Optical Qubits},
    pdfauthor = {Léo Van Damme, Zhao Zhang, Amit Devra, Steffen J. Glaser, and Andrea Alberti},
	colorlinks=true,linkcolor=blue,citecolor=blue,filecolor=blue,urlcolor=blue
  }

\DeclareMathOperator{\Ent}{Ent}
\DeclareMathOperator{\Target}{Tar}

\DeclareMathOperator{\Rec}{Rec}
\DeclareMathOperator{\Det}{Det}

\DeclareMathOperator{\Rab}{Rab}
\DeclareMathOperator{\Trap}{Trap}

\usepackage{xspace}

\def\TORF{TORF\xspace}

\DeclareMathOperator{\Sr88}{{}^{88}Sr}

\begin{document}
\begin{abstract}
In trapped-atom quantum computers, high-fidelity control of optical qubits is challenging due to the motion of atoms in the trap.
If not corrected, the atom motion gets entangled with the qubit degrees of freedom through two fundamental mechanisms, (i) photon recoil and (ii) thermal motion, both leading to a reduction of the gate fidelity.
We develop motion-insensitive pulses that suppress both sources of infidelity by modulating the phase of the driving laser field in time.
To eliminate photon recoil, we use bang-bang pulses---derived using time-optimal control---which shorten the gate duration by about 20 times compared to conventional pulses.
However, even when photon recoil is eliminated, we find that the gate error does not vanish, but is rather limited by a bound arising from thermal motion-induced entanglement.
Remarkably, this bound is independent of the Rabi frequency, meaning that, unlike for photon recoil, operating in the resolved sideband regime does not mitigate this source of infidelity.
To overcome this bound, we derive smooth-phase pulses, which allow for a further reduction of the gate error by more than an order of magnitude for typical thermal atoms.
Motion-insensitive pulses can be refined to compensate for laser inhomogeneities, enhancing the gate performance in practical situations.
Our results are validated through simulations of one-qubit gates operating on the optical clock transition of ${}^{88}$Sr atoms trapped in an optical tweezers array.
\end{abstract}
\title{Motion-Insensitive Time-Optimal Control of Optical Qubits}
\def\LMU{Fakultät für Physik, Ludwig-Maximilians-Universität München, 80799 München, Germany}
\def\MCQST{Munich Center for Quantum Science and Technology, 80799 München, Germany}
\def\MPQ{Max-Planck-Institut für Quantenoptik, 85748 Garching, Germany}
\def\TUM{School of Natural Sciences, Technical University of Munich, Lichtenbergstrasse 4, D-85747 Garching, Germany}
\date{\today}
\author{L\'eo Van Damme}
\email{leo.vandamme@gmx.fr}
\affiliation{\TUM}
\author{Zhao Zhang}
\thanks{${}^{\hspace{-1ex},*}$ These two authors contributed equally to this work.}
\affiliation{\MPQ}
\author{Amit Devra}
\author{Steffen J. Glaser}
\affiliation{\TUM}
\affiliation{\MCQST}
\author{Andrea Alberti}
\affiliation{\MPQ}
\affiliation{\MCQST}
\affiliation{\LMU}
\maketitle
\section{Introduction}
Trapped atoms are currently among the most promising platforms for quantum information processing due to their long coherence times, flexible connectivity, and fast gate times~\cite{winterspergerNeutralAtomQuantum2023,henrietQuantumComputingNeutral2020,bruzewiczTrappedionQuantumComputing2019,osti_1904273,Bluvstein:2024,Cong2022,PhysRevX.13.041051,PRXQuantum.5.020363,PRXQuantum.3.020326}.
These particles are confined within electromagnetic or optical fields, and their internal atomic states are manipulated using lasers, microwaves, or radiofrequency signals to achieve coherent control of quantum operations.

The quantum state of a trapped atom is defined by the qubit, which is used for quantum computation, and the motional states, which describe the spatial vibrations of the particle in the trap.
Quantum gates couple the qubit and the motional states, meaning that the vibration of the particle can affect the gate fidelity through qubit-motion entanglement.
The strength of this coupling is characterized by a constant called the Lamb-Dicke parameter, denoted as $\eta$.
In an ideal scenario where $\eta$ is vanishingly small, the motion is completely decoupled from the gate dynamics, so that the qubit behaves like a pure spin-1/2 particle.

In experiments, one can diminish $\eta$ by increasing the trap frequency.
This is the conventional approach in  trapped-ion quantum computers, where ions vibrate at frequencies up to the MHz range~\cite{RevModPhys.75.281}.
However, neutral atoms are confined in traps with much smaller trap frequencies, typically limited at about 100~kHz~\cite{kaufman2021quantum}, which sets a limit to the minimum $\eta$ that can be achieved.
This difference arises from several fundamental factors: limitation in the power of the trapping lasers, scattering of the trap photons by the atoms and, in set-ups driving Rydberg excitations, losses of atoms because of the anti-trapping potential in the Rydberg states \cite{Saffman:2005}.

In neutral atom-based quantum computers, the qubit information can be encoded into hyperfine states \cite{Xia:2015a,bluvstein2024logical}, nuclear spin~\cite{Ma:2022,Jenkins:2022}, fine-structure states~\cite{Pucher:2024a,PhysRevLett.132.150606}, and optical states~\cite{Young:2020,PhysRevX.9.041052}.
For hyperfine and nuclear qubits, the Lamb-Dicke parameter is negligible, meaning that the qubit can be driven independently of the motion.
This natural insensitivity to motional states allows driving these qubits with fidelities above $99.9\%$~\cite{Ma:2022,everedHighfidelityParallelEntangling2023} without any strategy to suppress recoil.
For fine structure qubits, the Lamb-Dicke parameter can effectively be made small, in the range $\eta\sim 10^{-2}$, by driving the qubits with copropagating Raman laser beams.
For such a small $\eta$, the system is in the Lamb-Dicke regime, where the effect of motion is strongly reduced (Mößbauer effect), but not completely suppressed.
For optical qubits, e.g., driving the optical-clock transition ${}^1\text{S}_0\leftrightarrow{}^3\text{P}_0$ in alkaline-earth atoms, the Lamb-Dicke parameter is typically larger than $0.1$.
When $\eta$ is in this range, the coupling between the qubit and the motional states is significant.
It causes undesired qubit-motion entanglement, resulting in a reduced gate fidelity.
Hence, developing motion-insensitive pulses is crucial to achieve high-fidelity optical qubit gates.

For a given value of the Lamb-Dicke parameter, the established way to mitigate photon recoil is to drive the optical qubit at Rabi frequencies much lower than the trap frequency to avoid motion-changing transitions (i.e., so-called motional sidebands). 
This is known as the resolved sideband regime~\cite{RevModPhys.75.281}.
This regime, however, implies relatively slow gates, typically lasting hundreds of microseconds~\cite{Finkelstein:2024}.

Away from this regime, photon recoil has a significant effect on the gate evolution.
Efforts have been made to quantify~\cite{robicheauxPhotonrecoilLaserfocusingLimits2021} and mitigate \cite{lisMidcircuitOperationsUsing2023} its impact on the gate fidelity.
In Ref.~\cite{lisMidcircuitOperationsUsing2023}, the authors found that applying two CORPSE pulses~\cite{cumminsTacklingSystematicErrors2003} at a specific ratio between the trap and the Rabi frequencies enables the execution of $\pi$-rotations while suppressing motional excitations.

In this work, we design time-optimal bang-bang pulses capable of implementing any arbitrary qubit gate while suppressing motional excitations.
Importantly, they can be applied at relatively high Rabi frequencies, comparable to the trap frequency, enabling significantly faster gates compared to constant pulses applied in the resolved sideband regime.
We call these time-optimal recoil-free (\TORF) pulses.

We find, however, that photon recoil is not the only source of motion-induced decoherence.
Due to imperfect cooling, atoms are in a thermal motional state where only a limited fraction (typically 0.90 to 0.99) occupy the ground state, leading to qubit-motion entanglement even when photon recoil is suppressed. 
This holds true deep in the resolved sideband regime, where photon recoil is naturally absent.
To address this additional source of decoherence, we use TOD (Time-Optimal Disentangling) pulses, which handle both photon recoil and thermal motion-induced entanglement.

These findings complement our recent work~\cite{Zhang2024}, where we have used quantum process tomography to design and analyze motion-insensitive optical qubit gates.
These gates use a particular phase-modulated pulse (which we called Mikado) interleaved with controllable $z$-rotations performed by off-resonant local addressing lasers.
Here, we do not restrict ourselves to Mikado pulses.

We develop a quantum optimal control approach to analyze the effects of atom motion and achieve high-fidelity control of any optical qubit gate. Photon recoil and thermal motion-induced entanglement are addressed within a mathematical framework based on Average Hamiltonian Theory~\cite{Haeberlen68,BrinkmannAVH}, which enables us to define clear control problems and gain a deeper insight into the underlying physical processes.

The Hamiltonian governing the dynamics of the system is introduced in Section~\ref{SecModel}, along with the tools to evaluate the gate fidelity for thermal atoms. 
Sections~\ref{SecLambDicke} and~\ref{SecLambDicke2} are dedicated to photon recoil within the Lamb-Dicke regime and beyond it, respectively.
Section~\ref{SecDisentangling} focuses on analyzing and mitigating the effects of thermal motion-induced entanglement.
Effects induced by laser inhomogeneities are addressed in Section~\ref{SecRobust}, and the performance of robust, motion-insensitive pulses is quantified through simulations in Section~\ref{SecSimu}.
We summarize the findings and outline possible extensions of the method in Section~\ref{SecDiscussion}.

\section{Model\label{SecModel}}
\paragraph*{\textbf{Dynamics of trapped atoms.}}
We consider neutral atoms trapped in an array of optical lattices.
Denoting $\ket{g}$ and $\ket{e}$ the internal states of an atom, used as ground and excited states of the qubit,
the Hilbert space of the system is the tensor product between the qubit and the motional states $\ket{0}$, $\ket{1}$, $\ket{2}$, $\dots$, which are in infinite number. A general state is given by:
\begin{equation}
\ket{\psi}=\sum_{m=0}^{\infty}\alpha_m\ket{g,m}+\beta_m\ket{e,m},
\label{EqPsiGen}
\end{equation}
where $m$ denotes the $m^{\text{th}}$ motional state. The laser pulse drives transitions between $\ket{g}$ and $\ket{e}$ at Rabi frequency $\Omega$, and its phase $\varphi(t)$ can be modulated over time to control the system. We can restrict the motion to one spatial dimension, aligned with the direction of the driving laser, so that it can be described by a quantum harmonic oscillator.
In a given rotating frame, the dynamics of an atom in the array is described by the Hamiltonian~\cite{RevModPhys.75.281,lizuainMotionalFrequencyShifts2007} ($\hbar=1$):
\begin{equation}
H(t)=\Delta\ket{e}\bra{e}+\left.\left.\frac{\Omega}{2}\right(\ket{e}\bra{g} e^{i\varphi(t)}e^{i\eta (a^{\dagger}+a)}+h.c.\right)+\omega a^{\dagger}a,\label{EqHTotal}
\end{equation}
where:
\begin{itemize}
\item $\ket{g}$ and $\ket{e}$ are the internal qubit states,
\item $\Omega$ is the Rabi frequency of the laser beam,
\item $\varphi(t)$ is the phase of the laser beam, used to control the system,
\item $\omega$ is the trap frequency, fixed to $\omega=2\pi\times(100~\mathrm{kHz})$ in this paper,
\item $a$ and $a^{\dagger}$ are the creation and annihilation operators,
\item $\eta=\sqrt{{\hbar k^2}/{(2\tilde{m}\omega)}}$ is the Lamb-Dicke parameter ($\tilde{m}$ is the atom's mass and $k=2\pi/\lambda$),
\item $\Delta$ is the detuning of the laser frequency from the qubit transition.
\end{itemize}
As explained in the introduction, the qubit is coupled to the motion through the Lamb-Dicke parameter $\eta$. Note that $\eta^2$ corresponds to the ratio between the recoil energy ${\hbar^2 k^2}/{2\tilde{m}}$ and the energy difference $\hbar\omega$ between two motional states.
The recoil energy depends on the atom species and the qubit transition.
For the sake of clarity, we focus on the optical transition ${}^1S_0\leftrightarrow{}^3P_0$ ($\lambda=698$~nm) of a $\Sr88$ atom, where $\eta\simeq 0.2156$ assuming a trap frequency of $100$~kHz.
The method applies to any other qubit where $\eta\lesssim 10^{-1}$.
\paragraph*{\textbf{Fidelity of qubit gates.}} 
The fidelity a quantum operation is measured as follows.
We first define a set of initial states $\ket{\psi_{km}}$ such that:
\begin{equation}
\begin{cases}
    \ket{\psi_{1m}}=\ket{g,m}\\
    \ket{\psi_{2m}}=\ket{e,m}\\
    \ket{\psi_{3m}}=\frac{\ket{g,m}+\ket{e,m}}{\sqrt{2}}\\
    \ket{\psi_{4m}}=\frac{\ket{g,m}+i\ket{e,m}}{\sqrt{2}},
\end{cases}\label{Eq4States}
\end{equation}
corresponding to four initial qubit states per motional state $m$.
Given the evolution operator $U(T)$, solution of the Schrödinger equation $\dot{U}=-iHU$, each one of these initial states leads to a final state of the form $\ket{\psi_{km}(T)}=U(T)\ket{\psi_{km}}$.
Since the goal is to realize a target gate $U_{\Target}\in SU(2)$ on the qubit, we define a set of target states as $\ket{\psi_{km}^{\Target}}=(U_{\Target}\otimes\mathbb{I}_{M\times M})\ket{\psi_{km}}$, where $M\to\infty$ is the number of motional states.
These target states correspond to ideal operations in the qubit subspace only.
Thus, we can define the gate fidelity associated to the motional state $m$ as:
\begin{equation}
    \mathcal{F}^{(m)}=\tfrac{1}{4}\sum_{k=1}^4|\braket{\psi_{km}^{\Target}|\psi_{km}(T)}|^2.\label{EqFidelity}
\end{equation}
An ideal qubit gate would achieve $\mathcal{F}^{(m)} = 1$ for all $m$, but this is extremely restrictive and unrealistic due to the infinite number of motional states.
An essential property of ultra-cold atoms is that they are cooled down to extremely low temperatures, on the order of $1~\mu$K~\cite{ediss26329,4319045,PhysRevLett.133.013401}, meaning that a limited number of motional states are involved in the dynamics.
Before applying the pulse, the motional states of a thermal atom follow a Boltzmann distribution, which depends on the probability $p_0$ of the motional ground state according to:
\begin{equation}
    p_m=\frac{(1-p_0)^m}{\sum_{k=0}^{M}{(1-p_0)^k}}, \label{EqBoltzmann}
\end{equation}
where the number of motional states $M$ is infinite in theory but can be truncated in the simulations.
The ground state probability is related to the temperature via $p_0=1-\exp[-\hbar\omega/k_B\hat{T}]$ where $k_B$ is the Boltzmann constant and $\hat{T}$ is the temperature.
For example, a temperature of $1~\mu$k corresponds to $p_0\simeq 0.99$ using a $100$~kHz trap frequency.
For thermal atoms, the gate fidelity can be constructed by weighting the contribution of each $\mathcal{F}^{(m)}$ by the Boltzmann coefficients according to the formula:
\begin{equation}
    \mathcal{F}=\sum_{m=0}^{M}p_m \mathcal{F}^{(m)}.\label{EqJGate}
\end{equation}
$M$ has to be sufficiently large to ensure that the simulation correctly describes the actual dynamics of the system. 
We choose $M=20$ in all our simulations, which is more than enough for the cases we will consider.

For the sake of clarity, all the results presented in this paper are derived using a target gate of the form: 
\begin{equation}
    U_{\Target}=e^{-\tfrac{i}{2}\sigma_x\theta_{\Target}},\label{UTarget}
\end{equation}
which is a rotation of angle $\theta_{\Target}$ along the $x$-axis of the Bloch sphere.
The results can be used to perform rotations along any other transverse axis of the Bloch sphere by adapting the initial phase $\varphi(0)$ of the pulse. Thus, all single qubit gates necessary to construct a universal set of gates can be obtained.

In all the sections concerning the analysis of photon recoil, we consider that 100\% of the atoms are initially in the ground state, i.e.,
that $p_0=1$, in which case $\mathcal{F}=\mathcal{F}^{(0)}$.
This is not true in practice because the temperature is not exactly zero.
In a realistic experiment, the atoms are around $90$ to $99\%$ in the ground state.
The choice $p_0=1$ aims to isolate the fundamental properties at the heart of the problem.
The effect of the temperature will be studied in detail in a dedicated part, and a realistic $p_0$ will be used for the final analysis of our findings.
\section{Recoil-free gates in the Lamb-Dicke regime\label{SecLambDicke}}
\begin{figure}[b]
\includegraphics[scale=0.36]{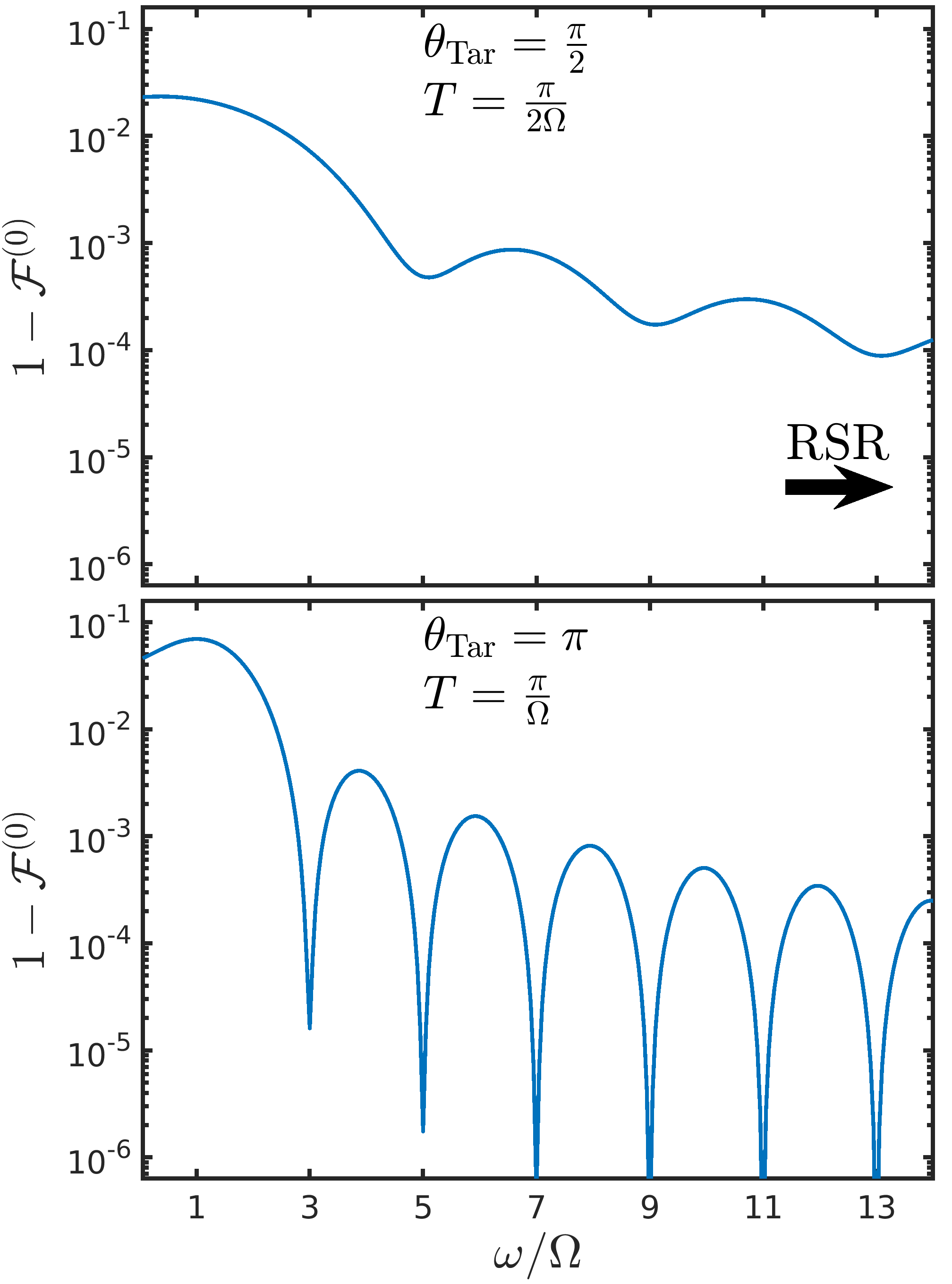}
\caption{\emph{Upper panel:} Error of a $\sqrt{\mathrm{NOT}}$ gate after applying a constant $\pi/2$-pulse as a function of $\omega/\Omega$.
\emph{Lower panel:} Error of a NOT gate after applying a constant $\pi$-pulse.
In all cases, the the system is simulated in the Lamb-Dicke regime (Eq.~\eqref{EqHLambDicke}) and the fidelity is computed for $p_0=1$.
The arrow indicates the resolved sideband regime (RSR) in which $\omega/\Omega$ is very large.
\label{FigRecoilLD}}
\end{figure}

\paragraph*{\textbf{Hamiltonian.}}
As a first approach, we consider a single atom evolving in the Lamb-Dicke regime, which occurs when $\eta\ll 1$, with the motion fully in the ground state ($p_0=1$).
We will see later that this level of approximation is not enough to describe accurately the optical transition described in the previous section because $\eta$ is not that small and the cooling is not perfect.
However, this regime embeds the most important properties, is less complex to analyze, provides numerous insights and gives more intuition on this kind of system.
The results can be applied to systems in which $\eta\lesssim 10^{-2}$, such as fine-structure qubit or trapped ions.
We focus exclusively on the effect of motion by assuming $\Delta=0$ (no detuning) and $\Omega=cst$ (negligible rising time).
The inhomogeneities of the system will be studied in a dedicated section.
The Hamiltonian is derived by using an expansion of the Hamiltonian~\eqref{EqHTotal} up to the first order in $\eta$, leading to:
\begin{equation}
\begin{aligned}
    H(t)&=\left.\left.\frac{\Omega}{2}\right(\ket{e}\bra{g} e^{i\varphi(t)}\big(\mathbb{I}+i\eta(a^{\dagger}+a)+o(\eta^2)\big)+h.c.\right)+\omega a^{\dagger}a\\
    &= h_q(t)+\eta h_p(t)(a^{\dagger}+a)+\omega a^{\dagger}a+o(\eta^2),\label{EqHLambDicke}
\end{aligned}
\end{equation}
where the second line is its expression in the Pauli basis, $h_q$ is a Hamiltonian of an ideal two-level quantum system, and $h_p$ is an operator that is orthogonal to it:
\begin{equation}
\begin{aligned}
  &h_q(t)=\left.\left.\tfrac{\Omega}{2}\right(\cos[\varphi(t)]\sigma_x+\sin[\varphi(t)]\sigma_y\right),\\
  &h_p(t)=\left.\left.\tfrac{\Omega}{2}\right(\cos[\varphi(t)]\sigma_y
  -\sin[\varphi(t)]\sigma_x\right),
\end{aligned}   \label{EqHQHP}
\end{equation}
where $\sigma_k$ are the Pauli matrices. For conciseness, we omit all tensor product symbols. For more rigorous expressions, any operator acting on the qubit, e.g. $h_q$, should be written as $h_q\otimes \mathbb{I}_{M\times M}$, and an operator in the motional subspace as, e.g., $\mathbb{I}_{2\times 2}\otimes a^{\dagger}$.
\paragraph*{\textbf{Effects of constant pulses.}}
To give a better intuition of the problem, we simulate the evolution operator $U(T)$ using the Hamiltonian $\eqref{EqHLambDicke}$ with $\varphi=0$, corresponding to a constant pulse along the $x$ axis of the Bloch sphere.
We study two target gates;
a rotation of $\theta_{\Target}=\pi/2$, corresponding to a $\sqrt{\mathrm{NOT}}$ gate, and of $\theta_{\Target}=\pi$, i.e., a NOT gate.
For an ideal qubit, these gates can be realized with pulses of duration $\pi/(2\Omega)$ and $\pi/\Omega$, respectively.
The figure~\ref{FigRecoilLD} shows the effect of these two simple pulses on the gate fidelity $\mathcal{F}^{(0)}$ as a function of the ratio $\omega/\Omega$.

Even though the particle is assumed to be in the Lamb-Dicke regime, figure~\ref{FigRecoilLD} already reveals a lot of features, giving an idea of the complexity of this system.
First, for all panels, the photon recoil and the gate error decrease globally as $\omega/\Omega$ increases.
This occurs because if $\omega\gg \Omega$, the system evolves deep in the resolved sideband regime, where motion-changing transitions are inhibited.
Since $\omega$ is limited in practice ($\sim 100$~kHz), this regime requires a relatively low Rabi frequency (few kHz) involving slow pulses of hundreds of microseconds.
Second, we can see that constant pulses perform very well for the NOT gate when $\omega/\Omega$ is odd.
This behavior is specific to NOT gates but, unfortunately, doesn't appear for any other qubit gate.
Nevertheless, it is very surprising that for these specific ratios, the gate fidelity is not limited by photon recoil.
It means that the atom can absorb a photon without experiencing a momentum kick in a regime where the motional sidebands are not inhibited.
In the following, we will see that this feature actually exists for any arbitrary qubit gate and for any ratio $\omega/\Omega$, but requires controls beyond the constant pulse scheme.
\paragraph*{\textbf{Recoil-free condition.}}
Since we are assuming that $\eta$ is small, the unitary operator $U(t)$, solution of $\dot{U}=-iHU$, can be approximated using the Average Hamiltonian Theory~\cite{Haeberlen68,BrinkmannAVH} applied in a toggling frame associated to a perfect qubit. More details are given in appendix~\ref{AppPerturbTheory}.
We obtain:
\begin{equation}
    U(t)= U_q(t) e^{-i\omega a^{\dagger}a t}e^{-i\eta\big[a^\dagger V_{\Rec}(t)+a V_{\Rec}^\dagger(t)\big]}+o(\eta^2),\label{EqOperatorO1}
\end{equation}
where $U_q$ and $V_{\Rec}$ are $2\times 2$ complex matrices such that:
\begin{equation}
\begin{aligned}
& \tfrac{d}{dt} U_q(t)=-ih_q(t)U_q(t),\\
& V_{\Rec}(t)=\int_0^tU_q^{\dagger}(t')h_p(t')U_q(t') e^{i\omega t'} dt',\label{EqUQURec}
\end{aligned}
\end{equation}
with $U_q(0)=\mathbb{I}$ and $V_{\Rec}(0)=0$ by definition.

The operator $e^{-i\omega a^{\dagger}a t}$ acts purely in the motional subspace; we can show that it has no effect on the gate fidelity so it can simply be ignored.
The matrix $U_q$ corresponds to the evolution operator of an ideal spin-1/2 particle driven by the Hamiltonian $h_q$. 
The matrix $V_{\Rec}$ is a non-unitary $2\times 2$ operator describing the effect of the qubit dynamics on the motional states.
Indeed, we can see from Eq.~\eqref{EqOperatorO1} that $V_{\Rec}$ causes transitions between the motional states through $a$ and $a^{\dagger}$, which tends to populate higher motional states. In other words, this operator induces photon recoil. 
This allows us to establish a recoil-free condition given by $V_{\Rec}(T)=0$. 
If this condition is satisfied, and if $h_q$ drives the qubit operator $U_q$ in a way that it reaches $U_{\Target}$, we obtain $U(T)=U_q(T)=U_{\Target}$, showing that the target gate is exactly reached, i.e., that $\mathcal{F}=1$.

Since $V_{\Rec}$ depends on the qubit dynamics, it is possible to control the system in a way that suppresses photon recoil by shaping the phase $\varphi(t)$ of the driving laser beam over time.
To achieve this, the phase shape $\varphi(t)$ must be a solution of the following control problem:
\begin{equation}
\begin{cases}
U_q(T)=U_{\Target}\\
V_{\Rec}(T) =0.
\end{cases}\label{EqRecFreeControl}
\end{equation}
The recoil-free condition adds complexity to the qubit control mechanisms compared to standard methods used for ideal two-level quantum systems. Indeed, driving the qubit with conventional pulses may not satisfy the recoil-free condition.
Note that, in the resolved sideband regime defined such that $\Omega\ll\omega$, the term $e^{i\omega t}$ oscillates very fast compared to $U_q^{\dagger}h_pU_q$.
Thus, the operator $V_{\Rec}$ averages out over several oscillations and approaches zero as $\omega/\Omega\to \infty$, meaning that photon recoil is naturally mitigated at low Rabi frequencies.

From a control perspective, it is highly beneficial that the original model governed by Eq.~\eqref{EqHLambDicke} has been simplified to the model of Eq.~\eqref{EqUQURec}.
Instead of working with an infinite-dimensional system that is complex to analyze and computationally intensive to simulate, we now work with two $2\times 2$ matrices under the constraint $V_{\Rec}(T)=0$.
This reduces the problem to that of a standard spin-1/2 particle with an additional constraint, making it much more suitable for optimization.
This problem can be solved using all types of pulses by leveraging the extensive research on optimal control of two-level quantum systems, including composite pulses~\cite{dridiRobustControlNot2020,PhysRevApplied.18.034062,wuCompositePulsesOptimal2023,genovCorrectionArbitraryField2014,jonesDesigningShortRobust2013,LEVITT198661}, inverse pulse engineering~\cite{PhysRevA.101.023822,PhysRevLett.111.050404}, optimization algorithms~\cite{Muller_2022,khanejaOptimalControlCoupled2005,skinnerOptimalControlDesign2010}, geometric curves~\cite{barnesDynamicallyCorrectedGates2022,nelsonDesigningDynamicallyCorrected2023}, maximum principle~\cite{PRXQuantum.2.030203,vandammeRobustOptimalControl2017,dionisTimeoptimalControlTwolevel2023}, and many more~\cite{Ansel_2024,Koch:909195}.
Here, we solve the problem \emph{in minimum time} by searching for the time-optimal shape $\varphi(t)$ solution of Eq.~\eqref{EqRecFreeControl}.
The solutions are referred to as Time-Optimal Recoil-Free (\TORF) pulses.

\paragraph*{\textbf{Time-optimal recoil-free pulses.}}
As a first approach, the recoil-free control problem given in Eq.~\eqref{EqRecFreeControl} is solved with a gradient algorithm based on GRAPE~\cite{khanejaOptimalControlCoupled2005}.
The time is discretized in very small intervals and $\varphi(t)$ is assumed to be constant over each interval.
We consider a longer and longer pulse duration $T$ until we find a numerical solution of the problem with a precision of~$\sim 10^{-10}$.
The solution is assumed to be the time-optimal pulse (multiple initializations are tested).

\begin{figure}[t]
\centering
\includegraphics[width=.6\textwidth]{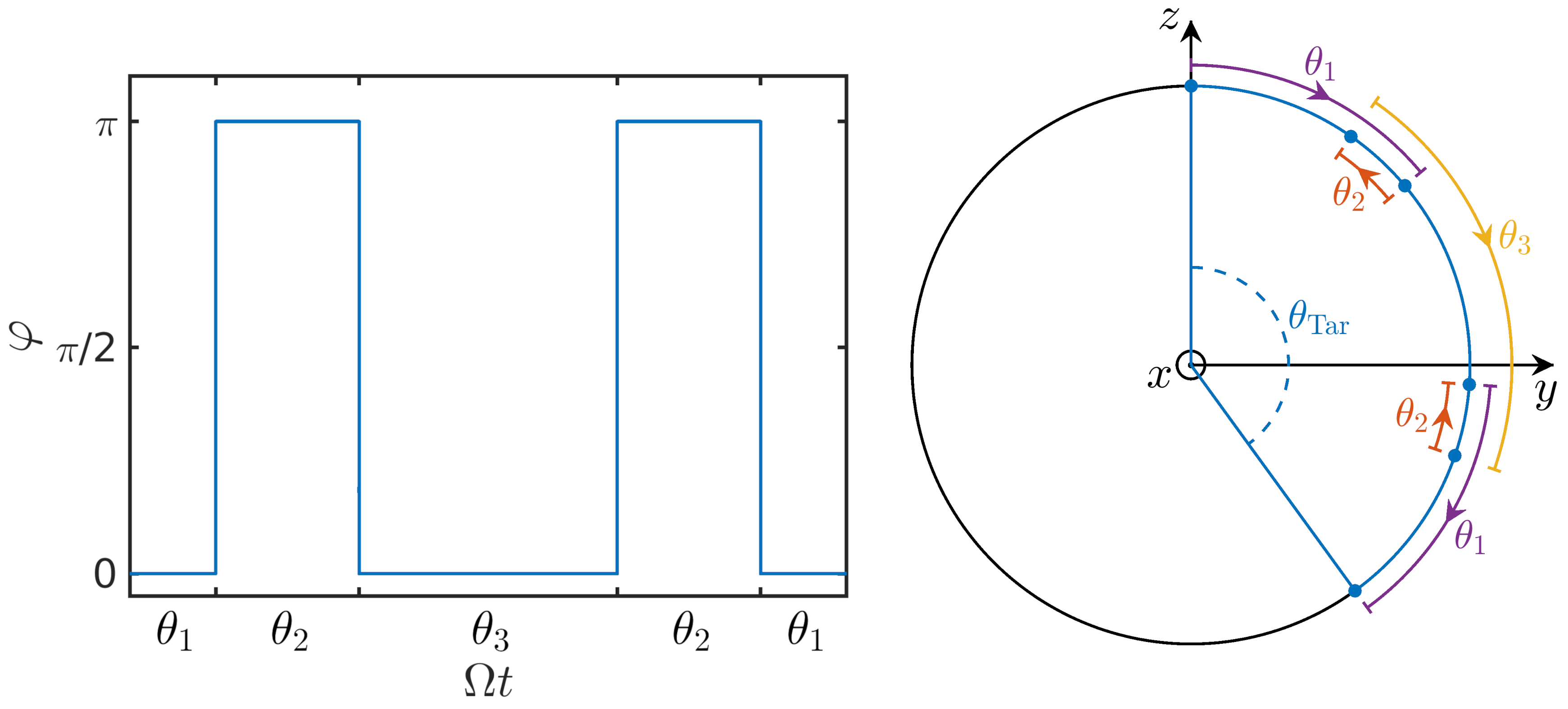}
\caption{\emph{Left panel:} Structure of a \TORF pulse. \emph{Right panel:} Bloch sphere representation of the qubit rotations.\label{FigSchemaTORF}}
\end{figure}

Remarkably, we obtain that the \TORF pulses always have the same structure.
For a given $\theta_{\Target}$ and ratio $\omega/\Omega$, we obtain symmetric bang-bang pulses made of segments of phase $\varphi=0$ and $\varphi=\pi$, as illustrated in Fig.~\ref{FigSchemaTORF}.
The~pulse is characterized by three parameters, $\theta_1$, $\theta_2$, and $\theta_3$, which represent the rotation angles produced by the different segments of the pulse on the Bloch sphere. These angles depend on the target gate and the ratio $\omega/\Omega$.

This result is empirical;
we don't have a mathematical proof that there is no shorter solution~\footnote{
When $\omega/\Omega\lesssim 1$, we find solutions that are very similar to the bang-bang pulse of Fig.~\ref{FigSchemaTORF}, but a bit smooth. 
They even become singular when $\omega/\Omega\to 0$, in the sense that they contain segments where the pulse amplitude vanishes.
However, these solutions are not relevant, as they arise in a regime where the motional sidebands are strongly excited, rendering the Average Hamiltonian Theory invalid, as briefly discussed in Appendix~\ref{AppPerturbTheory}.
}.
Explaining why the time-optimal solution takes this form would require further investigation.
A deeper mathematical analysis, such as applying the Pontryagin Maximum Principle~\cite{PontryaginOCT,10.5555/2209760,PRXQuantum.2.030203}, could offer more insight into the control mechanisms and the shape of these time-optimal pulses.
However, such an analysis is beyond the scope of this paper.
Therefore, we assume the given bang-bang pulse and use it to design recoil-free gates.

Using a \TORF pulse significantly simplifies the control problem.
Since the dynamics of the qubit is only made of rotations about the $x$-axis of the Bloch sphere, we can show using the symmetries that the problem consists in finding the parameters $\theta_k$ solution of a nonlinear system of equations (See Appendix~\ref{AppTORFproblem}).
One of these equations is obvious:
as the qubit must reach the target gate, we have $2\theta_1-2\theta_2+\theta_3=\theta_{\Target}$, which can be understood by looking at the second panel of Fig.~\ref{FigSchemaTORF}.
The two other equations are more complicated and we didn't find a solution without the help of a numerical algorithm.
For instance, a $\pi/2$-\TORF pulse designed for $\omega/\Omega=5$ is such that $\theta_1=0.0840\pi$, $\theta_2=0.0269\pi$ and $\theta_3=0.3858\pi$, leading to a pulse duration $T=0.6077\pi/\Omega$. $T$ corresponds to the physical minimum time to realize a recoil-free $\pi/2$-gate for this frequency ratio.

\begin{figure}[b]
\centering
\includegraphics[scale=0.4]{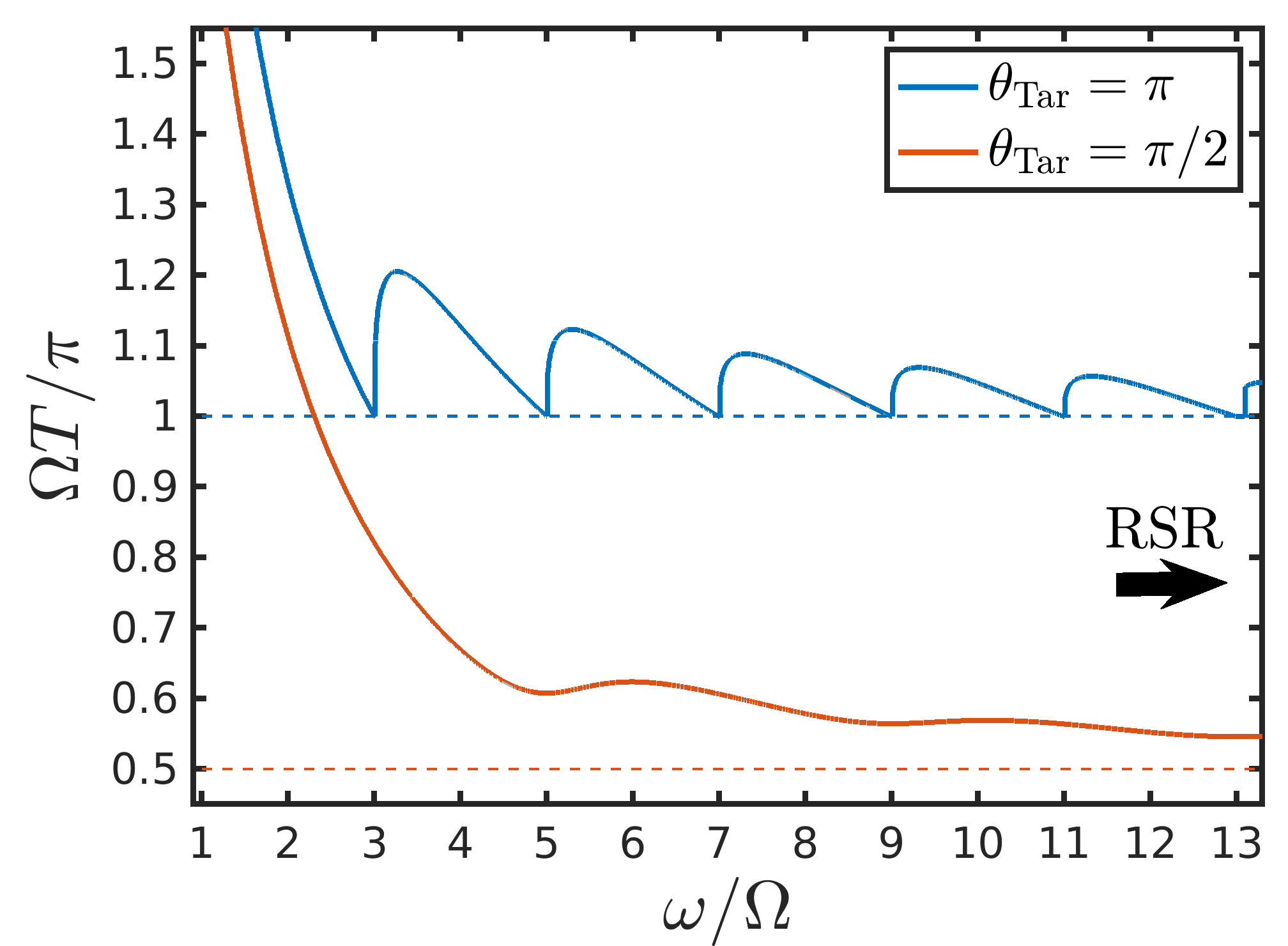}
\caption{Optimal pulse length $\Omega T$ of $\pi$-\TORF (blue) and $\pi/2$-\TORF (red) pulses as a function of $\omega/\Omega$. The dotted lines correspond to the quantum speed limit for realizing the corresponding gates if the motion is neglected, given by $\Omega T=\theta_{\Target}$. The black arrow indicates the resolved sideband regime (RSR) in which $\omega/\Omega\gg 1$.\label{FigToptLD}}
\end{figure}

The solution can be computed for all target gates and frequency ratios $\omega/\Omega$. We focus on the cases $\theta_{\Target}=\pi$ (NOT gate) and $\theta_{\Target}=\pi/2$ ($\sqrt{\mathrm{NOT}}$ gate). 
For the NOT gate, we obtain that when $\omega/\Omega$ is odd, $\theta_1=\theta_2=0$ while $\theta_3=\pi$, which is precisely a constant $\pi$-pulse.
This is also evident in Fig.~\ref{FigToptLD}, which shows the optimal pulse duration as a function of $\omega/\Omega$. Indeed, we can see that $T$ reaches the quantum speed limit of $\pi/\Omega$ under these conditions.

This result matches beautifully the empirical findings outlined in Fig.~\ref{FigRecoilLD}.
Indeed, the peaks appear precisely when $\omega/\Omega$ is odd because, in these conditions, a constant $\pi$-pulse is a \TORF pulse.
If $\omega/\Omega$ is not odd, the phase flips are necessary and the pulse is a bit longer than $\pi/\Omega$.

This characteristic never occurs for $\pi/2$ pulses.
While we do observe minima of the pulse duration at $\omega/\Omega=5$, $9$, $13$, and so forth, it remains always higher than the quantum speed limit.
This means that phase flips are always necessary to achieve recoil-free $\sqrt{\mathrm{NOT}}$ gates.
They are imperative for any qubit target gate such that $\theta_{\Target}\neq n\pi$, meaning that constant pulses cannot be used to construct a universal set of recoil-free qubit gates.

Globally, in all cases, $T$ decreases toward the quantum speed limit of $\theta_{\Target}/\Omega$ when $\omega/\Omega\to \infty$.
This is expected since it corresponds to pulses operating deep in resolved sideband regime, where photon recoil is naturally suppressed using constant pulses. However, it implies low Rabi frequencies and thus slow gates in practice.

\section{Recoil-free gates beyond the Lamb-Dicke regime\label{SecLambDicke2}}
\paragraph*{\textbf{Hamiltonian.}}
Since the parameter $\eta$ is not very small in our model, the Lamb-Dicke regime does not capture all the complexity of the original Hamiltonian defined in Eq.~\eqref{EqHTotal}.
It is thus necessary to extend the method beyond the Lamb-Dicke regime by expanding the Hamiltonian up to the second order in $\eta$. In this section, we still assume that $p_0=1$ and focus on photon recoil.
For the sake of clarity, many details about the derivation of the following results are reported in Appendix~\ref{AppLambDicke2}.
We present here only the insightful formulas.
The second-order Hamiltonian can be written as:
\begin{equation}
H(t)=\underbrace{h_q(t)\big(1-\tfrac{\eta^2}{2}\big)}_{\textbf{Qubit}}+\underbrace{\eta h_p(t)(a^{\dagger}+a)-\tfrac{\eta^2}{2}h_q(t)\big({a^{\dagger}}^2+a^2\big)}_{\textbf{Recoil}}-\underbrace{\eta^2 h_q(t) a^{\dagger}a}_{\textbf{Entanglement}} +\underbrace{\omega a^{\dagger}a}_{\textbf{Motion}},\label{EqHamO2}
\end{equation}
where $h_q$ and $h_p$ are given in Eq.~\eqref{EqHQHP}.
The Hamiltonian can be decomposed into three parts:
one part describing the dynamics in the qubit subspace, another describing the coupling between the qubit and the motion, and a third acting only on the motional state.
The coupling part can be subdivided into a recoil contribution, mixing the motional states through $a$, $a^2$, ${a^{\dagger}}$ and ${a^{\dagger}}^2$, and a an additional term mixing the qubit and the motional states through $h_q$ and $a^{\dagger}a$.

\begin{figure}[h]
    \centering
    \includegraphics[scale=0.4]{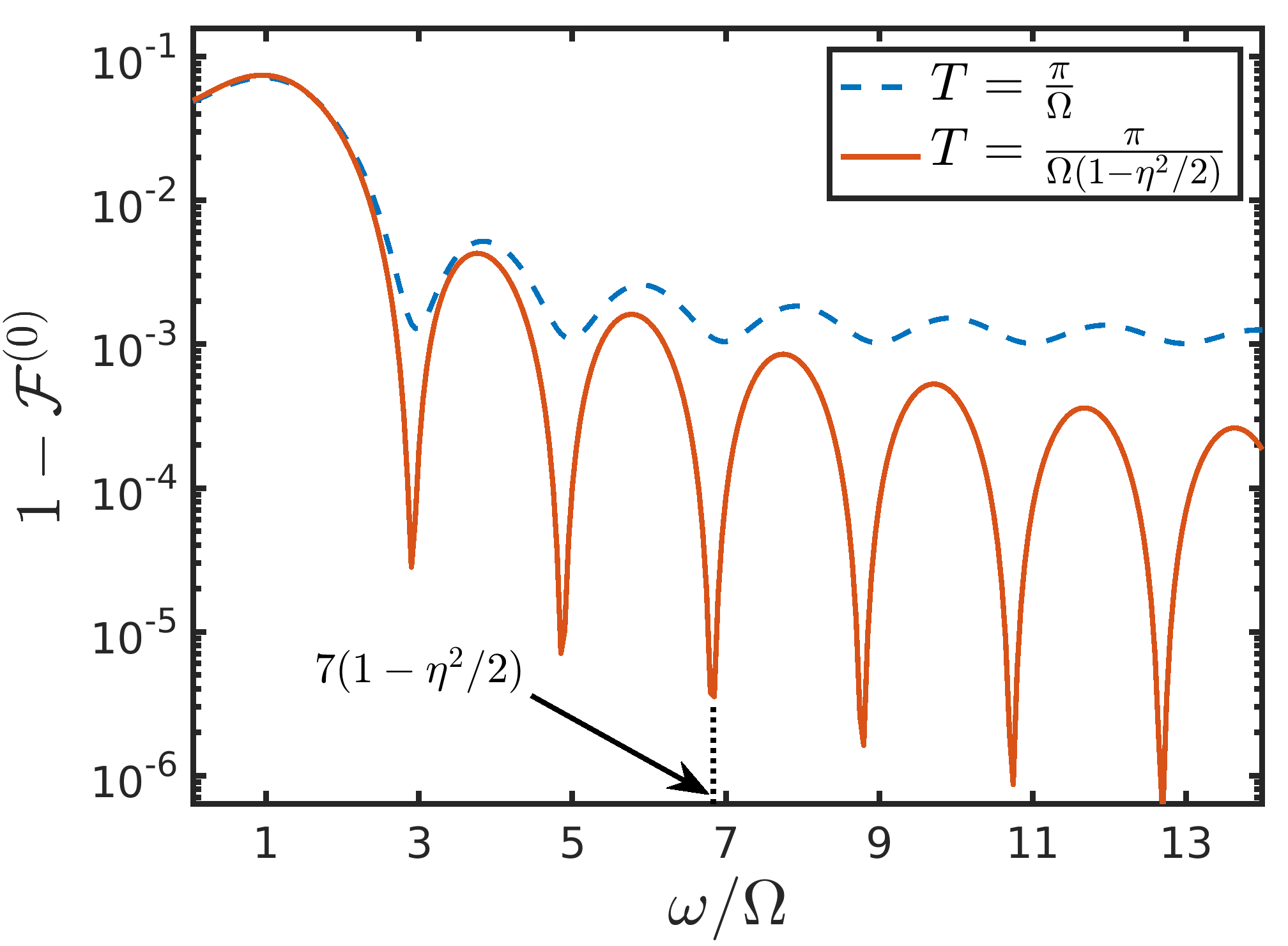}
 \caption{Error of a NOT gate after applying a constant pulse of duration $T=\pi/\Omega$ (blue dashed line) and of duration $T=\pi/[\Omega(1-\eta^2/2)]$ (solid red line) as a function of $\omega/\Omega$.
The peaks appear at frequency ratios $\omega/\Omega=(2k+1)(1-\eta^2/2)$ with $k$ an integer, which is illustrated by the black dotted line for $k=3$.
In all cases, the system is simulated using the Hamiltonian~\eqref{EqHamO2} and the fidelity is computed for $p_0=1$.}
 \label{FigRecoilO2}
\end{figure}

Remarkably, the qubit part is scaled by a factor $(1-\eta^2/2)$, which slows down the gate speed in the qubit space regardless of the motional distribution.
In other words, it is a direct effect of the motion on the qubit dynamics.
A general expression of this factor can be found in Ref.~\cite{RevModPhys.75.281}.
\paragraph*{\textbf{Effects of constant pulses.}} Let's analyze the effect of constant pulses on the gate fidelity for a target NOT gate, simulated using the Hamiltonian~\eqref{EqHamO2}.
We consider a standard constant $\pi$-pulse of duration $T=\pi/\Omega$ and a slightly longer one of duration $T=\pi/[(1-\eta^2)\Omega]$.
The figure~\ref{FigRecoilO2} shows the gate error $1-\mathcal{F}^{(0)}$ as a function of $\omega/\Omega$.

We can see that using a slightly longer duration is necessary to obtain a good fidelity, as it compensates for the lack of speed caused by the factor $(1-\eta^2/2)$ on the qubit part of the Hamiltonian. 
This remains true even when the pulse is applied infinitely deep in the resolved sideband regime.
Note also that we still have the peaks of very low error as in the Lamb-Dicke regime, but shifted by a factor $(1-\eta^2/2)$.
Like in the Lamb-Dicke regime, this property does not appear for any other qubit gate.
\paragraph*{\textbf{Evolution operator.}}
By applying the Average Hamiltonian Theory, we can show that the evolution operator of the system under the Hamiltonian~\eqref{EqHamO2} is given by (see Appendix~\ref{AppLambDicke2}):
\begin{equation}
    U(t)=U_q(t)e^{-i\omega a^{\dagger}a t}e^{-i\left[\bar{H}_{\Rec}^{(1)}(t)+\bar{H}_{\Rec}^{(1)}(t)+\bar{H}_{\Ent}(t)\right]}+o(\eta^3),\label{EqUSecondOrder}
\end{equation}
with $\bar{H}_{\Rec}^{(1)}(t)=\eta a^{\dagger}V_{\Rec}^{(1)}(t)+h.c.$, $\bar{H}_{\Rec}^{(2)}(t)=-\tfrac{\eta^2}{2} {a^{\dagger}}^2V_{\Rec}^{(2)}(t)+h.c.$, $\bar{H}_{\Ent}(t)=-\eta^2a^{\dagger}a V_{\Ent}(t)$, and with:
\begin{equation}
\begin{aligned}
&\displaystyle\dot{U}_q(t)=h_q(t)\big(1-\tfrac{\eta^2}{2}\big)U_q(t),\\
&\displaystyle V_{\Rec}^{(1)}(t)=\int_0^TU_q^{\dagger}(t')h_p(t')U_q(t') e^{i\omega t'} dt,\\
&\displaystyle V_{\Rec}^{(2)}(t)=\int_{0}^t U_q^{\dagger}(t')h_q(t')U_q(t')e^{2i\omega t'} \; dt',\\
&\displaystyle V_{\Ent}(t)=\int_{0}^t U_q^{\dagger}(t')h_q(t')U_q(t')\; dt'.
\end{aligned}\label{EqOperators}
\end{equation}
In this scenario, the qubit-motion entanglement is not only induced by photon recoil but also by $a^{\dagger}a V_{\Ent}$.
Unlike $V_{\Rec}^{(1)}$ and $V_{\Rec}^{(2)}$, $V_{\Ent}$ does not vanish when $\omega/\Omega\to\infty$.
However, we will see in the next section that under the assumption $p_0=1$, it doesn't affect the gate fidelity. We can thus ignore it at the moment.

\begin{figure}[b]
    \centering
    \includegraphics[scale=0.4]{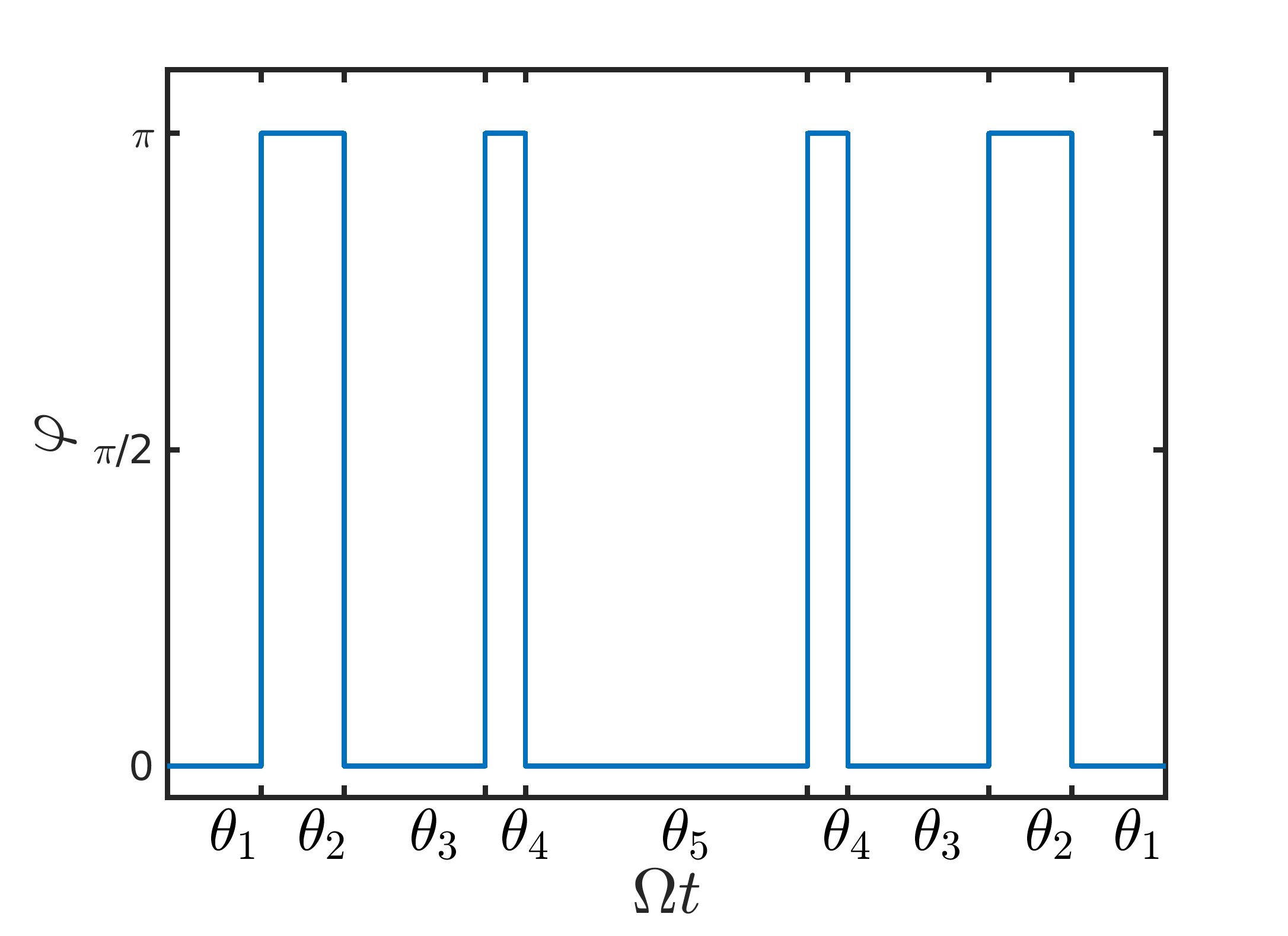}
    \includegraphics[scale=0.4]{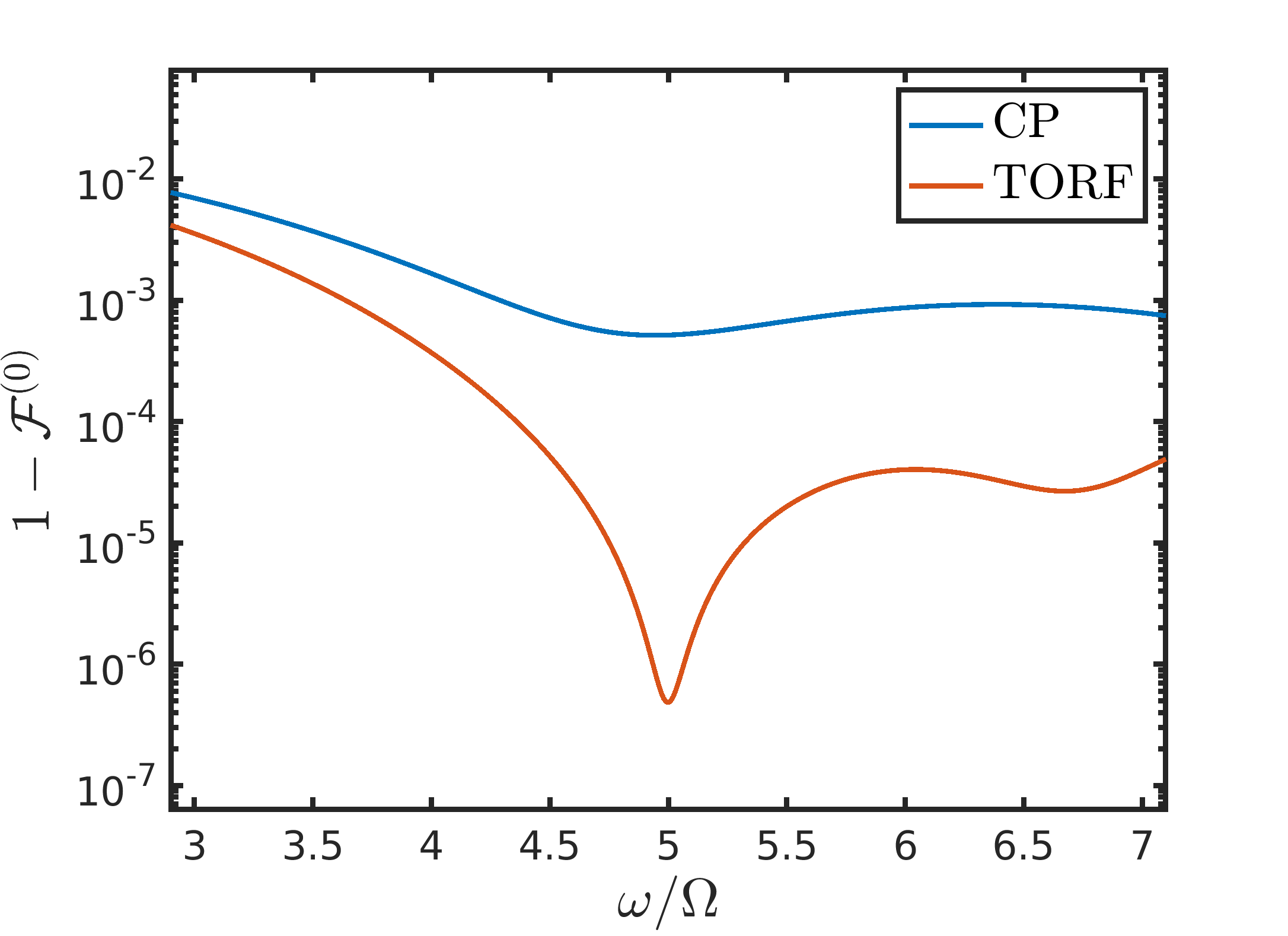}
    \caption{\emph{Left panel:} Structure of a second-order \TORF pulse. \emph{Right panel:} Error of a $\sqrt{\mathrm{NOT}}$ obtained using a second-order $\pi/2$-\TORF pulse designed for a ratio $\omega/\Omega=5$ as given in the text (red line), and of a constant pulse (CP) of duration $T=\pi/[2\Omega(1-\eta^2/2)]$ (blue line), assuming $p_0=1$.}
    \label{FigTORFo2}
\end{figure}

\paragraph*{\textbf{Second-order \TORF pulses.}}
A second-order recoil-free control problem is defined as:
\begin{equation}
\begin{cases}
U_q(T)=U_{\Target}\\
V_{\Rec}^{(1)}(T) =0\\
V_{\Rec}^{(2)}(T) =0.
\end{cases}\label{EqRecFreeControlO2}
\end{equation}
By solving the problem in minimum time using a gradient algorithm, we obtain the surprising result that the pulse is bang-bang and symmetric, but with more segments than in the Lanb-Dicke regime, as depicted in Fig.~\ref{FigTORFo2}.
For this kind of pulse, the duration is such that $\Omega T=2(\theta_1+\theta_2+\theta_3+\theta_4)+\theta_5$.
It is important to note that the qubit operations are slower than for a standard spin-$1/2$ by a factor of $1-\eta^2/2$.
As a consequence, to ensure $U_q(T)=U_{\Target}$, the following relation holds:
\[2(\theta_1+\theta_3-\theta_2-\theta_4)+\theta_5=\frac{\theta_{\Target}}{1-\frac{\eta^2}{2}}.\]
The value of these parameters depends on the target gate and the frequency ratio $\omega/\Omega$.
Remarkably, we can show that for a target NOT gate,
constant pulses of duration $T=\pi/[\Omega(1-\eta^2/2)]$ are solutions of the problem~\eqref{EqRecFreeControlO2} when $\omega/[\Omega(1-\eta^2/2)]$ is an odd integer. This implies that, under these conditions, constant pulses are recoil-free up to the second order in $\eta$, which explains the peaks visible in Fig.~\ref{FigRecoilO2}. However, as in the Lamb-Dicke regime, this property does not hold for any other qubit gate.

The right panel of figure~\ref{FigTORFo2} shows the fidelity of a $\sqrt{\mathrm{NOT}}$ gate using a second-order \TORF pulse designed for $\omega/\Omega=5$ compared to a constant pulse.
The \TORF pulse is such that $\theta_k=\{0.0589\pi,0.0313\pi,0.1015\pi,0.0097\pi,0.2729\pi\}$.
We can see that while the qubit gate error is $\sim 10^{-3}$ for the constant pulse, it reaches $\sim 10^{-6}$ using the \TORF one.
An error of $\sim 10^{-6}$ can also be reached with a constant pulse, but it must be applied very deep in the resolved sideband regime, i.e., using a frequency ratio of $\omega/\Omega\simeq 130$.

Let us convert this into proper units.
For a trap frequency of $\omega/(2\pi)=100~\mathrm{kHz}$, a ratio of $130$ corresponds to a Rabi frequency of $\Omega/(2\pi)=770~\mathrm{Hz}$, resulting in a duration of $T=\pi/[2\Omega(1-\eta^2/2)]=332~\mu$s for the constant pulse.
In comparison, the \TORF pulse is designed for $\Omega/(2\pi)=20~$kHz and has a duration of $T=0.6751\pi/\Omega=16.89~\mu$s.
In other words, the \TORF pulse allows one to reach the same fidelity with a $\sim 20$ times shorter pulse! 

\section{Thermal motion-induced entanglement\label{SecDisentangling}}
\paragraph*{\textbf{Consequences of imperfect cooling.}} Until now, we have considered the atom to be fully in the motional ground state by setting $p_0=1$ in the definition~\eqref{EqJGate}.
This assumption was made to capture the main effects coming from photon recoil but is not true in practice because of cooling imperfections. The probability $p_0$ is related to the temperature of the atoms through the formula:
\begin{equation}
    p_0=1-e^{-\frac{\hbar\omega}{k_B\hat{T}}},\label{Eqp02t}
\end{equation}
where $k_B$ is the Boltzmann constant and $\hat{T}$ is the temperature.
Currently, $\Sr88$ atoms are cooled up to $p_0\simeq 0.98$~\cite{PhysRevLett.133.013401}, corresponding to a temperature of $1.2~\mu$K using a $100$~kHz trap frequency.

\begin{figure}[b]
    \centering
    \includegraphics[scale=0.4]{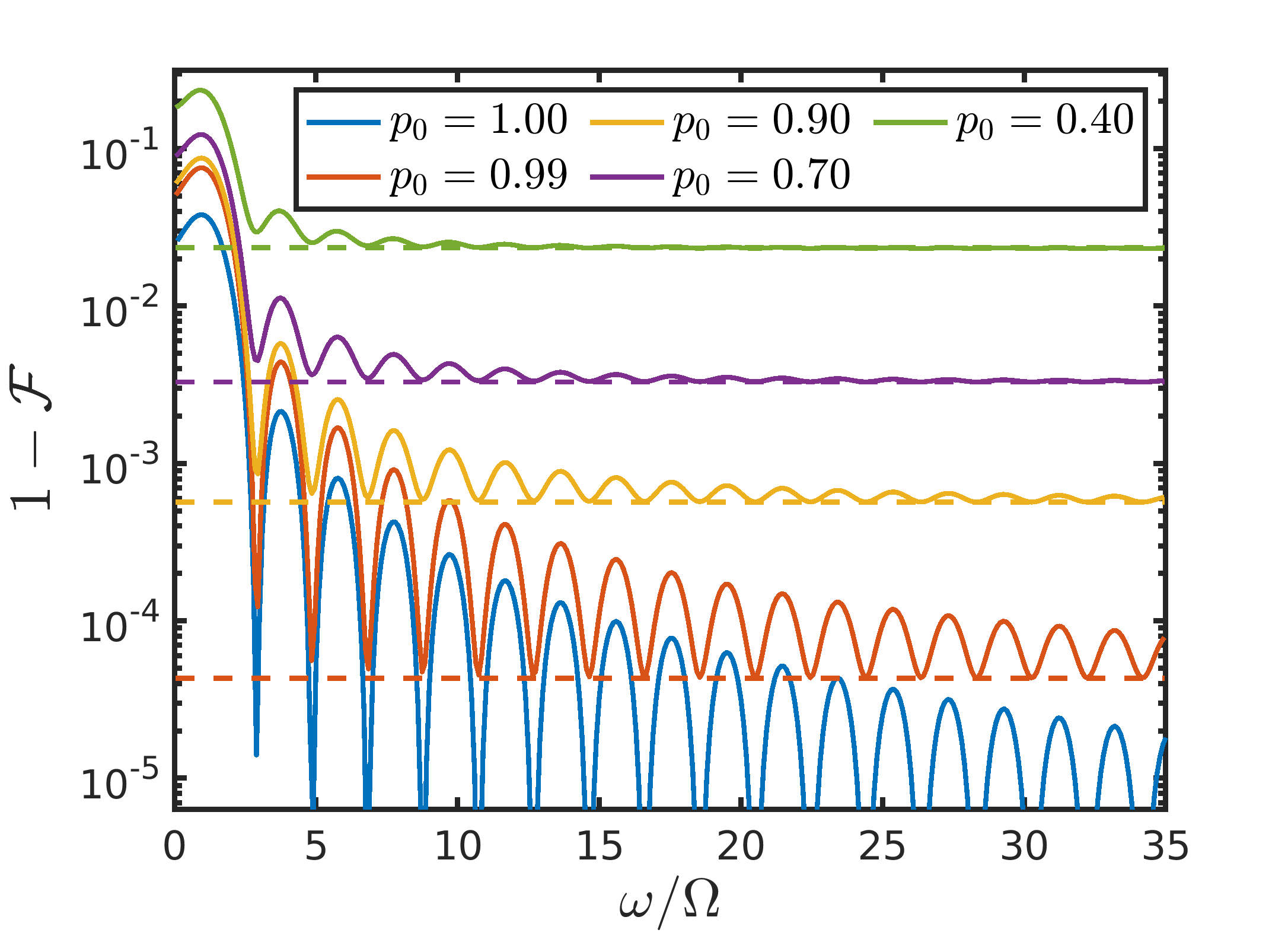}
    \includegraphics[scale=0.4]{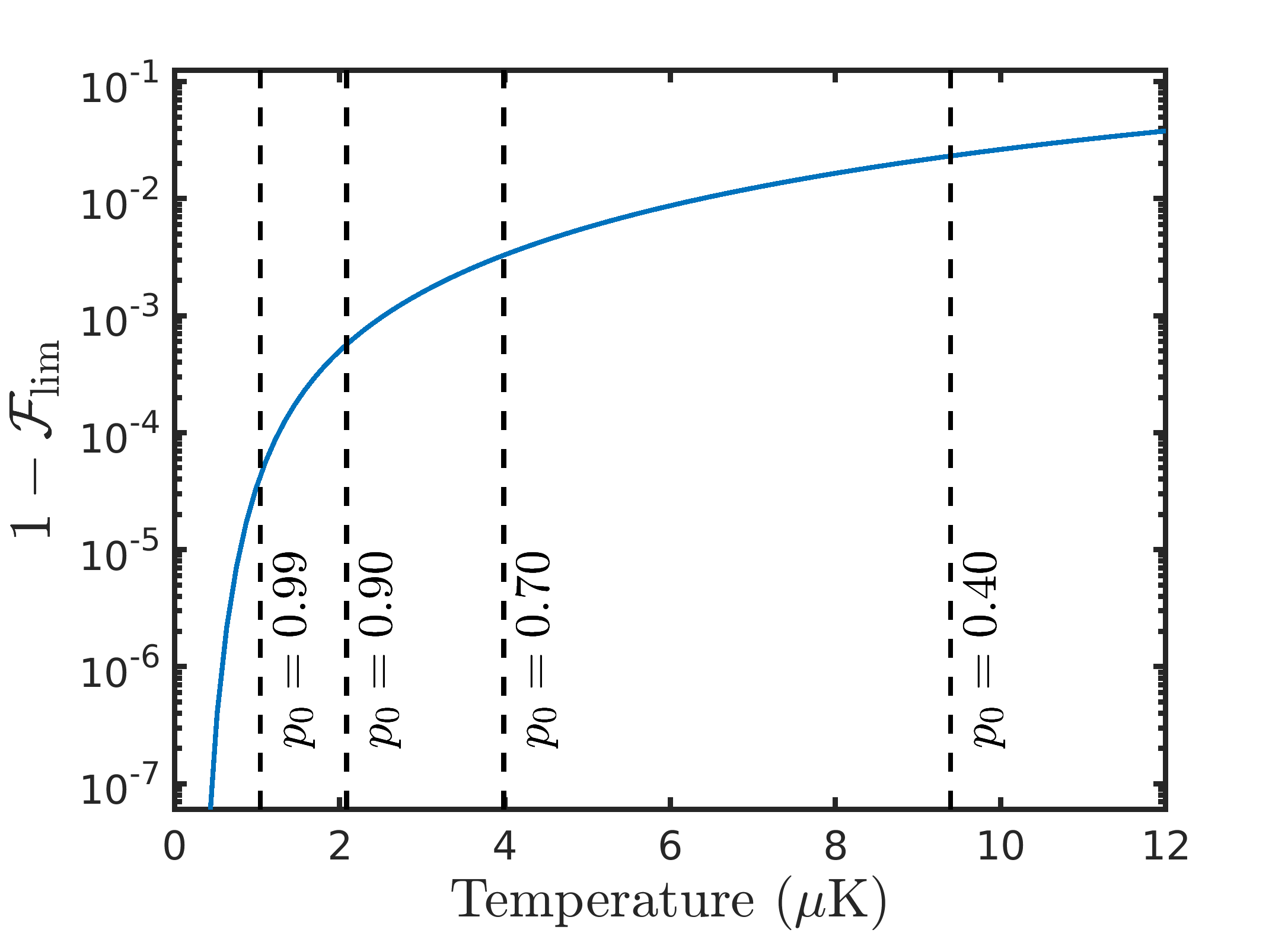}
    \caption{\emph{Left panel:} Error of a NOT gate using a constant pulse of duration $T=\pi/[\Omega(1-\eta^2/2)]$ as a function of $\omega/\Omega$ for various ground state probabilities. The dashed lines depict the limit $1-\mathcal{F}_{\lim}$ given in Eq.~\eqref{EqJlim2Temp}. The system is simulated using the Hamiltonian~\eqref{EqHamO2}. \emph{Right panel:} 
    Limit $1-\mathcal{F}_{\lim}$ computed for a target NOT gate as a function of the temperature using a $100$~kHz trap frequency.\label{FigRecoilTemp}}
\end{figure}

Figure~\ref{FigRecoilTemp} shows the gate error $1-\mathcal{F}$ for a target NOT gate using a constant pulse of duration $T=\pi/[\Omega(1-\eta^2/2)]$ as a function of $\omega/\Omega$ and for various $p_0$. Remember that this pulse is recoil free when  $\omega/[\Omega(1-\eta^2/2)]$ is an odd number.

We can see that the gate error doesn't drop below a threshold, which is reached when the recoil is suppressed, i.e. for $\omega/\Omega\simeq 2k+1$. 
Remarkably, operating infinitely deep in the resolved sideband regime doesn't mitigate this additional error.
Nevertheless, it vanishes when $p_0\to 1$, i.e., for an atom cooled to absolute zero temperature.

This limit arises from entanglement induced by thermal motion of the atom.
It can be derived by expressing the evolution operator~\eqref{EqUSecondOrder} under a recoil-free pulse, such as a second-order \TORF pulse or a constant one applied deep in the resolved sideband regime, and then deriving the corresponding fidelity.
If the two recoil-free conditions are fulfilled and if the target is reached in the qubit subspace,
the evolution operator at the final time becomes:
\begin{equation}
    U(T)= U_{\Target}e^{-i\eta^2a^{\dagger}a V_{\Ent}(T)}+o(\eta^3),\label{EqUTEnt}
\end{equation}
where we ignored the term $e^{-i\omega a^{\dagger}a T}$ since it does not affect the fidelity. 
Bang-bang or constant controls involve rotations only about the $x$ axis in the qubit subspace, implying that $h_q(t)$ and $U_q(t)$ commute. 
Therefore, $V_{\Ent}$ can be derived analytically using $V_{\Ent}(T)=\int_0^Th_q(t)dt$. We obtain:
\[V_{\Ent}(T)=\frac{\theta_{\Target}}{1-\frac{\eta^2}{2}}\frac{\sigma_x}{2}.\]
This equation holds for any of the recoil-free pulses mentioned above. 
As shown in the appendix~\ref{AppFlim}, substituting this expression in Eq.~\eqref{EqUTEnt} and computing the fidelity using Eq.~\eqref{EqJGate}, we obtain that the gate error is bounded by:
\begin{equation}
    1-\mathcal{F}_{\lim}=\frac{3}{16}\frac{(1-p_0)(2-p_0)\eta^4\theta_{\Target}^2}{p_0^2}+o(\eta^8).\label{EqJlim2Temp}
\end{equation}

For a given target gate and Lamb-Dicke parameter, $\mathcal{F}_{\lim}$ is fully determined by the ground state probability. 
Crucially, it is independent of the Rabi frequency, meaning that operating in the resolved sideband regime doesn't mitigate entanglement induced by thermal motion.
Furthermore, it always holds when the qubit Hamiltonian $h_q(t)$ commutes with the qubit operator $U_q(t)$, which is true for all pulses applied along a single direction of the Bloch sphere.
Therefore, an advanced phase modulation of the driving laser is essential to overcome this limit.
Besides, $1-\mathcal{F}_{\lim}$ is proportional to $\eta^4$, indicating that qubits with smaller $\eta$, like hyperfine or fine-structure qubits, are significantly less affected by thermal motion.

\begin{figure}[b]
    \centering
    \includegraphics[scale=0.36]{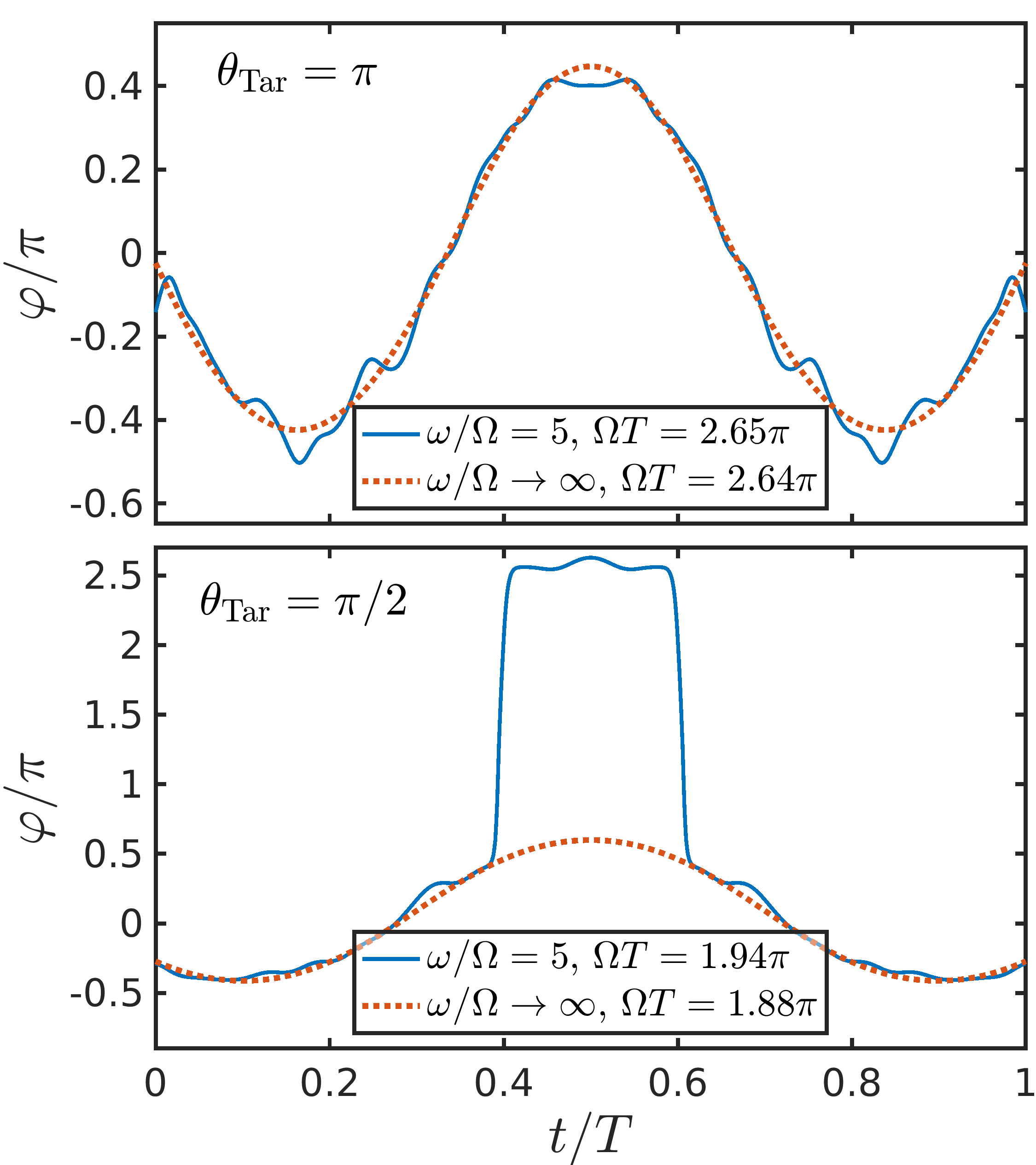}
    \caption{\emph{Upper panel:} TOD pulses realizing a NOT gate. \emph{Lower panel:} TOD pulses realizing a $\sqrt{\mathrm{NOT}}$ gate. The case $\omega/\Omega\to\infty$ is optimized by ignoring the recoil-free constraints in the problem~\eqref{EqDisentControl}. It can be used deep in the resolved sideband regime.
    \label{FigTOD}}
\end{figure}

We can see the good accuracy of this bound on the left panel of Fig.~\ref{FigRecoilTemp}, even for relatively low $p_0$. 
The right panel displays it as a function of the temperature, which is obtained by substituting~\eqref{Eqp02t} in~\eqref{EqJlim2Temp}. It shows that, at $2~\mu$K, the gate error is already bounded by $\sim 10^{-3}$, which is significantly high, especially for a theoretical limit.
\paragraph*{\textbf{Disentangling pulses.}} 
A pulse suppresses all sources of entanglement if it is solution of the control problem:
\begin{equation}
    \begin{cases}
        U_q(T)=U_{\Target}\\
        V_{\Rec}^{(1)}(T)=0\\
        V_{\Rec}^{(2)}(T)=0\\
        V_{\Ent}(T)=0,\label{EqDisentControl}
    \end{cases}
\end{equation}
where these operators are given in Eq.~\eqref{EqOperators}.
The time-optimal solutions of this problem consist of pulses with smoothly shaped phases that are difficult to understand intuitively.
We call them TOD (Time Optimal Disentangling) pulses.
The figure~\ref{FigTOD} shows TOD pulses realizing NOT and $\sqrt{\mathrm{NOT}}$ gates as a function of time, designed for $\omega/\Omega=5$ and $\omega/\Omega=\infty$.
In the latter case, we ignored the constraints $V_{\Rec}^{(1)}$ and $V_{\Rec}^{(2)}$ in the control problem~\eqref{EqDisentControl} in order to highlight their role in the pulse shape.
These solutions can be applied deep in the resolved sideband regime where these operators naturally vanish.

Note that these pulses have a much higher duration than the \TORF pulses.
For example, for $\omega/(2\pi)=100$~kHz and $\Omega/(2\pi)=20$~kHz, a $\pi/2$-\TORF pulse requires a duration $T=16.89~\mu$s, while a $\pi/2$-TOD pulse lasts $T=48.25~\mu$s, which is almost $3$ times longer. It is, however, significantly shorter than a constant pulse applied deep in the resolved sideband regime, while this latter doesn't eliminate entanglement induced by thermal motion.

\begin{figure}[h]
    \centering
    \includegraphics[scale=0.35]{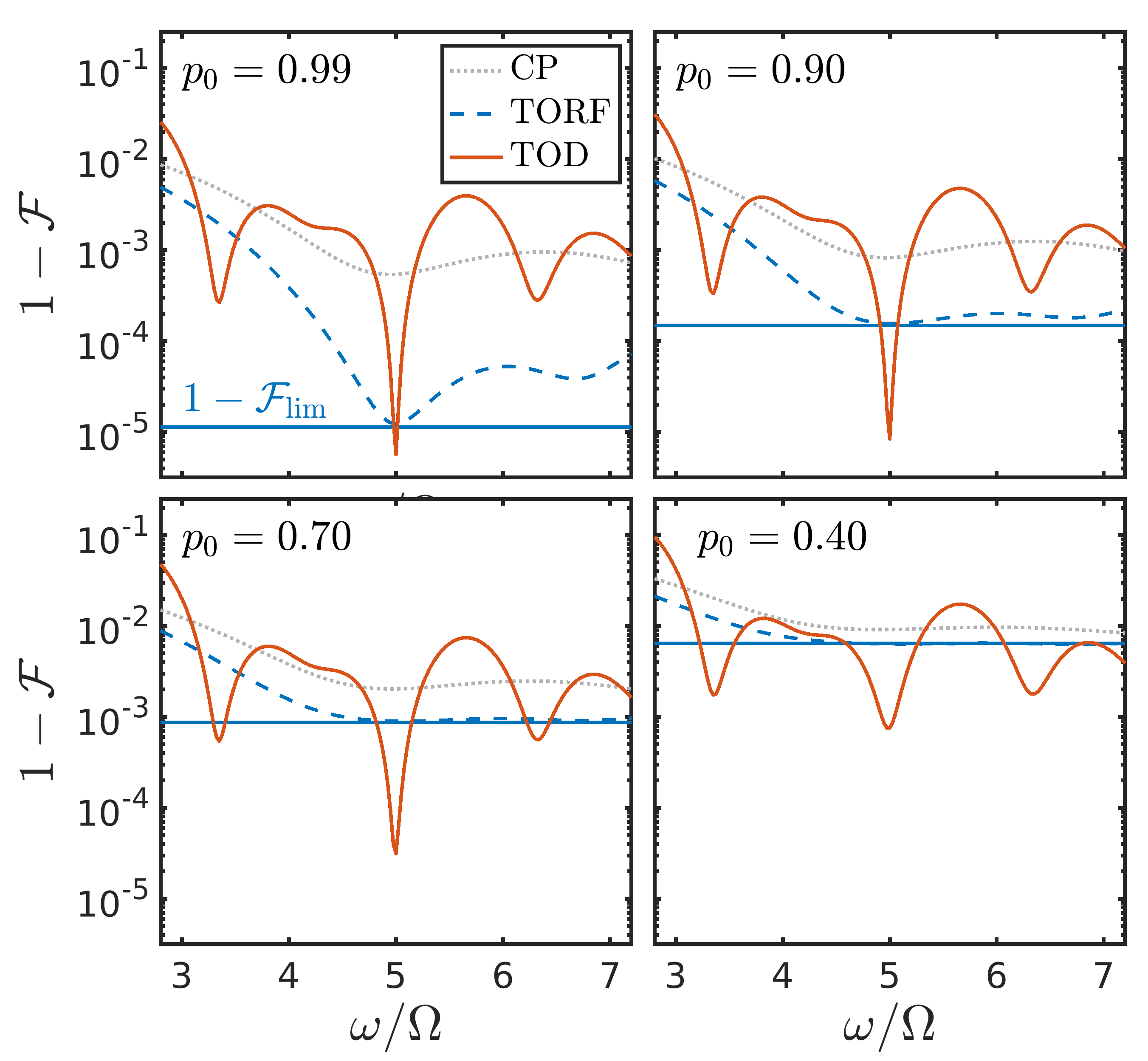}
    \caption{Error of a $\sqrt{\mathrm{NOT}}$ gate using a constant $\pi/2$-pulse (CP) of duration $T=\pi/[2\Omega(1-\eta^2/2)]$ (light dotted line), a second-order $\pi/2$-\TORF pulse designed for $\omega/\Omega=5$ (dashed blue line), and the $\pi/2$-TOD pulse depicted in Fig~\ref{FigTOD}, also designed for $\omega/\Omega=5$  (solid red line), for various ground state probabilities. The solid blue line shows the limit of recoil-free gates given in Eq.~\eqref{EqJlim2Temp}.}
    \label{FigFidTOD}
\end{figure}

Figure~\ref{FigFidTOD} shows the performance of a TOD and a second-order \TORF pulse for various valued of $p_0$. 
We observe that when $p_0$ is close to $1$, the TOD pulse barely outperforms the \TORF one. This is expected because, in this case, the qubit-motion entanglement is primarily attributed to photon recoil, which is already suppressed by the \TORF pulse.
When $p_0$ decreases, however, they allow for a significant improvement of the gate fidelity. For a typical thermal atom, where $p_0\simeq 0.9$, they improve the fidelity by more than one order of magnitude beyond the limit $\mathcal{F}_{\lim}$ reached by recoil-free pulses.
\section{Robust motion-insensitive pulses\label{SecRobust}}
In practice, the driving laser field is not homogeneous across the array of atoms.
It mainly involves detuning induced by probe shift and Rabi frequency deviations, which have a significant impact on the gate fidelity.
They can be compensated by shaping the pulse phase $\varphi(t)$ in order to improve its robustness.
Robust control applied to inhomogeneous ensembles of spin-1/2 particles has been thoroughly studied in the past decades using all kinds of pulse engineering methods~\cite{dridiRobustControlNot2020,khanejaOptimalControlCoupled2005, skinnerOptimalControlDesign2010, dionisTimeoptimalControlTwolevel2023,genovCorrectionArbitraryField2014,vandammeRobustOptimalControl2017, wuCompositePulsesOptimal2023, barnesDynamicallyCorrectedGates2022, nelsonDesigningDynamicallyCorrected2023,Ansel_2024,Koch:909195}.
All these methods do not apply directly here because we need to account for the effects of atom motion, but they can be adapted by including the constraints given in Eq.~\eqref{EqDisentControl}. 
An interesting technique consists of designing composite-\TORF pulses.
The idea is to use composite pulses from the literature (or to derive them), designed for ideal spin-1/2 particles, and to replace the segments causing recoil by \TORF pulses, with the correct phases.
The method is efficient and very handy, but composite pulses are inherently slow compared to other optimal control techniques;
we prefer to highlight a method for designing smooth-shaped pulses based on time-optimal control. 

The detuning and Rabi frequency deviations act on the qubit subspace and can be taken into account in the model by redefining the qubit Hamiltonian $h_q$ given in Eq.~\eqref{EqHQHP} as:
\begin{equation}
\tilde{h}_q(t)=\delta\Delta\tfrac{\sigma_z}{2}+\left(1+\tfrac{\delta\Omega}{\Omega}\right)h_q(t).\label{EqHQInhom}
\end{equation}
To compensate for these inhomogeneities, we once again use Average Hamiltonian Theory~\cite{Haeberlen68,BrinkmannAVH}, which allows us to design pulses that are insensitive to first-order effects of detuning and Rabi frequency deviations.
We can show that the qubit operator under $\tilde{h}_q$ is given by:
\begin{equation}
    \tilde{U}_q(t)=U_q(t)e^{-i\tfrac{\delta\Delta}{\Omega} V_{\Det}(t)}e^{-i\tfrac{\delta\Omega}{\Omega}V_{\Rab}(t)}+o\left(\left(\tfrac{\delta\Delta}{\Omega}\right)^2,\left(\tfrac{\delta\Omega}{\Omega}\right)^2,\left(\tfrac{\delta\Delta\delta\Omega}{\Omega^2}\right)\right),
\end{equation}
where $U_q$ is the homogeneous qubit operator given in Eq.~\eqref{EqOperators}, while $V_{\Det}$ and $V_{\Rab}$ describe the perturbations induced by the inhomogeneities.
They are given by: 
\begin{equation}
    \begin{aligned}
    & V_{\Det}(t)=\tfrac{\Omega}{2}\int_0^tU_q^{\dagger}(t')\sigma_zU_q(t') dt',\\
    & V_{\Rab}(t)=\int_0^tU_q^{\dagger}(t')h_q(t')U_q(t') dt'\equiv V_{\Ent}(t).
    \end{aligned}\label{EqUDetUAmp}
\end{equation}
Surprisingly, the expression of $V_{\Rab}$ is exactly the same as $V_{\Ent}$ (c.f Eq.~\eqref{EqOperators}).
It implies that a robust pulse, designed for solving $V_{\Rab}(T)=0$, naturally mitigates entanglement induced by thermal motion.

In total, a robust motion-insensitive pulse can be designed by shaping the phase $\varphi(t)$ of the driving laser in order to solve the control problem:
\begin{equation}
\begin{cases}
U_q(T)=U_{\Target}\\
V_{\Rec}^{(1)}(T)=0\\ 
V_{\Rec}^{(2)}(T)=0\\ 
V_{\Det}(T)=0\\
V_{\Rab}(T)=0\equiv V_{\Ent}(T),\label{EqControlRobust}
\end{cases}
\end{equation}
where $V_{\Rec}^{(k)}$ and $U_q$ are given in Eq.~\eqref{EqOperators}. 

In this context, $U_q(T)=U_{\Target}$ ensures that the desired qubit gate is realized, $V_{\Rec}(T)=0$ and $V_{\Rec}^{(2)}(T) = 0$ ensure that the gate is recoil-free, $V_{\Det}(T)=0$ ensures robustness against detuning, and $V_{\Rab}(T) = V_{\Ent}(T) = 0$ ensures robustness against Rabi frequency deviations and insensitivity to thermal motion-induced entanglement. We use time-optimal control to solve this problem. The solutions are referred to as Robust-MI (Motion-Insensitive) pulses.

If we ignore the recoil-free constraints, the problem is already well-known;
the expression of $V_{\Det}$ and $V_{\Rab}$ can be found, for example, in Ref.~\cite{barnesDynamicallyCorrectedGates2022}. Several strategies have been developed to compensate for these types of inhomogeneities in two-level quantum systems.
In our framework, this problem corresponds to the scenario where $\omega/\Omega\to\infty$, since in this case the recoil operators are naturally vanishing.
A pulse optimized by ignoring the recoil-free constraints can thus be applied deep in the resolved sideband regime with good accuracy.
We will refer to this type of pulse as classical robust pulses---despite not finding the exact phase shapes in the literature, as this specific problem may not have been solved using time-optimal control---because they address a well-known quantum optimal control problem.

\section{Simulations on an array of trapped Strontium atoms\label{SecSimu}} This section is aimed at evaluating the performances of our techniques using the full Hamiltonian given in Eq.~\eqref{EqHTotal} for the optical transition of $\Sr88$ atoms discussed in the model section.
The detuning and the Rabi frequency deviations are assumed to be constant over time.
The laser being inhomogeneous across the array of atoms, each of them experiences a certain detuning and Rabi frequency.
We assume that the trapping lasers are homogeneous.
The dynamics of an atom in the array is described by the Hamiltonian:
\begin{equation}
H(\delta\Delta,\delta\Omega,t)=\delta\Delta\ket{e}\bra{e}+\left.\left.\frac{\Omega+\delta\Omega}{2}\right(\ket{e}\bra{g} e^{i\eta (a^{\dagger}+a)+i\varphi(t)}+h.c.\right)+\omega a^{\dagger}a,\label{EqHTotInhom}
\end{equation}
where $\omega/{2\pi}=100$~kHz and $\eta=0.2156$. 

\begin{figure}[t]
\includegraphics[width=0.85\linewidth]{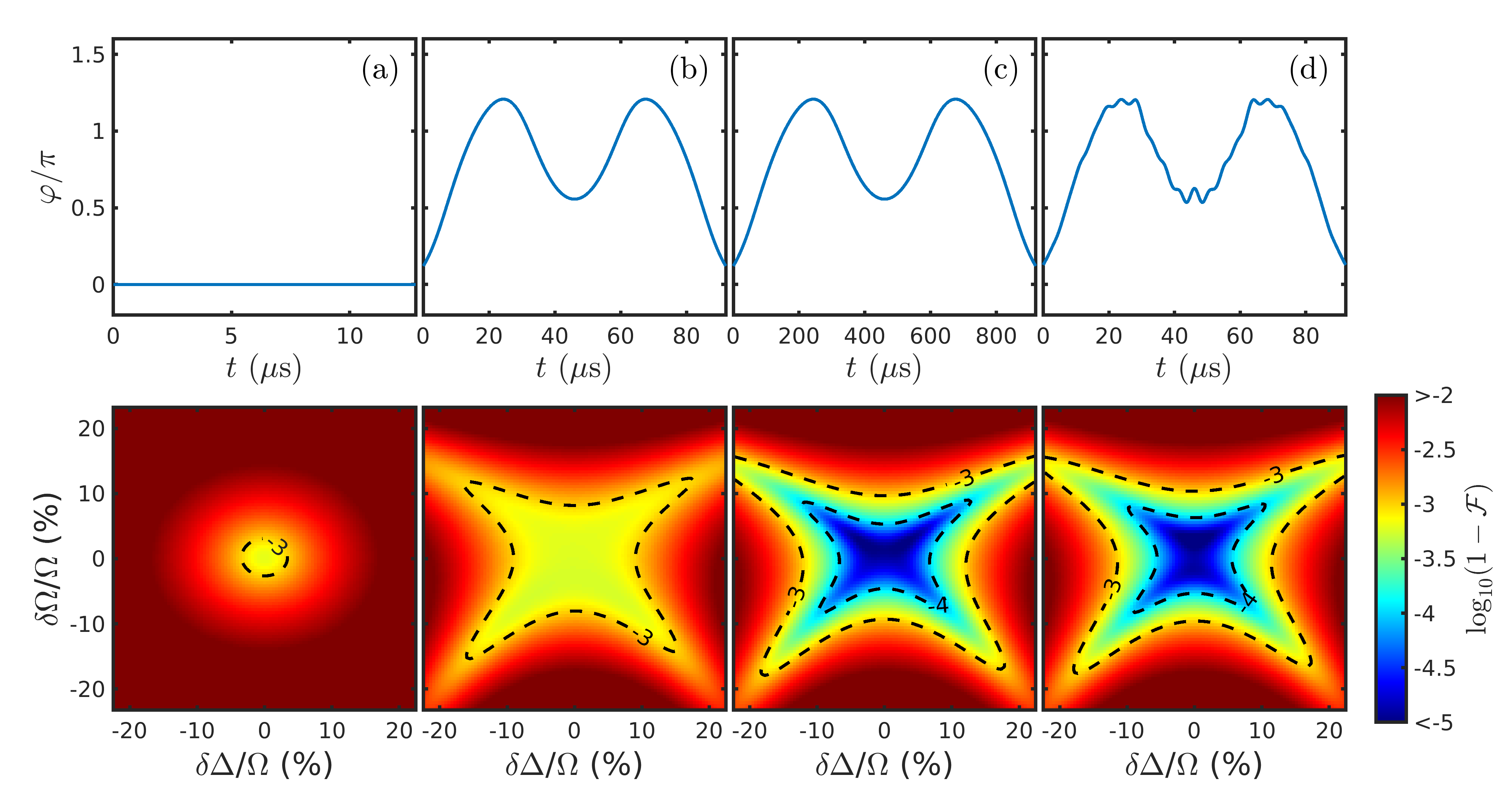}
\caption{Pulse phase as a function of time (upper panels) and error computed for a target $\sqrt{\mathrm{NOT}}$ gate (lower panels) using (a) a constant pulse of duration $T=\pi/[2\Omega(1-\eta^2/2)]$ applied at Rabi frequency $\Omega/(2\pi)=20$~kHz, (b) a classical robust $\pi/2$-pulse also applied at $\Omega/(2\pi)=20$~kHz, (c) a classical robust $\pi/2$-pulse applied deep in the resolved sideband regime at $\Omega/(2\pi)=2$~kHz, and (d) a robust-MI $\pi/2$-pulse solution of the problem~\eqref{EqControlRobust} applied at $\Omega/(2\pi)=20$~kHz. The system is simulated using the Hamiltonian~\eqref{EqHTotInhom} and the gate error is computed using $p_0=0.95$.\label{FigRobustExc}}
\end{figure}

The robustness of a pulse is evaluated by measuring the gate error $1-\mathcal{F}$ as a function of the detuning and the Rabi frequency deviations.
Ideally, the error should be zero for all pairs $(\delta\Delta,\delta\Omega)$, but this is unrealistic.
A robust pulse is such that $1-\mathcal{F}\simeq 0$ for a relatively wide range of inhomogeneities, which can be visualised by plotting $1-\mathcal{F}$ in the plane $(\delta\Delta,\delta\Omega)$.
In all the following results, $\mathcal{F}$ is calculated using a ground state probability $p_0=0.95$.
Figure~\ref{FigRobustExc} shows the error of a $\sqrt{{\mathrm{NOT}}}$ gate using various pulses.

First of all, we observe that the constant pulse (column (a)) exhibits a relatively high gate error ($\sim 10^{-3}$), even at the center of the plane where the system is homogeneous. 
This error arises from photon recoil and thermal motion-induced entanglement, which are not mitigated by such a pulse. Furthermore, the region where the pulse performs relatively well ($\leq 10^{-3}$) is small, as this pulse is not robust.

In column (b), we apply a classical robust pulse, but not deep in the resolved sideband regime.
Compared to the constant pulse, we observe that it is much more robust, but the fidelity is still limited by qubit-motion entanglement.
When applied deep in the resolved sideband regime (column (c)), the classical robust pulse performs significantly better because, in this case, photon recoil is naturally eliminated, and thermal motion-induced decoherence is mitigated by the pulse's robustness against Rabi frequency deviations.
However, it is excessively long ($T=920.5\mu s$).
Moreover, this representation is advantageous for this pulse because the range of inhomogeneities is normalized by $\Omega$, which is smaller than that of the other pulses.
It means that it is evaluated over smaller Rabi frequency deviations and detuning in absolute value.

In contrast, the robust-MI pulse (column (d)) shows the same performance with a duration that is 10 times shorter. 
Note that pulses (b) and (d) are very similar in shape and duration.
It is quite remarkable that tiny differences in the pulse shape play such a crucial role in suppressing the photon recoil.

\begin{figure}[t]
\includegraphics[scale=0.4]{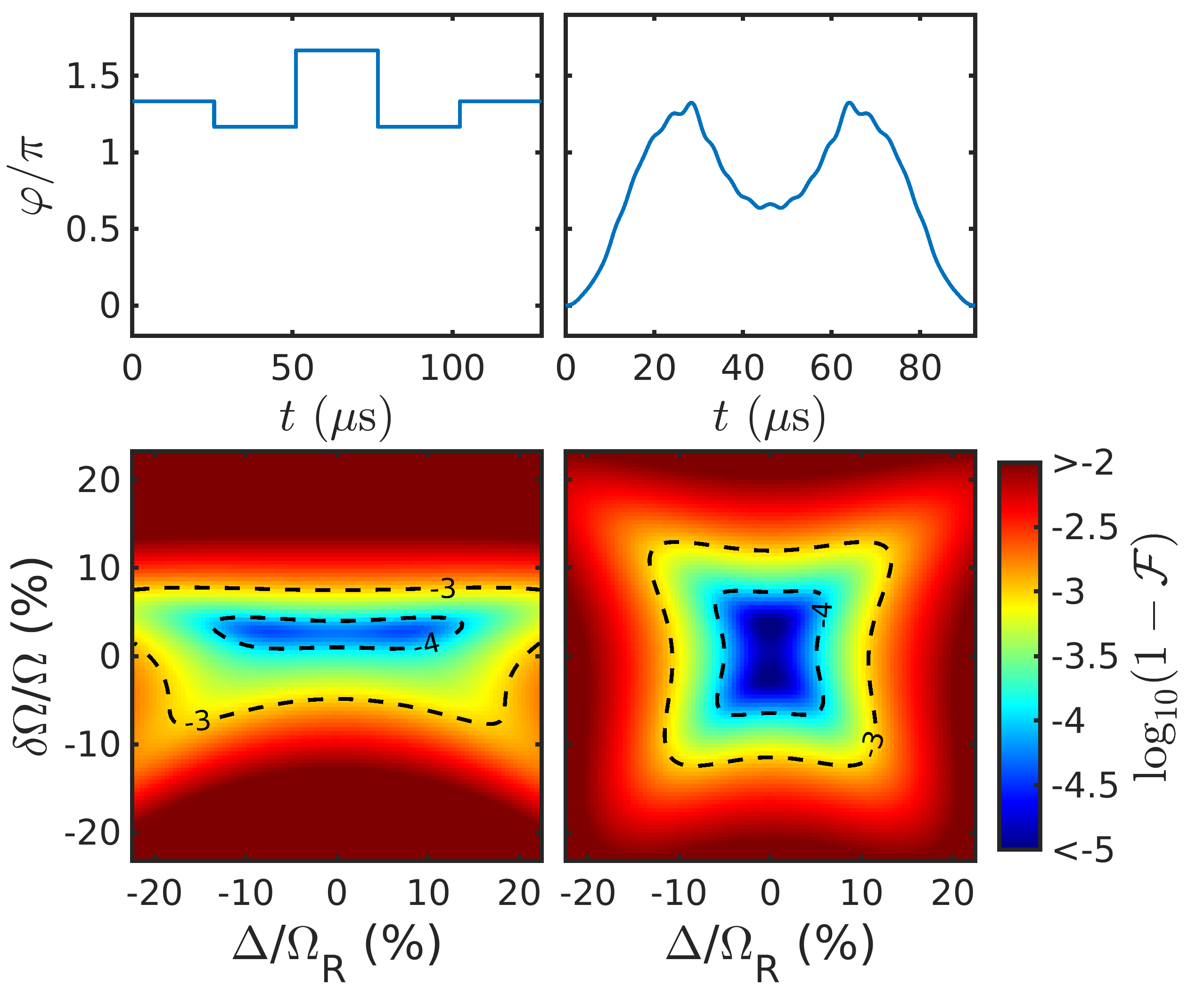}
\caption{Pulse phase as a function of time (upper panels) and error computed for a target NOT gate (lower panels) using the composite pulse mentioned in the text (left), and a robust-MI $\pi$-pulse (right). The robust-MI pulse is applied at $20$~kHz Rabi frequency, and the composite pulse is applied at $20.48$~kHz.
\label{FigRobustInv}}
\end{figure}

In the particular case of a NOT gate, and for a ratio of $\omega/\Omega=5$ ($\Omega=2\pi\times(20$~kHz)), the control problem~\eqref{EqControlRobust} can be solved using composite pulses made of constant $\pi$-subpulses with different phases.
Indeed, as we have shown e.g.
in Fig.~\ref{FigRecoilO2}, photon recoil is suppressed in these conditions, meaning that each subpulse is recoil-free.
However, we need to compensate for the factor $1-\eta^2/2$ on the qubit subspace (see Sec.~\ref{SecLambDicke2}), which can be done by using a slightly higher Rabi frequency $\tilde{\Omega}=\Omega/[1-\eta^2/2]$, i.e.
$\tilde{\Omega}/(2\pi)=20.48$~kHz, and a duration of $\pi/\Omega=25~\mu$s for each subpulse.
We can then use any composite $\pi$-pulse from the literature that improves the robustness against detuning and Rabi frequency deviations.
One of the pulses of~Ref.~\cite{jonesDesigningShortRobust2013} is precisely designed to satisfy the robustness conditions $V_{\Det}$ and $V_{\Rab}$ (Eq.~\eqref{EqUDetUAmp}).
It is given in their table 1 (row 5(a)).
It is made of 5 $\pi$-subpulses of phase $\phi_{k}=\{240^{\circ},210^{\circ},300^{\circ},210^{\circ},240^{\circ}\}$ and has a duration of $T=5\pi/\Omega=125~\mu$s. In contrast, a robust-MI $\pi$-pulse has a duration
$T=92.5~\mu$s when applied at $20$~kHz Rabi frequency.
Figure~\ref{FigRobustInv} shows the two pulses and their performances.

We observe that the composite pulse performs relatively well.
The gate error remains below $10^{-4}$ over a wide range of $\delta\Delta/\Omega$, though not so much along the vertical axis.
In contrast, the fidelity of the robust-MI pulse is well balanced and reaches an error of $10^{-5}$ in the center of the plane, while the pulse is shorter.
It shows that our robust, time optimal motion-insensitive pulses, even if they have more complex shapes, perform much better than other intuitive approaches.
\section{Discussion\label{SecDiscussion}}
In trapped atom hardware, the atom motion affects the gate fidelity via two mechanisms: photon recoil and thermal motion-induced entanglement.
They are particularly limiting for controlling optical qubits with high fidelity because the coupling between the qubit and the motion of the particle is significant in these systems.

We have developed a mathematical framework based on Average Hamiltonian Theory to formulate motion-insensitive control problems involving recoil-free (Eqs.~\eqref{EqRecFreeControl} and~\eqref{EqRecFreeControlO2}) and disentangling (Eq.~\eqref{EqDisentControl}) constraints, which we have solved using time-optimal control.

When only photon-recoil is considered, the optimal solutions consist of fast bang-bang pulses.
Compared to ordinary pulses, which achieve high gate fidelity only deep in the resolved sideband regime, these pulses can operate at relatively high Rabi frequencies, reducing the gate duration by up to a factor 20 while maintaining the same fidelity.

However, recoil-free pulses do not reduce the gate error below a fundamental threshold caused by thermal motion-induced entanglement. 
This threshold can be significantly high for optical qubits if they are not cooled very close to $p_0=1$ (where $p_0$ is the probability of the motional ground state).
It indicates that for optical qubits, cooling is significantly more critical compared to hyperfine or fine-structure qubits.

Nevertheless, smooth-phase pulses can lower this limit further if they fulfill the disentangling constraints.
The gain obtained with these pulses can be significant depending on the ground state population. For a typical thermal atom, i.e., such that $p_0\simeq 0.9$, the gate error can be reduced by more than an order of magnitude below this limit.
Despite this, enhancing cooling is a more effective solution for optical qubits because, in this case, photon recoil is the primary source of error and can be addressed very efficiently.

The method is very flexible and can be used to design motion-insensitive pulses capable of compensating for laser inhomogeneities.
They show excellent performances in the simulations.

This work can be expanded in several directions.

First, several other types of inhomogeneities can be taken into account.
For example, one can derive an additional condition to enhance the robustness against inhomogeneities of the trap frequency (see appendix~\ref{AppMiscellaneous}).
It becomes particularly meaningful when the pulse duration becomes too long, as it may render the pulse efficient within a narrow range around the desired trap frequency.
Subsequently, the method easily accommodates additional experimental effects, such as the Stark shift and time-varying Rabi frequencies.

Second, all motion-insensitive and robust control problems can be solved using various pulse schemes based on different techniques, such as composite pulses, inverse engineering, or analytically shaped pulses. The latter may be more suitable for designing phase profiles that are easily implementable by the hardware.

Third, the duration of robust pulses can be considerably reduced using local addressing laser beams performing rotations around the $z$-axis of the Bloch sphere, together with global Mikado pulses applied to all the atoms in the array~\cite{Zhang2024}.

Finally, the method may also be valuable for two-qubit gates, where our mathematical framework can be used to isolate the two-qubit space and express the sources of error as perturbative operators.
\newpage

\begin{acknowledgments}
The project is part of the Munich Quantum Valley, supported by the Bavarian state government with funds from the Hightech Agenda Bayern Plus. We aknowledge support from the Munich Center for Quantum Science and Technology. S.J.G. and L.V.D. acknowledge support from the Deutsche Forschungsgemeinschaft (DFG, German Research Foundation) under Germanys Excellence Strategy, Grant No. EXC-2111390814868. L.V.D. aknowledges Josepe del Rio Vega for insightful discussions.
\end{acknowledgments}

\appendix
\section{Average Hamiltonian theory in the toggling frame\label{AppPerturbTheory}}
\paragraph*{\textbf{Basics.}}
The average Hamiltonian theory (AVH)~\cite{Haeberlen68,BrinkmannAVH} allows one to approximate the dynamics of an evolution operator under a time-dependent Hamiltonian.
Given $\dot{U}=-iHU$ where $H$ is a given hermitian Hamiltonian, the AVH states that $U(t)$ is given by:
\[U(t)=e^{-i\bar{H}(t)},\]
where $\bar{H}$ is the average (or effective) Hamiltonian given by an infinite series whose two first terms are given by:
\[\bar{H}(t)=\int_0^tH(t')dt'+\tfrac{1}{2i}\int_0^t\int_{0}^{t'}[H(t'),H(t'')]dt''dt'+\cdots \]
The first order terms is enough to approximate the exact evolution operator if $|H|T\ll \pi$, where $|\bullet|$ denotes the Frobenius norm.
\paragraph*{\textbf{AVH in the toggling frame.}}
When the Hamiltonian is the sum of two parts such that:
\begin{equation}
    H(t)=H_0(t)+H_1(t),
    \label{EqAppH0pH1}
\end{equation}
it is common to apply the AVH in the \emph{Toggling Frame}, which represents a frame associated to an operator $U_0$ solution of $\dot{U}_0=-iH_0U_0$.
To express the system in this new frame, we define an interaction operator $U_{\mathcal{I}}=U_0^{\dagger}U$.
Expressing the derivative of $U_{\mathcal{I}}$ leads to:
\[\begin{aligned}
   \dot{U}_{\mathcal{I}}(t)&=-i U_0^{\dagger}(t)H_{1}(t)U_0(t)U_{\mathcal{I}}(t)\\
   &\equiv-iH_{\mathcal{I}}(t)U_{\mathcal{I}}(t),
\end{aligned}\]
where $H_{\mathcal{I}}(t)$ is the interaction Hamiltonian.
$U_{\mathcal{I}}$ can thus be approximated using the AVH.
In this paper, we always use the first order of the AVH, leading to:
\begin{equation}
    U_{\mathcal{I}}(t)=e^{-i\int_0^t H_{\mathcal{I}}(t')dt'}.\label{EqAppUI}
\end{equation}
In total, the evolution operator of the system is thus $U(t)=U_0(t)U_{\mathcal{I}}(t)$, and is valid if $|H_{\mathcal{I}}|T\ll \pi$.
\paragraph*{\textbf{Validity for trapped particles.}}
Using the AVH in the toggling frame is more relevant since $|H_{\mathcal{I}}|=|H_1|$ is proportional to the perturbation ($\eta$ in our case).
Note that in theory, however, the number of motional states is infinite, involving $|a|\to \infty$ and therefore $|H_{\mathcal{I}}|\to\infty$. 
Nevertheless, for ultra cold atoms, only few low energy states are populated, so the system behaves as if $a$ and $a^{\dagger}$ were truncated, thus the AVH can be applied. 

In our model, the Eq.~\eqref{EqAppUI} is true if (i) $p_0$ is sufficiently close to $1$, (ii) $\omega/\Omega\gtrsim 2$ to prevent strong excitation of the motion during the application of the pulse, and (iii) the pulse duration is sufficiently short.
The validity of the method is confirmed by our results, which are simulated without approximation.

\subsection{Application to an atom in the Lamb-Dicke regime}
The Hamiltonian expressed in the Lamb-Dicke regime, given by Eq.~\eqref{EqHLambDicke}, can be express as $H=H_0+H_1$ with:
\[H_0(t)=h_q(t)+\omega a^{\dagger}a,\quad H_1(t)=\eta h_p(t)(a^{\dagger}+a).\]
$H_0$ is the sum of two terms acting in independent subspaces. We can show that the operator $U_0$ solution of $\dot{U}_0=-iH_0U_0$ is the product of two operators:
\[
\begin{aligned}
   U_0&=(U_q\otimes \mathbb{I}_{M\times M})~(\mathbb{I}_{2\times 2}e^{-ia^{\dagger}a\omega t})\\
   &\equiv U_q e^{-ia^{\dagger}a\omega t}, 
\end{aligned}
\]
where $M\to \infty$ is the number of motional states, and with $\dot{U}_q=-ih_qU_q$.
Using the commutation relation $U_qe^{-ia^{\dagger}a\omega t}=e^{-ia^{\dagger}a\omega t}U_q$ and the property $(e^{-i\omega t a^{\dagger}a})^{\dagger}(a^{\dagger}+a)e^{-i\omega t a^{\dagger}a}=(a^{\dagger}e^{i\omega t}+ae^{-i\omega t})$, we can show that the interaction Hamiltonian is given by:
\[H_{\mathcal{I}}(t)=U_0^{\dagger}(t)H_{1}(t)U_0(t)=\eta U_q^{\dagger}(t)h_p(t)(a^{\dagger}e^{i\omega t}+ae^{-i\omega t})U_q(t).\]
Since $h_p$ and $U_q$ commute with $a$ and $a^{\dagger}$, we obtain $H_{\mathcal{I}}=\eta U_q^{\dagger}h_pU_q(a^{\dagger}e^{i\omega t}+ae^{-i\omega t})$, and the interaction operator $U_{\mathcal{I}}$ defined by~\eqref{EqAppUI} becomes:
\[U_{\mathcal{I}}(t)=e^{-i\eta \left[a^{\dagger}V_{\Rec}(t)+aV_{\Rec}^{\dagger}(t)\right]}+o(\eta^2)\]
with $V_{\Rec}=\int_0^TU_q^{\dagger}(t')h_p(t')U_q(t')e^{i\omega t'}dt'$.
Thus, using $U=U_0U_{\mathcal{I}}$, we obtain Eq.~\eqref{EqOperatorO1}.
\subsection{Application beyond the Lamb-Dicke regime\label{AppLambDicke2}}
The Hamiltonian~\eqref{EqHTotal} ($\Delta=0$) can be rewritten in the Pauli basis as:
\[H=h_q(t)\cos[\eta(a+a^\dagger)]+h_p(t)\sin[\eta(a+a^\dagger)]+\omega a^\dagger a,\]
where $h_q$ and $h_p$ are given in Eq.~\eqref{EqHQHP}.
Up to the second order in $\eta$, we obtain:
\[H=h_q(t)\big(1-\tfrac{1}{2}\eta^2(a+a^\dagger)^2\big)+h_p(t)\eta(a+a^\dagger)+\omega a^\dagger a+o(\eta^3).\]
Expanding $(a+a^\dagger)^2$ and using $[a,a^{\dagger}]=\mathbb{I}\Rightarrow a a^{\dagger}=\mathbb{I}+a^{\dagger}a$, we obtain the expression~\eqref{EqHamO2}.

The Hamiltonian~\eqref{EqHamO2} can be decomposed into $H=H_0+H_1$ with:
\[\begin{aligned} 
&H_0(t)=h_q(t)\left(1-\tfrac{\eta^2}{2}\right)+\omega a^{\dagger}a,\\
&H_1(t)=\eta h_p(t)(a^{\dagger}+a)-\tfrac{\eta^2}{2}h_q(t)\big({a^{\dagger}}^2+a^2\big)-\eta^2 h_q(t) a^{\dagger}a.
\end{aligned}\]
Applying the AVH in the toggling frame, we obtain:
\[
    U(t)=U_q(t)e^{-i\omega a^{\dagger}a t}e^{-i\left(\bar{H}_{\Rec}^{(1)}(t)+\bar{H}_{\Rec}^{(2)}(t)+\bar{H}_{\Ent}(t)\right)}+o(\eta^3)
\]
with 
\[\begin{aligned}
    & \bar{H}_{\Rec}^{(1)}(t)=\eta \left[a^{\dagger}V_{\Rec}^{(1)}(t)+aV_{\Rec}^{(1)\dagger}(t)\right],\\
    & \bar{H}_{\Rec}^{(2)}(t)=-\tfrac{\eta^2}{2} \left[{a^{\dagger}}^2V_{\Rec}^{(2)}(t)+aV_{\Rec}^{(2)\dagger}(t)\right],\\
    & \bar{H}_{\Ent}(t)=-\eta^2a^{\dagger}a V_{\Ent}(t),
\end{aligned}\]
where $V_{\Rec}^{(1)}$, $V_{\Rec}^{(2)}$ and $V_{\Ent}$ are given in Eq.~\eqref{EqOperators}.
\section{Simplification of the problem in the Lamb-Dicke regime\label{AppTORFproblem}}
\paragraph*{\textbf{System of equations}.}
In this section, we show that under a \TORF pulse displayed in Fig.~\ref{FigSchemaTORF}, the control problem~\eqref{EqRecFreeControl} simplifies into the search of a set of parameters $\theta_1$, $\theta_2$, and $\theta_3$, solution of the nonlinear system of equations:
\begin{equation}
\begin{cases}
r_0=2\theta_1-2\theta_2+\theta_3-\theta_{\Target}=0,\\
r_1=(1+\lambda)A_1-2\lambda(A_2)+2\lambda A_3=0,\\
r_2=(1-\lambda)B_1+2\lambda(B_2)-2\lambda B_3=0,
\end{cases}\label{EqRoots}
\end{equation}
with:
\[
\begin{aligned}
& A_1=\sin\left[\theta_1(\lambda-1)+\theta_2(\lambda+1)+\tfrac{\theta_3}{2}(\lambda-1)\right]\\
& A_2=\sin\left[\theta_2(\lambda+1)+\tfrac{\theta_3}{2}(\lambda-1)\right]\\
& A_3=\sin\left[\tfrac{\theta_3}{2}(\lambda-1)\right]
\end{aligned}\quad 
\begin{aligned}
& B_1=\sin\left[\theta_1(\lambda+1)+\theta_2(\lambda-1)+\tfrac{\theta_3}{2}(\lambda+1)\right]\\
& B_2=\sin\left[\theta_2(\lambda-1)+\tfrac{\theta_3}{2}(\lambda+1)\right]\\
& B_3=\sin\left[\tfrac{\theta_3}{2}(\lambda+1)\right],
\end{aligned}
\]
and with $\lambda=\omega/\Omega$. Multiple solutions to this problem are given in Table~\ref{TabL3L5}. 
\begin{table}[h!]
\begin{tabular}{|c||ccccc|}
\hline 
\multicolumn{6}{|c|}{$\theta_{\Target}=45^\circ$}\\
\hline
$\omega/\Omega$ & 2 & 3 & 4 & 5 & 6\\\cline{2-6}
$\theta_1$ & 26.36 & 21.28 & 18.17 & 15.98 & 14.21 \\ 
$\theta_2$ & 30.11 & 19.74 & 13.77 & 9.93 & 7.28 \\ 
$\theta_3$ & 52.51 & 41.93 & 36.18 & 32.90 & 31.13 \\
\hline
\hline
\multicolumn{6}{|c|}{$\theta_{\Target}=90^\circ$}\\
\hline
$\omega/\Omega$ & 2 & 3 & 4 & 5 & 6\\\cline{2-6}
$\theta_1$ & 37.66 & 30.04 & 23.88 & 15.12 & 11.3052 \\ 
$\theta_2$ & 27.55 & 14.42 & 7.63 & 4.85 & 5.57 \\ 
$\theta_3$ & 69.79 & 58.75 & 57.50 & 69.45 & 78.54 \\
\hline
\hline
\multicolumn{6}{|c|}{$\theta_{\Target}=180^\circ$}\\
\hline
$\omega/\Omega$ & 2 & 3 & 4 & 5 & 6\\\cline{2-6}
$\theta_1$ & 67.43 & 0 & 31.17 & 0 & 20.54 \\ 
$\theta_2$ & 14.83 & 0 & 5.72 & 0 & 3.66 \\ 
$\theta_3$ & 74.80 & 180 & 129.11 & 180 & 146.23 \\
\hline
\end{tabular}
\caption{\TORF pulse parameters $\theta_1$, $\theta_2$ and $\theta_3$ solutions of~\eqref{EqRoots} (in degrees) for various target gates of the form $e^{-i\frac{\sigma_x}{2}\theta_{\Target}}$.
\emph{\textbf{Remark:}} These solutions can also be used beyond the Lamb-Dicke regime with good performances by applying a slightly higher Rabi frequency $\tilde{\Omega}=\Omega/[1-\eta^2/2]$, to compensate for the lack of speed in the qubit subspace caused by higher order effects.\label{TabL3L5}}
\end{table}

\paragraph*{\textbf{Demonstration.}} 
Using a bang-bang pulse made of $5$ segments (Fig.~\ref{FigSchemaTORF}), the operator $V_{\Rec}$ given in Eq.~\eqref{EqUQURec} becomes:
\[
    V_{\Rec}(T)=\sum_{n=1}^5\int_{t_{n-1}}^{t_{n}}U_q^{\dagger}(t)h_p(t)U_q(t)e^{i\omega t}dt,
\]
with $t_0=0$ and $t_5=T$.
Since the pulse is only about the $x$-axis, we get that $h_q$ is an operator along $\sigma_x$, $h_p$ is an operator along $\sigma_y$ and $U_q$ is made of rotations about $\sigma_x$.
Thus, we obtain the commutation relation $h_p(t)U_q(t)=U_q^{\dagger}(t)h_p(t)$.
This can be shown by expliciting $U$ as a $SU(2)$ rotation along $x$, i.e.
$U=\mathbb{I}\cos(\gamma)-i\sigma_x\sin(\gamma)$ ($\gamma$ is generic), and using the fact that $\sigma_x$ anticommutes with $\sigma_y$ and thus with $h_p$.
We obtain:
\[
    V_{\Rec}(T)=\sum_{n=1}^5\int_{t_{n-1}}^{t_{n}}(U_q^{\dagger}(t))^2h_p(t)e^{i\omega t}dt
\]
$h_q$ and $h_p$ are piecewise constant, such that if $t\in[t_{n-1},t_n]$, we have $h_p=h_{p_n}$ and $h_q=h_{q_n}$.
The operator $U_q(t)$ between $t_{n-1}$ and $t_n$ is thus given by:
\[\begin{aligned}
    U_q(t)&=e^{-ih_{q_n}(t-t_{n-1})}e^{-ih_{q_{n-1}}(t_{n-1}-t_{n-2})}\cdots e^{-ih_{q_1}(t_1-t_{0})}\\
    &\equiv e^{-ih_{q_n}(t-t_{n-1})}U_{q_{n-1}}.
\end{aligned}\]
More explicitely, we have $h_{q_n}=(-1)^{n-1}\Omega\sigma_x/2$, $h_{p_n}=(-1)^{n-1}\Omega\sigma_y/2$.
Denoting $\lambda=\omega/\Omega$ and $P_n=e^{-i h_{q_n}(t_n-t_{n-1})}$, we have:
\[
   \begin{aligned} 
   V_{\Rec}(T)& =\sum_{n=1}^5(U_{q_{n-1}}^{\dagger})^2h_{p_n}\int_{t_{n-1}}^{t_{n}}e^{-ih_{q_n}(t-t_{n-1})}e^{i\omega t}dt\\
    &=\sum_{n=1}^5(U_{q_{n-1}}^{\dagger})^2\tfrac{(-1)^{n-1}\lambda\sigma_y-i\sigma_z}{1-\lambda^2}\left[P_n^2e^{i\omega t_{n}}-\mathbb{I}e^{i\omega t_{n-1}}\right]\\
    &=\sum_{n=1}^5\tfrac{(-1)^{n-1}\lambda\sigma_y-i\sigma_z}{1-\lambda^2}U_{q_{n-1}}^2\left[P_ne^{i\omega t_{n}}-\mathbb{I}e^{i\omega t_{n-1}}\right]\\
    &=\sum_{n=1}^5\tfrac{(-1)^{n-1}\lambda\sigma_y-i\sigma_z}{1-\lambda^2}\left[U_{q_{n}}^2e^{i\omega t_{n}}-U_{q_{n-1}}^2e^{i\omega t_{n-1}}\right]
    \end{aligned}
\]
where we used $(U_{q_{n-1}}^{\dagger})^2\tfrac{(-1)^{n-1}\lambda\sigma_y-i\sigma_z}{1-\lambda^2}=\tfrac{(-1)^{n-1}\lambda\sigma_y-i\sigma_z}{1-\lambda^2}U_{q_{n-1}}^2$.
The parameters of the \TORF pulse are defined as $\theta_1=\Omega(t_1-t_{0})$, $\theta_2=\Omega(t_3-t_2)$ and $\theta_3=\Omega(t_4-t_3)$.
To reach the target in the qubit subspace, we must have $U_T=e^{-i\frac{\sigma_x}{2}\theta_{\Target}}=e^{-i\tfrac{\sigma_x}{2}(2\theta_1-2\theta_2+\theta_3)}\equiv R_x(2\theta_1-2\theta_2+\theta_3)$, which implies the first equation of the system~\eqref{EqRoots}.
Using the symmetries, we can show the following relations:
\[
\begin{aligned}
& U_{q_0}^2U_T^{\dagger}e^{i\omega(t_0-T/2)}=R_x(-2\theta_1+2\theta_2-\theta_3)e^{-i\tfrac{\lambda}{2}\big(2\theta_1+2\theta_2+\theta_3\big)}\equiv V_0\\
&U_{q_1}^2U_T^{\dagger}e^{i\omega(t_1-T/2)}=R_x(2\theta_2-\theta_3)e^{-i\tfrac{\lambda}{2}\big(2\theta_2+\theta_3\big)}\equiv V_1\\
&U_{q_2}^2U_T^{\dagger}e^{i\omega(t_2-T/2)}=R_x(-\theta_3)e^{-i\tfrac{\lambda}{2}\big(\theta_3\big)}\equiv V_2\\
&U_{q_3}^2U_T^{\dagger}e^{i\omega(t_3-T/2)}=R_x(\theta_3)e^{i\tfrac{\lambda}{2}\big(\theta_3\big)}\equiv V_3\\
&U_{q_4}^2U_T^{\dagger}e^{i\omega(t_4-T/2)}=R_x(-2\theta_2+\theta_3)e^{i\tfrac{\lambda}{2}\big(2\theta_2+\theta_3\big)}\equiv V_4\\
& U_{q_5}^2U_T^{\dagger}e^{i\omega(t_5-T/2)}=R_x(2\theta_1-2\theta_2+\theta_3)e^{i\tfrac{\lambda}{2}\big(2\theta_1+2\theta_2+\theta_3\big)}\equiv V_5.
\end{aligned}
\]
The recoil-free condition $V_{\Rec}(t)=0$ is equivalent to the equation  $\tilde{V}_{\Rec}=i(1-\lambda^2)\sigma_z V_{\Rec}(T)U_T^\dagger e^{-i\tfrac{\lambda\Omega T}{2}}=0$, which gives explicitly:
\[\tilde{V}_{\Rec}=B_+\Big[(V_5-V_0)-(V_4-V_1)+(V_3-V_2)\Big]+B_-\Big[(V_4-V_1)-(V_3-V_2)\Big]=0\]
with $B_+=\mathbb{I}+\lambda\sigma_x$ and $B_-=\mathbb{I}-\lambda\sigma_x$.
However:
\[
\begin{aligned}
&V_5-V_0=2i\Big\{\cos\left[\tfrac{1}{2}\big(2\theta_1-2\theta_2+\theta_3\big)\right]\sin\left[\tfrac{\lambda}{2}\big(2\theta_1+2\theta_2+\theta_3\big)\right]\mathbb{I}-\sin\left[\tfrac{1}{2}\big(2\theta_1-2\theta_2+\theta_3\big)\right]\cos\left[\tfrac{\lambda}{2}\big(2\theta_1+2\theta_2+\theta_3\big)\right]\sigma_x\Big\}\\
&V_4-V_1=2i\Big\{\cos\left[\tfrac{1}{2}\big(-2\theta_2+\theta_3\big)\right]\sin\left[\tfrac{\lambda}{2}\big(2\theta_2+\theta_3\big)\right]\mathbb{I}-\sin\left[\tfrac{1}{2}\big(-2\theta_2+\theta_3\big)\right]\cos\left[\tfrac{\lambda}{2}\big(2\theta_2+\theta_3\big)\right]\sigma_x\Big\}\\
&V_3-V_2=2i\Big\{\cos\left[\tfrac{1}{2}\big(\theta_3\big)\right]\sin\left[\tfrac{\lambda}{2}\big(\theta_3\big)\right]\mathbb{I}-\sin\left[\tfrac{1}{2}\big(\theta_3\big)\right]\cos\left[\tfrac{\lambda}{2}\big(\theta_3\big)\right]\sigma_x\Big\}.
\end{aligned}
\]
Identifying the terms as:
\[
\begin{aligned}
&V_5-V_0=2i\Big\{a_1\mathbb{I}+b_1\sigma_x\Big\}\\
&V_4-V_1=2i\Big\{a_2\mathbb{I}+b_2\sigma_x\Big\}\\
&V_3-V_2=2i\Big\{a_3\mathbb{I}+b_3\sigma_x\Big\}
\end{aligned}
\]
we have to solve:
\[
\begin{cases}
R_1:a_1+\lambda b_1-2\lambda b_2+2\lambda b_3=0\\
R_2:b_1+\lambda a_1-2\lambda a_2+2\lambda a_3=0.
\end{cases}
\]
Defining $A_n$ and $B_n$ such that:
\[
\begin{aligned}
& A_1\equiv a_1+b_1=\sin\left[\theta_1(\lambda-1)+\theta_2(\lambda+1)+\tfrac{\theta_3}{2}(\lambda-1)\right]\\
& A_2\equiv a_2+b_2=\sin\left[\theta_2(\lambda+1)+\tfrac{\theta_3}{2}(\lambda-1)\right]\\
& A_3\equiv a_3+b_3=\sin\left[\tfrac{\theta_3}{2}(\lambda-1)\right]
\end{aligned}\quad 
\begin{aligned}
& B_1\equiv a_1-b_1=\sin\left[\theta_1(\lambda+1)+\theta_2(\lambda-1)+\tfrac{\theta_3}{2}(\lambda+1)\right]\\
& B_2\equiv a_2-b_2=\sin\left[\theta_2(\lambda-1)+\tfrac{\theta_3}{2}(\lambda+1)\right]\\
& B_3\equiv a_3-b_3=\sin\left[\tfrac{\theta_3}{2}(\lambda+1)\right],
\end{aligned}
\]
one can solve $r_1:R_1+R_2=0$ and $r_2:R_1-R_2=0$ leading to the two additional equations of the system~\eqref{EqRoots}.
\section{Limit of recoil-free gates\label{AppFlim}}
This section aims at demonstrating the limit given in Eq.~\eqref{EqJlim2Temp}.
Assuming the operator~\eqref{EqUTEnt} and
using the formula $e^{-i\eta^2a^{\dagger}a V_{\Ent}(T)}\ket{\psi_{km}}=e^{-im\eta^2 V_{\Ent}(T)}\ket{\psi_{km}}$, the contribution $\mathcal{F}^{(m)}$ to the gate fidelity defined in~\eqref{EqJGate} becomes:
\begin{equation}
    \mathcal{F}^{(m)}=\tfrac{1}{4}\sum_{k=1}^4|\braket{\psi_{km}|e^{-im\eta^2 V_{\Ent}(T)}|\psi_{km}}|^2.
    \label{EqJGateEnt}
\end{equation}
Moreover, under \TORF or constant pulses, we have $\sin(\varphi(t))=0$ and thus $[U_q,h_q]=0$, leading to:
\[V_{\Ent}(t)=\int_0^Th_q(t)dt=\int_0^T\Omega\cos[\varphi(t)]dt\tfrac{\sigma_x}{2}.\]
To achieve $U_q(T)=U_{\Target}=e^{-i\theta_{\Target}\frac{\sigma_x}{2}}$ in the qubit subspace, one can explicit $U_q$ (Eq.~\eqref{EqOperators}). For a constant or TORF pulse, we have $U_q(T)=e^{-i\int_0^T(1-\eta^2)h_q(t)dt}$, and thus $(1-\eta^2)\int_0^Th_q(t)dt=\theta_{\Target}\frac{\sigma_x}{2}$. The operator $V_{\Ent}$ becomes:
 \[V_{\Ent}(T)=\frac{\theta_{\Target}}{1-\eta^2/2}\frac{\sigma_x}{2}.\]
Substituting this expression in Eq.~\eqref{EqJGateEnt} and being careful that $\braket{\psi_{km}|\sigma_x|\psi_{km}}=1$ if $k=4$ and $0$ otherwise, we obtain:
\[
\begin{aligned}
    \mathcal{F}_{\lim}^{(m)}&=1-\tfrac{3}{4}\sin^2\left(\tfrac{m\eta^2\theta_{\Target}}{2\left(1-\frac{\eta^2}{2}\right)}\right)\\
    &\equiv 1-\tfrac{3}{4}\sin^2\left(m\tfrac{\gamma}{2}\right)
\end{aligned}
\]
with $\gamma=\tfrac{\eta^2\theta_{\Target}}{1-\frac{\eta^2}{2}}$.
Using the full expression $\mathcal{F}_{\lim}=\sum p_m\mathcal{F}_{\lim}^{(m)}$ and $\sum_0^{\infty}p_m=1$, we obtain:
\[
    \mathcal{F}_{\lim}= 1-\tfrac{3}{4}\sum_{m=0}^{\infty}p_m\sin^2\left(m\tfrac{\gamma}{2}\right).
\]
The sine can be rewritten using $\sin^2(m\gamma/2)=(1-\cos(m\gamma))/2$.
Given that $\sum_{k=0}^{\infty}(1-p_0)^k=1/p_0$, we get
$p_m=p_0(1-p_0)^m$ and thus:
\[\begin{aligned}
\mathcal{F}_{\lim}&=1-\frac{3}{8}\sum_{m=0}^{\infty}p_m[1-\cos(m\gamma)]\\
&=1-\frac{3}{8}\sum_{m=0}^{\infty}p_m\left[1-\tfrac{1}{2}\big(e^{im\gamma}+e^{-im\gamma}\big)\right]\\
&=\frac{5}{8}+\frac{3}{16}\sum_{m=0}^{\infty}p_m\big(e^{im\gamma}+e^{-im\gamma}\big)\\
&=\frac{5}{8}+\frac{3}{16}\sum_{m=0}^{\infty}p_0(1-p_0)^m\left(e^{im\gamma}+e^{-im\gamma}\right)\\
&=\frac{5}{8}+\frac{3p_0}{16}\sum_{m=0}^{\infty} (z_A^m+z_B^m)
\end{aligned}\]
with $z_A=(1-p_0)e^{i\gamma}$ and $z_B=(1-p_0)e^{-i\gamma}$.
Since the sum corresponds to a geometric series, we can apply the formula $\sum_{k=0}^{\infty}x^k=\tfrac{1}{1-x}$, which holds because $|z_a|$ and $|z_b|$ are smaller that $1$, leading to:
\[
\mathcal{F}_{\lim}=\frac{5}{8}+\frac{3p_0}{16}\left(\frac{1}{1-z_A}+\frac{1}{1-z_B}\right).\]
Expliciting $\gamma$ and simplifying, we obtain:
\[
    1-\mathcal{F}_{\lim}=\frac{3}{8}\left(\frac{(1-p_0)(2-p_0)\left[1-\cos\left(\dfrac{\eta^2\theta_{\Target}}{1-\sfrac{\eta^2}{2}}\right)\right]}{1-2(1-p_0)\cos\left(\tfrac{\eta^2\theta_{\Target}}{1-\sfrac{\eta^2}{2}}\right)+(1-p_0)^2}\right).
\]
Finally, applying a series exapansion in terms of $\eta$ leads to Eq.~\eqref{EqJlim2Temp}.
\section{Trap robust condition\label{AppMiscellaneous}} The pulse can be made robust against deviations of the trap frequency by substituting $\omega\rightarrow\omega+\delta\omega$ in the recoil-free condition as:
\[
V_{\Rec}(\omega+\delta\omega)=\int_0^TU_q^{\dagger}(t)h_p(t)U_q(t) e^{i(\omega+\delta\omega) t} dt.
\]
Assuming $\delta\omega/\omega\ll 1$, an expansion up to the first order leads to:
\[\begin{aligned}
V_{\Rec}(\omega+\delta\omega)&=\int_0^TU_q^{\dagger}(t)h_p(t)U_q(t) e^{i\omega t}(1+i\delta\omega t)dt\\
&=V_{\Rec}+i\delta\omega\int_0^TU_q^{\dagger}(t)h_p(t)U_q(t)e^{i\omega t}\\
&=V_{\Rec}+i\delta\omega V_{\Trap},
\end{aligned}\]
with:
\begin{equation}
    V_{\Trap}=\int_0^TtU_q^{\dagger}(t)h_p(t)U_q(t)e^{i\omega t}
\end{equation}
(note the factor $t$ in the integral).
Adding the constraint $V_{\Trap}=0$ into any of the control problems allows one to design trap robust gates.
The method can also be applied to the second-order recoil-free condition, leading to:
\begin{equation}
    V_{\Trap}^{(2)}=\int_0^TtU_q^{\dagger}(t)h_q(t)U_q(t)e^{2i\omega t}.
\end{equation}
\bibliographystyle{apsrev4-2}
\bibliography{Biblio}

\begin{thebibliography}{53}%
\makeatletter
\providecommand \@ifxundefined [1]{%
 \@ifx{#1\undefined}
}%
\providecommand \@ifnum [1]{%
 \ifnum #1\expandafter \@firstoftwo
 \else \expandafter \@secondoftwo
 \fi
}%
\providecommand \@ifx [1]{%
 \ifx #1\expandafter \@firstoftwo
 \else \expandafter \@secondoftwo
 \fi
}%
\providecommand \natexlab [1]{#1}%
\providecommand \enquote  [1]{``#1''}%
\providecommand \bibnamefont  [1]{#1}%
\providecommand \bibfnamefont [1]{#1}%
\providecommand \citenamefont [1]{#1}%
\providecommand \href@noop [0]{\@secondoftwo}%
\providecommand \href [0]{\begingroup \@sanitize@url \@href}%
\providecommand \@href[1]{\@@startlink{#1}\@@href}%
\providecommand \@@href[1]{\endgroup#1\@@endlink}%
\providecommand \@sanitize@url [0]{\catcode `\\12\catcode `\$12\catcode
  `\&12\catcode `\#12\catcode `\^12\catcode `\_12\catcode `\%12\relax}%
\providecommand \@@startlink[1]{}%
\providecommand \@@endlink[0]{}%
\providecommand \url  [0]{\begingroup\@sanitize@url \@url }%
\providecommand \@url [1]{\endgroup\@href {#1}{\urlprefix }}%
\providecommand \urlprefix  [0]{URL }%
\providecommand \Eprint [0]{\href }%
\providecommand \doibase [0]{https://doi.org/}%
\providecommand \selectlanguage [0]{\@gobble}%
\providecommand \bibinfo  [0]{\@secondoftwo}%
\providecommand \bibfield  [0]{\@secondoftwo}%
\providecommand \translation [1]{[#1]}%
\providecommand \BibitemOpen [0]{}%
\providecommand \bibitemStop [0]{}%
\providecommand \bibitemNoStop [0]{.\EOS\space}%
\providecommand \EOS [0]{\spacefactor3000\relax}%
\providecommand \BibitemShut  [1]{\csname bibitem#1\endcsname}%
\let\auto@bib@innerbib\@empty
\bibitem [{\citenamefont {Wintersperger}\ \emph {et~al.}(2023)\citenamefont
  {Wintersperger}, \citenamefont {Dommert}, \citenamefont {Ehmer},
  \citenamefont {Hoursanov}, \citenamefont {Klepsch}, \citenamefont {Mauerer},
  \citenamefont {Reuber}, \citenamefont {Strohm}, \citenamefont {Yin},\ and\
  \citenamefont {Luber}}]{winterspergerNeutralAtomQuantum2023}%
  \BibitemOpen
  \bibfield  {author} {\bibinfo {author} {\bibfnamefont {K.}~\bibnamefont
  {Wintersperger}}, \bibinfo {author} {\bibfnamefont {F.}~\bibnamefont
  {Dommert}}, \bibinfo {author} {\bibfnamefont {T.}~\bibnamefont {Ehmer}},
  \bibinfo {author} {\bibfnamefont {A.}~\bibnamefont {Hoursanov}}, \bibinfo
  {author} {\bibfnamefont {J.}~\bibnamefont {Klepsch}}, \bibinfo {author}
  {\bibfnamefont {W.}~\bibnamefont {Mauerer}}, \bibinfo {author} {\bibfnamefont
  {G.}~\bibnamefont {Reuber}}, \bibinfo {author} {\bibfnamefont
  {T.}~\bibnamefont {Strohm}}, \bibinfo {author} {\bibfnamefont
  {M.}~\bibnamefont {Yin}},\ and\ \bibinfo {author} {\bibfnamefont
  {S.}~\bibnamefont {Luber}},\ }\bibfield  {title} {\enquote {\bibinfo {title}
  {{Neutral Atom Quantum Computing Hardware: Performance and End-User
  Perspective}},}\ }\href {https://doi.org/10.1140/epjqt/s40507-023-00190-1}
  {\bibfield  {journal} {\bibinfo  {journal} {EPJ Quantum Technol.}\ }\textbf
  {\bibinfo {volume} {10}},\ \bibinfo {pages} {1} (\bibinfo {year}
  {2023})}\BibitemShut {NoStop}%
\bibitem [{\citenamefont {Henriet}\ \emph {et~al.}(2020)\citenamefont
  {Henriet}, \citenamefont {Beguin}, \citenamefont {Signoles}, \citenamefont
  {Lahaye}, \citenamefont {Browaeys}, \citenamefont {Reymond},\ and\
  \citenamefont {Jurczak}}]{henrietQuantumComputingNeutral2020}%
  \BibitemOpen
  \bibfield  {author} {\bibinfo {author} {\bibfnamefont {L.}~\bibnamefont
  {Henriet}}, \bibinfo {author} {\bibfnamefont {L.}~\bibnamefont {Beguin}},
  \bibinfo {author} {\bibfnamefont {A.}~\bibnamefont {Signoles}}, \bibinfo
  {author} {\bibfnamefont {T.}~\bibnamefont {Lahaye}}, \bibinfo {author}
  {\bibfnamefont {A.}~\bibnamefont {Browaeys}}, \bibinfo {author}
  {\bibfnamefont {G.-O.}\ \bibnamefont {Reymond}},\ and\ \bibinfo {author}
  {\bibfnamefont {C.}~\bibnamefont {Jurczak}},\ }\bibfield  {title} {\enquote
  {\bibinfo {title} {{Quantum computing with neutral atoms}},}\ }\href
  {https://doi.org/10.22331/q-2020-09-21-327} {\bibfield  {journal} {\bibinfo
  {journal} {Quantum}\ }\textbf {\bibinfo {volume} {4}},\ \bibinfo {pages}
  {327} (\bibinfo {year} {2020})}\BibitemShut {NoStop}%
\bibitem [{\citenamefont {Bruzewicz}\ \emph {et~al.}(2019)\citenamefont
  {Bruzewicz}, \citenamefont {Chiaverini}, \citenamefont {McConnell},\ and\
  \citenamefont {Sage}}]{bruzewiczTrappedionQuantumComputing2019}%
  \BibitemOpen
  \bibfield  {author} {\bibinfo {author} {\bibfnamefont {C.~D.}\ \bibnamefont
  {Bruzewicz}}, \bibinfo {author} {\bibfnamefont {J.}~\bibnamefont
  {Chiaverini}}, \bibinfo {author} {\bibfnamefont {R.}~\bibnamefont
  {McConnell}},\ and\ \bibinfo {author} {\bibfnamefont {J.~M.}\ \bibnamefont
  {Sage}},\ }\bibfield  {title} {\enquote {\bibinfo {title} {{Trapped-ion
  quantum computing: Progress and challenges}},}\ }\href
  {https://doi.org/10.1063/1.5088164} {\bibfield  {journal} {\bibinfo
  {journal} {Appl. Phys. Rev.}\ }\textbf {\bibinfo {volume} {6}},\ \bibinfo
  {pages} {021314} (\bibinfo {year} {2019})}\BibitemShut {NoStop}%
\bibitem [{\citenamefont {Bluvstein}\ \emph {et~al.}(2022)\citenamefont
  {Bluvstein}, \citenamefont {Levine}, \citenamefont {Semeghini}, \citenamefont
  {Wang}, \citenamefont {Ebadi}, \citenamefont {Kalinowski}, \citenamefont
  {Keesling}, \citenamefont {Maskara}, \citenamefont {Pichler}, \citenamefont
  {Greiner}, \citenamefont {Vuleti{\'c}},\ and\ \citenamefont
  {Lukin}}]{osti_1904273}%
  \BibitemOpen
  \bibfield  {author} {\bibinfo {author} {\bibfnamefont {D.}~\bibnamefont
  {Bluvstein}}, \bibinfo {author} {\bibfnamefont {H.}~\bibnamefont {Levine}},
  \bibinfo {author} {\bibfnamefont {G.}~\bibnamefont {Semeghini}}, \bibinfo
  {author} {\bibfnamefont {T.~T.}\ \bibnamefont {Wang}}, \bibinfo {author}
  {\bibfnamefont {S.}~\bibnamefont {Ebadi}}, \bibinfo {author} {\bibfnamefont
  {M.}~\bibnamefont {Kalinowski}}, \bibinfo {author} {\bibfnamefont
  {A.}~\bibnamefont {Keesling}}, \bibinfo {author} {\bibfnamefont
  {N.}~\bibnamefont {Maskara}}, \bibinfo {author} {\bibfnamefont
  {H.}~\bibnamefont {Pichler}}, \bibinfo {author} {\bibfnamefont
  {M.}~\bibnamefont {Greiner}}, \bibinfo {author} {\bibfnamefont
  {V.}~\bibnamefont {Vuleti{\'c}}},\ and\ \bibinfo {author} {\bibfnamefont
  {M.~D.}\ \bibnamefont {Lukin}},\ }\bibfield  {title} {\enquote {\bibinfo
  {title} {{A quantum processor based on coherent transport of entangled atom
  arrays}},}\ }\href {https://doi.org/10.1038/s41586-022-04592-6} {\bibfield
  {journal} {\bibinfo  {journal} {Nature}\ }\textbf {\bibinfo {volume} {604}},\
  \bibinfo {pages} {451} (\bibinfo {year} {2022})}\BibitemShut {NoStop}%
\bibitem [{\citenamefont {Bluvstein}\ \emph
  {et~al.}(2024{\natexlab{a}})\citenamefont {Bluvstein}, \citenamefont
  {Evered}, \citenamefont {Geim}, \citenamefont {Li}, \citenamefont {Zhou},
  \citenamefont {Manovitz}, \citenamefont {Ebadi}, \citenamefont {Cain},
  \citenamefont {Kalinowski}, \citenamefont {Hangleiter}, \citenamefont
  {Bonilla~Ataides}, \citenamefont {Maskara}, \citenamefont {Cong},
  \citenamefont {Gao}, \citenamefont {Sales~Rodriguez}, \citenamefont
  {Karolyshyn}, \citenamefont {Semeghini}, \citenamefont {Gullans},
  \citenamefont {Greiner}, \citenamefont {Vuleti{\'c}},\ and\ \citenamefont
  {Lukin}}]{Bluvstein:2024}%
  \BibitemOpen
  \bibfield  {author} {\bibinfo {author} {\bibfnamefont {D.}~\bibnamefont
  {Bluvstein}}, \bibinfo {author} {\bibfnamefont {S.~J.}\ \bibnamefont
  {Evered}}, \bibinfo {author} {\bibfnamefont {A.~A.}\ \bibnamefont {Geim}},
  \bibinfo {author} {\bibfnamefont {S.~H.}\ \bibnamefont {Li}}, \bibinfo
  {author} {\bibfnamefont {H.}~\bibnamefont {Zhou}}, \bibinfo {author}
  {\bibfnamefont {T.}~\bibnamefont {Manovitz}}, \bibinfo {author}
  {\bibfnamefont {S.}~\bibnamefont {Ebadi}}, \bibinfo {author} {\bibfnamefont
  {M.}~\bibnamefont {Cain}}, \bibinfo {author} {\bibfnamefont {M.}~\bibnamefont
  {Kalinowski}}, \bibinfo {author} {\bibfnamefont {D.}~\bibnamefont
  {Hangleiter}}, \bibinfo {author} {\bibfnamefont {J.~P.}\ \bibnamefont
  {Bonilla~Ataides}}, \bibinfo {author} {\bibfnamefont {N.}~\bibnamefont
  {Maskara}}, \bibinfo {author} {\bibfnamefont {I.}~\bibnamefont {Cong}},
  \bibinfo {author} {\bibfnamefont {X.}~\bibnamefont {Gao}}, \bibinfo {author}
  {\bibfnamefont {P.}~\bibnamefont {Sales~Rodriguez}}, \bibinfo {author}
  {\bibfnamefont {T.}~\bibnamefont {Karolyshyn}}, \bibinfo {author}
  {\bibfnamefont {G.}~\bibnamefont {Semeghini}}, \bibinfo {author}
  {\bibfnamefont {M.~J.}\ \bibnamefont {Gullans}}, \bibinfo {author}
  {\bibfnamefont {M.}~\bibnamefont {Greiner}}, \bibinfo {author} {\bibfnamefont
  {V.}~\bibnamefont {Vuleti{\'c}}},\ and\ \bibinfo {author} {\bibfnamefont
  {M.~D.}\ \bibnamefont {Lukin}},\ }\bibfield  {title} {\enquote {\bibinfo
  {title} {{Logical quantum processor based on reconfigurable atom arrays}},}\
  }\href {https://doi.org/10.1038/s41586-023-06927-3} {\bibfield  {journal}
  {\bibinfo  {journal} {Nature}\ }\textbf {\bibinfo {volume} {626}},\ \bibinfo
  {pages} {58} (\bibinfo {year} {2024}{\natexlab{a}})}\BibitemShut {NoStop}%
\bibitem [{\citenamefont {Cong}\ \emph {et~al.}(2022)\citenamefont {Cong},
  \citenamefont {Levine}, \citenamefont {Keesling}, \citenamefont {Bluvstein},
  \citenamefont {Wang},\ and\ \citenamefont {Lukin}}]{Cong2022}%
  \BibitemOpen
  \bibfield  {author} {\bibinfo {author} {\bibfnamefont {I.}~\bibnamefont
  {Cong}}, \bibinfo {author} {\bibfnamefont {H.}~\bibnamefont {Levine}},
  \bibinfo {author} {\bibfnamefont {A.}~\bibnamefont {Keesling}}, \bibinfo
  {author} {\bibfnamefont {D.}~\bibnamefont {Bluvstein}}, \bibinfo {author}
  {\bibfnamefont {S.-T.}\ \bibnamefont {Wang}},\ and\ \bibinfo {author}
  {\bibfnamefont {M.~D.}\ \bibnamefont {Lukin}},\ }\bibfield  {title} {\enquote
  {\bibinfo {title} {{Hardware-Efficient, Fault-Tolerant Quantum Computation
  with Rydberg Atoms}},}\ }\href {https://doi.org/10.1103/PhysRevX.12.021049}
  {\bibfield  {journal} {\bibinfo  {journal} {Phys. Rev. X}\ }\textbf {\bibinfo
  {volume} {12}},\ \bibinfo {pages} {021049} (\bibinfo {year}
  {2022})}\BibitemShut {NoStop}%
\bibitem [{\citenamefont {Graham}\ \emph {et~al.}(2023)\citenamefont {Graham},
  \citenamefont {Phuttitarn}, \citenamefont {Chinnarasu}, \citenamefont {Song},
  \citenamefont {Poole}, \citenamefont {Jooya}, \citenamefont {Scott},
  \citenamefont {Scott}, \citenamefont {Eichler},\ and\ \citenamefont
  {Saffman}}]{PhysRevX.13.041051}%
  \BibitemOpen
  \bibfield  {author} {\bibinfo {author} {\bibfnamefont {T.~M.}\ \bibnamefont
  {Graham}}, \bibinfo {author} {\bibfnamefont {L.}~\bibnamefont {Phuttitarn}},
  \bibinfo {author} {\bibfnamefont {R.}~\bibnamefont {Chinnarasu}}, \bibinfo
  {author} {\bibfnamefont {Y.}~\bibnamefont {Song}}, \bibinfo {author}
  {\bibfnamefont {C.}~\bibnamefont {Poole}}, \bibinfo {author} {\bibfnamefont
  {K.}~\bibnamefont {Jooya}}, \bibinfo {author} {\bibfnamefont
  {J.}~\bibnamefont {Scott}}, \bibinfo {author} {\bibfnamefont
  {A.}~\bibnamefont {Scott}}, \bibinfo {author} {\bibfnamefont
  {P.}~\bibnamefont {Eichler}},\ and\ \bibinfo {author} {\bibfnamefont
  {M.}~\bibnamefont {Saffman}},\ }\bibfield  {title} {\enquote {\bibinfo
  {title} {{Midcircuit Measurements on a Single-Species Neutral Alkali Atom
  Quantum Processor}},}\ }\href {https://doi.org/10.1103/PhysRevX.13.041051}
  {\bibfield  {journal} {\bibinfo  {journal} {Phys. Rev. X}\ }\textbf {\bibinfo
  {volume} {13}},\ \bibinfo {pages} {041051} (\bibinfo {year}
  {2023})}\BibitemShut {NoStop}%
\bibitem [{\citenamefont {Li}\ and\ \citenamefont
  {Thompson}(2024)}]{PRXQuantum.5.020363}%
  \BibitemOpen
  \bibfield  {author} {\bibinfo {author} {\bibfnamefont {Y.}~\bibnamefont
  {Li}}\ and\ \bibinfo {author} {\bibfnamefont {J.~D.}\ \bibnamefont
  {Thompson}},\ }\bibfield  {title} {\enquote {\bibinfo {title} {{High-Rate and
  High-Fidelity Modular Interconnects between Neutral Atom Quantum
  Processors}},}\ }\href {https://doi.org/10.1103/PRXQuantum.5.020363}
  {\bibfield  {journal} {\bibinfo  {journal} {PRX Quantum}\ }\textbf {\bibinfo
  {volume} {5}},\ \bibinfo {pages} {020363} (\bibinfo {year}
  {2024})}\BibitemShut {NoStop}%
\bibitem [{\citenamefont {Burgers}\ \emph {et~al.}(2022)\citenamefont
  {Burgers}, \citenamefont {Ma}, \citenamefont {Saskin}, \citenamefont
  {Wilson}, \citenamefont {Alarc{\'o}n}, \citenamefont {Greene},\ and\
  \citenamefont {Thompson}}]{PRXQuantum.3.020326}%
  \BibitemOpen
  \bibfield  {author} {\bibinfo {author} {\bibfnamefont {A.~P.}\ \bibnamefont
  {Burgers}}, \bibinfo {author} {\bibfnamefont {S.}~\bibnamefont {Ma}},
  \bibinfo {author} {\bibfnamefont {S.}~\bibnamefont {Saskin}}, \bibinfo
  {author} {\bibfnamefont {J.}~\bibnamefont {Wilson}}, \bibinfo {author}
  {\bibfnamefont {M.~A.}\ \bibnamefont {Alarc{\'o}n}}, \bibinfo {author}
  {\bibfnamefont {C.~H.}\ \bibnamefont {Greene}},\ and\ \bibinfo {author}
  {\bibfnamefont {J.~D.}\ \bibnamefont {Thompson}},\ }\bibfield  {title}
  {\enquote {\bibinfo {title} {{Controlling Rydberg Excitations Using Ion-Core
  Transitions in Alkaline-Earth Atom-Tweezer Arrays}},}\ }\href
  {https://doi.org/10.1103/PRXQuantum.3.020326} {\bibfield  {journal} {\bibinfo
   {journal} {PRX Quantum}\ }\textbf {\bibinfo {volume} {3}},\ \bibinfo {pages}
  {020326} (\bibinfo {year} {2022})}\BibitemShut {NoStop}%
\bibitem [{\citenamefont {Leibfried}\ \emph {et~al.}(2003)\citenamefont
  {Leibfried}, \citenamefont {Blatt}, \citenamefont {Monroe},\ and\
  \citenamefont {Wineland}}]{RevModPhys.75.281}%
  \BibitemOpen
  \bibfield  {author} {\bibinfo {author} {\bibfnamefont {D.}~\bibnamefont
  {Leibfried}}, \bibinfo {author} {\bibfnamefont {R.}~\bibnamefont {Blatt}},
  \bibinfo {author} {\bibfnamefont {C.}~\bibnamefont {Monroe}},\ and\ \bibinfo
  {author} {\bibfnamefont {D.}~\bibnamefont {Wineland}},\ }\bibfield  {title}
  {\enquote {\bibinfo {title} {{Quantum dynamics of single trapped ions}},}\
  }\href {https://doi.org/10.1103/RevModPhys.75.281} {\bibfield  {journal}
  {\bibinfo  {journal} {Rev. Mod. Phys.}\ }\textbf {\bibinfo {volume} {75}},\
  \bibinfo {pages} {281} (\bibinfo {year} {2003})}\BibitemShut {NoStop}%
\bibitem [{\citenamefont {Kaufman}\ and\ \citenamefont
  {Ni}(2021)}]{kaufman2021quantum}%
  \BibitemOpen
  \bibfield  {author} {\bibinfo {author} {\bibfnamefont {A.~M.}\ \bibnamefont
  {Kaufman}}\ and\ \bibinfo {author} {\bibfnamefont {K.-K.}\ \bibnamefont
  {Ni}},\ }\bibfield  {title} {\enquote {\bibinfo {title} {{Quantum science
  with optical tweezer arrays of ultracold atoms and molecules}},}\ }\href
  {https://doi.org/10.1038/s41567-021-01357-2} {\bibfield  {journal} {\bibinfo
  {journal} {Nat. Phys.}\ }\textbf {\bibinfo {volume} {17}},\ \bibinfo {pages}
  {1324} (\bibinfo {year} {2021})}\BibitemShut {NoStop}%
\bibitem [{\citenamefont {Saffman}\ and\ \citenamefont
  {Walker}(2005)}]{Saffman:2005}%
  \BibitemOpen
  \bibfield  {author} {\bibinfo {author} {\bibfnamefont {M.}~\bibnamefont
  {Saffman}}\ and\ \bibinfo {author} {\bibfnamefont {T.~G.}\ \bibnamefont
  {Walker}},\ }\bibfield  {title} {\enquote {\bibinfo {title} {{Analysis of a
  quantum logic device based on dipole-dipole interactions of optically trapped
  Rydberg atoms}},}\ }\href {https://doi.org/10.1103/PhysRevA.72.022347}
  {\bibfield  {journal} {\bibinfo  {journal} {Phys. Rev. A}\ }\textbf {\bibinfo
  {volume} {72}},\ \bibinfo {pages} {022347} (\bibinfo {year}
  {2005})}\BibitemShut {NoStop}%
\bibitem [{\citenamefont {Xia}\ \emph {et~al.}(2015)\citenamefont {Xia},
  \citenamefont {Lichtman}, \citenamefont {Maller}, \citenamefont {Carr},
  \citenamefont {Piotrowicz}, \citenamefont {Isenhower},\ and\ \citenamefont
  {Saffman}}]{Xia:2015a}%
  \BibitemOpen
  \bibfield  {author} {\bibinfo {author} {\bibfnamefont {T.}~\bibnamefont
  {Xia}}, \bibinfo {author} {\bibfnamefont {M.}~\bibnamefont {Lichtman}},
  \bibinfo {author} {\bibfnamefont {K.}~\bibnamefont {Maller}}, \bibinfo
  {author} {\bibfnamefont {A.~W.}\ \bibnamefont {Carr}}, \bibinfo {author}
  {\bibfnamefont {M.~J.}\ \bibnamefont {Piotrowicz}}, \bibinfo {author}
  {\bibfnamefont {L.}~\bibnamefont {Isenhower}},\ and\ \bibinfo {author}
  {\bibfnamefont {M.}~\bibnamefont {Saffman}},\ }\bibfield  {title} {\enquote
  {\bibinfo {title} {{Randomized Benchmarking of Single-Qubit Gates in a 2D
  Array of Neutral-Atom Qubits}},}\ }\href
  {https://doi.org/10.1103/PhysRevLett.114.100503} {\bibfield  {journal}
  {\bibinfo  {journal} {Phys. Rev. Lett.}\ }\textbf {\bibinfo {volume} {114}},\
  \bibinfo {pages} {100503} (\bibinfo {year} {2015})}\BibitemShut {NoStop}%
\bibitem [{\citenamefont {Bluvstein}\ \emph
  {et~al.}(2024{\natexlab{b}})\citenamefont {Bluvstein}, \citenamefont
  {Evered}, \citenamefont {Geim}, \citenamefont {Li}, \citenamefont {Zhou},
  \citenamefont {Manovitz}, \citenamefont {Ebadi}, \citenamefont {Cain},
  \citenamefont {Kalinowski}, \citenamefont {Hangleiter}, \citenamefont
  {Bonilla~Ataides}, \citenamefont {Maskara}, \citenamefont {Cong},
  \citenamefont {Gao}, \citenamefont {Sales~Rodriguez}, \citenamefont
  {Karolyshyn}, \citenamefont {Semeghini}, \citenamefont {Gullans},
  \citenamefont {Greiner}, \citenamefont {Vuleti{\'c}},\ and\ \citenamefont
  {Lukin}}]{bluvstein2024logical}%
  \BibitemOpen
  \bibfield  {author} {\bibinfo {author} {\bibfnamefont {D.}~\bibnamefont
  {Bluvstein}}, \bibinfo {author} {\bibfnamefont {S.~J.}\ \bibnamefont
  {Evered}}, \bibinfo {author} {\bibfnamefont {A.~A.}\ \bibnamefont {Geim}},
  \bibinfo {author} {\bibfnamefont {S.~H.}\ \bibnamefont {Li}}, \bibinfo
  {author} {\bibfnamefont {H.}~\bibnamefont {Zhou}}, \bibinfo {author}
  {\bibfnamefont {T.}~\bibnamefont {Manovitz}}, \bibinfo {author}
  {\bibfnamefont {S.}~\bibnamefont {Ebadi}}, \bibinfo {author} {\bibfnamefont
  {M.}~\bibnamefont {Cain}}, \bibinfo {author} {\bibfnamefont {M.}~\bibnamefont
  {Kalinowski}}, \bibinfo {author} {\bibfnamefont {D.}~\bibnamefont
  {Hangleiter}}, \bibinfo {author} {\bibfnamefont {J.~P.}\ \bibnamefont
  {Bonilla~Ataides}}, \bibinfo {author} {\bibfnamefont {N.}~\bibnamefont
  {Maskara}}, \bibinfo {author} {\bibfnamefont {I.}~\bibnamefont {Cong}},
  \bibinfo {author} {\bibfnamefont {X.}~\bibnamefont {Gao}}, \bibinfo {author}
  {\bibfnamefont {P.}~\bibnamefont {Sales~Rodriguez}}, \bibinfo {author}
  {\bibfnamefont {T.}~\bibnamefont {Karolyshyn}}, \bibinfo {author}
  {\bibfnamefont {G.}~\bibnamefont {Semeghini}}, \bibinfo {author}
  {\bibfnamefont {M.~J.}\ \bibnamefont {Gullans}}, \bibinfo {author}
  {\bibfnamefont {M.}~\bibnamefont {Greiner}}, \bibinfo {author} {\bibfnamefont
  {V.}~\bibnamefont {Vuleti{\'c}}},\ and\ \bibinfo {author} {\bibfnamefont
  {M.~D.}\ \bibnamefont {Lukin}},\ }\bibfield  {title} {\enquote {\bibinfo
  {title} {{Logical quantum processor based on reconfigurable atom arrays}},}\
  }\href {https://doi.org/10.1038/s41586-023-06927-3} {\bibfield  {journal}
  {\bibinfo  {journal} {Nature}\ }\textbf {\bibinfo {volume} {626}},\ \bibinfo
  {pages} {58} (\bibinfo {year} {2024}{\natexlab{b}})}\BibitemShut {NoStop}%
\bibitem [{\citenamefont {Ma}\ \emph {et~al.}(2022)\citenamefont {Ma},
  \citenamefont {Burgers}, \citenamefont {Liu}, \citenamefont {Wilson},
  \citenamefont {Zhang},\ and\ \citenamefont {Thompson}}]{Ma:2022}%
  \BibitemOpen
  \bibfield  {author} {\bibinfo {author} {\bibfnamefont {S.}~\bibnamefont
  {Ma}}, \bibinfo {author} {\bibfnamefont {A.~P.}\ \bibnamefont {Burgers}},
  \bibinfo {author} {\bibfnamefont {G.}~\bibnamefont {Liu}}, \bibinfo {author}
  {\bibfnamefont {J.}~\bibnamefont {Wilson}}, \bibinfo {author} {\bibfnamefont
  {B.}~\bibnamefont {Zhang}},\ and\ \bibinfo {author} {\bibfnamefont {J.~D.}\
  \bibnamefont {Thompson}},\ }\bibfield  {title} {\enquote {\bibinfo {title}
  {{Universal Gate Operations on Nuclear Spin Qubits in an Optical Tweezer
  Array of ${}^{171}$Yb Atoms}},}\ }\href
  {https://doi.org/10.1103/PhysRevX.12.021028} {\bibfield  {journal} {\bibinfo
  {journal} {Phys. Rev. X}\ }\textbf {\bibinfo {volume} {12}},\ \bibinfo
  {pages} {021028} (\bibinfo {year} {2022})}\BibitemShut {NoStop}%
\bibitem [{\citenamefont {Jenkins}\ \emph {et~al.}(2022)\citenamefont
  {Jenkins}, \citenamefont {Lis}, \citenamefont {Senoo}, \citenamefont
  {McGrew},\ and\ \citenamefont {Kaufman}}]{Jenkins:2022}%
  \BibitemOpen
  \bibfield  {author} {\bibinfo {author} {\bibfnamefont {A.}~\bibnamefont
  {Jenkins}}, \bibinfo {author} {\bibfnamefont {J.~W.}\ \bibnamefont {Lis}},
  \bibinfo {author} {\bibfnamefont {A.}~\bibnamefont {Senoo}}, \bibinfo
  {author} {\bibfnamefont {W.~F.}\ \bibnamefont {McGrew}},\ and\ \bibinfo
  {author} {\bibfnamefont {A.~M.}\ \bibnamefont {Kaufman}},\ }\bibfield
  {title} {\enquote {\bibinfo {title} {{Ytterbium Nuclear-Spin Qubits in an
  Optical Tweezer Array}},}\ }\href
  {https://doi.org/10.1103/PhysRevX.12.021027} {\bibfield  {journal} {\bibinfo
  {journal} {Phys. Rev. X}\ }\textbf {\bibinfo {volume} {12}},\ \bibinfo
  {pages} {021027} (\bibinfo {year} {2022})}\BibitemShut {NoStop}%
\bibitem [{\citenamefont {Pucher}\ \emph {et~al.}(2024)\citenamefont {Pucher},
  \citenamefont {Kl{\"u}sener}, \citenamefont {Spriestersbach}, \citenamefont
  {Geiger}, \citenamefont {Schindewolf}, \citenamefont {Bloch},\ and\
  \citenamefont {Blatt}}]{Pucher:2024a}%
  \BibitemOpen
  \bibfield  {author} {\bibinfo {author} {\bibfnamefont {S.}~\bibnamefont
  {Pucher}}, \bibinfo {author} {\bibfnamefont {V.}~\bibnamefont
  {Kl{\"u}sener}}, \bibinfo {author} {\bibfnamefont {F.}~\bibnamefont
  {Spriestersbach}}, \bibinfo {author} {\bibfnamefont {J.}~\bibnamefont
  {Geiger}}, \bibinfo {author} {\bibfnamefont {A.}~\bibnamefont {Schindewolf}},
  \bibinfo {author} {\bibfnamefont {I.}~\bibnamefont {Bloch}},\ and\ \bibinfo
  {author} {\bibfnamefont {S.}~\bibnamefont {Blatt}},\ }\bibfield  {title}
  {\enquote {\bibinfo {title} {{Fine-Structure Qubit Encoded in Metastable
  Strontium Trapped in an Optical Lattice}},}\ }\href
  {https://doi.org/10.1103/PhysRevLett.132.150605} {\bibfield  {journal}
  {\bibinfo  {journal} {Phys. Rev. Lett.}\ }\textbf {\bibinfo {volume} {132}},\
  \bibinfo {pages} {150605} (\bibinfo {year} {2024})}\BibitemShut {NoStop}%
\bibitem [{\citenamefont {Unnikrishnan}\ \emph {et~al.}(2024)\citenamefont
  {Unnikrishnan}, \citenamefont {Ilzh{\"o}fer}, \citenamefont {Scholz},
  \citenamefont {H{\"o}lzl}, \citenamefont {G{\"o}tzelmann}, \citenamefont
  {Gupta}, \citenamefont {Zhao}, \citenamefont {Krauter}, \citenamefont
  {Weber}, \citenamefont {Makki}, \citenamefont {B{\"u}chler}, \citenamefont
  {Pfau},\ and\ \citenamefont {Meinert}}]{PhysRevLett.132.150606}%
  \BibitemOpen
  \bibfield  {author} {\bibinfo {author} {\bibfnamefont {G.}~\bibnamefont
  {Unnikrishnan}}, \bibinfo {author} {\bibfnamefont {P.}~\bibnamefont
  {Ilzh{\"o}fer}}, \bibinfo {author} {\bibfnamefont {A.}~\bibnamefont
  {Scholz}}, \bibinfo {author} {\bibfnamefont {C.}~\bibnamefont {H{\"o}lzl}},
  \bibinfo {author} {\bibfnamefont {A.}~\bibnamefont {G{\"o}tzelmann}},
  \bibinfo {author} {\bibfnamefont {R.~K.}\ \bibnamefont {Gupta}}, \bibinfo
  {author} {\bibfnamefont {J.}~\bibnamefont {Zhao}}, \bibinfo {author}
  {\bibfnamefont {J.}~\bibnamefont {Krauter}}, \bibinfo {author} {\bibfnamefont
  {S.}~\bibnamefont {Weber}}, \bibinfo {author} {\bibfnamefont
  {N.}~\bibnamefont {Makki}}, \bibinfo {author} {\bibfnamefont {H.~P.}\
  \bibnamefont {B{\"u}chler}}, \bibinfo {author} {\bibfnamefont
  {T.}~\bibnamefont {Pfau}},\ and\ \bibinfo {author} {\bibfnamefont
  {F.}~\bibnamefont {Meinert}},\ }\bibfield  {title} {\enquote {\bibinfo
  {title} {{Coherent Control of the Fine-Structure Qubit in a Single
  Alkaline-Earth Atom}},}\ }\href
  {https://doi.org/10.1103/PhysRevLett.132.150606} {\bibfield  {journal}
  {\bibinfo  {journal} {Phys. Rev. Lett.}\ }\textbf {\bibinfo {volume} {132}},\
  \bibinfo {pages} {150606} (\bibinfo {year} {2024})}\BibitemShut {NoStop}%
\bibitem [{\citenamefont {Young}\ \emph {et~al.}(2020)\citenamefont {Young},
  \citenamefont {Eckner}, \citenamefont {Milner}, \citenamefont {Kedar},
  \citenamefont {Norcia}, \citenamefont {Oelker}, \citenamefont {Schine},
  \citenamefont {Ye},\ and\ \citenamefont {Kaufman}}]{Young:2020}%
  \BibitemOpen
  \bibfield  {author} {\bibinfo {author} {\bibfnamefont {A.~W.}\ \bibnamefont
  {Young}}, \bibinfo {author} {\bibfnamefont {W.~J.}\ \bibnamefont {Eckner}},
  \bibinfo {author} {\bibfnamefont {W.~R.}\ \bibnamefont {Milner}}, \bibinfo
  {author} {\bibfnamefont {D.}~\bibnamefont {Kedar}}, \bibinfo {author}
  {\bibfnamefont {M.~A.}\ \bibnamefont {Norcia}}, \bibinfo {author}
  {\bibfnamefont {E.}~\bibnamefont {Oelker}}, \bibinfo {author} {\bibfnamefont
  {N.}~\bibnamefont {Schine}}, \bibinfo {author} {\bibfnamefont
  {J.}~\bibnamefont {Ye}},\ and\ \bibinfo {author} {\bibfnamefont {A.~M.}\
  \bibnamefont {Kaufman}},\ }\bibfield  {title} {\enquote {\bibinfo {title}
  {{Half-minute-scale atomic coherence and high relative stability in a tweezer
  clock}},}\ }\href {https://doi.org/10.1038/s41586-020-3009-y} {\bibfield
  {journal} {\bibinfo  {journal} {Nature}\ }\textbf {\bibinfo {volume} {588}},\
  \bibinfo {pages} {408} (\bibinfo {year} {2020})}\BibitemShut {NoStop}%
\bibitem [{\citenamefont {Madjarov}\ \emph {et~al.}(2019)\citenamefont
  {Madjarov}, \citenamefont {Cooper}, \citenamefont {Shaw}, \citenamefont
  {Covey}, \citenamefont {Schkolnik}, \citenamefont {Yoon}, \citenamefont
  {Williams},\ and\ \citenamefont {Endres}}]{PhysRevX.9.041052}%
  \BibitemOpen
  \bibfield  {author} {\bibinfo {author} {\bibfnamefont {I.~S.}\ \bibnamefont
  {Madjarov}}, \bibinfo {author} {\bibfnamefont {A.}~\bibnamefont {Cooper}},
  \bibinfo {author} {\bibfnamefont {A.~L.}\ \bibnamefont {Shaw}}, \bibinfo
  {author} {\bibfnamefont {J.~P.}\ \bibnamefont {Covey}}, \bibinfo {author}
  {\bibfnamefont {V.}~\bibnamefont {Schkolnik}}, \bibinfo {author}
  {\bibfnamefont {T.~H.}\ \bibnamefont {Yoon}}, \bibinfo {author}
  {\bibfnamefont {J.~R.}\ \bibnamefont {Williams}},\ and\ \bibinfo {author}
  {\bibfnamefont {M.}~\bibnamefont {Endres}},\ }\bibfield  {title} {\enquote
  {\bibinfo {title} {{An Atomic-Array Optical Clock with Single-Atom
  Readout}},}\ }\href {https://doi.org/10.1103/PhysRevX.9.041052} {\bibfield
  {journal} {\bibinfo  {journal} {Phys. Rev. X}\ }\textbf {\bibinfo {volume}
  {9}},\ \bibinfo {pages} {041052} (\bibinfo {year} {2019})}\BibitemShut
  {NoStop}%
\bibitem [{\citenamefont {Evered}\ \emph {et~al.}(2023)\citenamefont {Evered},
  \citenamefont {Bluvstein}, \citenamefont {Kalinowski}, \citenamefont {Ebadi},
  \citenamefont {Manovitz}, \citenamefont {Zhou}, \citenamefont {Li},
  \citenamefont {Geim}, \citenamefont {Wang}, \citenamefont {Maskara},
  \citenamefont {Levine}, \citenamefont {Semeghini}, \citenamefont {Greiner},
  \citenamefont {Vuleti{\'c}},\ and\ \citenamefont
  {Lukin}}]{everedHighfidelityParallelEntangling2023}%
  \BibitemOpen
  \bibfield  {author} {\bibinfo {author} {\bibfnamefont {S.~J.}\ \bibnamefont
  {Evered}}, \bibinfo {author} {\bibfnamefont {D.}~\bibnamefont {Bluvstein}},
  \bibinfo {author} {\bibfnamefont {M.}~\bibnamefont {Kalinowski}}, \bibinfo
  {author} {\bibfnamefont {S.}~\bibnamefont {Ebadi}}, \bibinfo {author}
  {\bibfnamefont {T.}~\bibnamefont {Manovitz}}, \bibinfo {author}
  {\bibfnamefont {H.}~\bibnamefont {Zhou}}, \bibinfo {author} {\bibfnamefont
  {S.~H.}\ \bibnamefont {Li}}, \bibinfo {author} {\bibfnamefont {A.~A.}\
  \bibnamefont {Geim}}, \bibinfo {author} {\bibfnamefont {T.~T.}\ \bibnamefont
  {Wang}}, \bibinfo {author} {\bibfnamefont {N.}~\bibnamefont {Maskara}},
  \bibinfo {author} {\bibfnamefont {H.}~\bibnamefont {Levine}}, \bibinfo
  {author} {\bibfnamefont {G.}~\bibnamefont {Semeghini}}, \bibinfo {author}
  {\bibfnamefont {M.}~\bibnamefont {Greiner}}, \bibinfo {author} {\bibfnamefont
  {V.}~\bibnamefont {Vuleti{\'c}}},\ and\ \bibinfo {author} {\bibfnamefont
  {M.~D.}\ \bibnamefont {Lukin}},\ }\bibfield  {title} {\enquote {\bibinfo
  {title} {{High-fidelity parallel entangling gates on a neutral-atom quantum
  computer}},}\ }\href {https://doi.org/10.1038/s41586-023-06481-y} {\bibfield
  {journal} {\bibinfo  {journal} {Nature}\ }\textbf {\bibinfo {volume} {622}},\
  \bibinfo {pages} {268} (\bibinfo {year} {2023})}\BibitemShut {NoStop}%
\bibitem [{\citenamefont {Finkelstein}\ \emph {et~al.}(2024)\citenamefont
  {Finkelstein}, \citenamefont {Bing-Shiun~Tsai}, \citenamefont {Sun},
  \citenamefont {Scholl}, \citenamefont {Direkci}, \citenamefont {Gefen},
  \citenamefont {Choi}, \citenamefont {Shaw},\ and\ \citenamefont
  {Endres}}]{Finkelstein:2024}%
  \BibitemOpen
  \bibfield  {author} {\bibinfo {author} {\bibfnamefont {R.}~\bibnamefont
  {Finkelstein}}, \bibinfo {author} {\bibfnamefont {R.}~\bibnamefont
  {Bing-Shiun~Tsai}}, \bibinfo {author} {\bibfnamefont {X.}~\bibnamefont
  {Sun}}, \bibinfo {author} {\bibfnamefont {P.}~\bibnamefont {Scholl}},
  \bibinfo {author} {\bibfnamefont {S.}~\bibnamefont {Direkci}}, \bibinfo
  {author} {\bibfnamefont {T.}~\bibnamefont {Gefen}}, \bibinfo {author}
  {\bibfnamefont {J.}~\bibnamefont {Choi}}, \bibinfo {author} {\bibfnamefont
  {A.~L.}\ \bibnamefont {Shaw}},\ and\ \bibinfo {author} {\bibfnamefont
  {M.}~\bibnamefont {Endres}},\ }\href@noop {} {\enquote {\bibinfo {title}
  {{Universal quantum operations and ancilla-based readout for tweezer
  clocks}},}\ } (\bibinfo {year} {2024}),\ \Eprint
  {https://arxiv.org/abs/2402.16220} {arXiv:2402.16220 [quant-ph]} \BibitemShut
  {NoStop}%
\bibitem [{\citenamefont {Robicheaux}\ \emph {et~al.}(2021)\citenamefont
  {Robicheaux}, \citenamefont {Graham},\ and\ \citenamefont
  {Saffman}}]{robicheauxPhotonrecoilLaserfocusingLimits2021}%
  \BibitemOpen
  \bibfield  {author} {\bibinfo {author} {\bibfnamefont {F.}~\bibnamefont
  {Robicheaux}}, \bibinfo {author} {\bibfnamefont {T.~M.}\ \bibnamefont
  {Graham}},\ and\ \bibinfo {author} {\bibfnamefont {M.}~\bibnamefont
  {Saffman}},\ }\bibfield  {title} {\enquote {\bibinfo {title} {{Photon-recoil
  and laser-focusing limits to Rydberg gate fidelity}},}\ }\href
  {https://doi.org/10.1103/PhysRevA.103.022424} {\bibfield  {journal} {\bibinfo
   {journal} {Phys. Rev. A}\ }\textbf {\bibinfo {volume} {103}},\ \bibinfo
  {pages} {022424} (\bibinfo {year} {2021})}\BibitemShut {NoStop}%
\bibitem [{\citenamefont {Lis}\ \emph {et~al.}(2023)\citenamefont {Lis},
  \citenamefont {Senoo}, \citenamefont {McGrew}, \citenamefont {R{\"o}nchen},
  \citenamefont {Jenkins},\ and\ \citenamefont
  {Kaufman}}]{lisMidcircuitOperationsUsing2023}%
  \BibitemOpen
  \bibfield  {author} {\bibinfo {author} {\bibfnamefont {J.~W.}\ \bibnamefont
  {Lis}}, \bibinfo {author} {\bibfnamefont {A.}~\bibnamefont {Senoo}}, \bibinfo
  {author} {\bibfnamefont {W.~F.}\ \bibnamefont {McGrew}}, \bibinfo {author}
  {\bibfnamefont {F.}~\bibnamefont {R{\"o}nchen}}, \bibinfo {author}
  {\bibfnamefont {A.}~\bibnamefont {Jenkins}},\ and\ \bibinfo {author}
  {\bibfnamefont {A.~M.}\ \bibnamefont {Kaufman}},\ }\bibfield  {title}
  {\enquote {\bibinfo {title} {{Midcircuit Operations Using the omg
  Architecture in Neutral Atom Arrays}},}\ }\href
  {https://doi.org/10.1103/PhysRevX.13.041035} {\bibfield  {journal} {\bibinfo
  {journal} {Phys. Rev. X}\ }\textbf {\bibinfo {volume} {13}},\ \bibinfo
  {pages} {041035} (\bibinfo {year} {2023})}\BibitemShut {NoStop}%
\bibitem [{\citenamefont {Cummins}\ \emph {et~al.}(2003)\citenamefont
  {Cummins}, \citenamefont {Llewellyn},\ and\ \citenamefont
  {Jones}}]{cumminsTacklingSystematicErrors2003}%
  \BibitemOpen
  \bibfield  {author} {\bibinfo {author} {\bibfnamefont {H.~K.}\ \bibnamefont
  {Cummins}}, \bibinfo {author} {\bibfnamefont {G.}~\bibnamefont {Llewellyn}},\
  and\ \bibinfo {author} {\bibfnamefont {J.~A.}\ \bibnamefont {Jones}},\
  }\bibfield  {title} {\enquote {\bibinfo {title} {{Tackling systematic errors
  in quantum logic gates with composite rotations}},}\ }\href
  {https://doi.org/10.1103/PhysRevA.67.042308} {\bibfield  {journal} {\bibinfo
  {journal} {Phys. Rev. A}\ }\textbf {\bibinfo {volume} {67}},\ \bibinfo
  {pages} {042308} (\bibinfo {year} {2003})}\BibitemShut {NoStop}%
\bibitem [{\citenamefont {Zhang}\ \emph {et~al.}(2024)\citenamefont {Zhang},
  \citenamefont {Van~Damme}, \citenamefont {Rossignolo}, \citenamefont {Festa},
  \citenamefont {Melchner}, \citenamefont {Eberhard}, \citenamefont {Tsevas},
  \citenamefont {Mours}, \citenamefont {Reches}, \citenamefont {Zeiher},
  \citenamefont {Blatt}, \citenamefont {Bloch}, \citenamefont {Glaser},\ and\
  \citenamefont {Alberti}}]{Zhang2024}%
  \BibitemOpen
  \bibfield  {author} {\bibinfo {author} {\bibfnamefont {Z.}~\bibnamefont
  {Zhang}}, \bibinfo {author} {\bibfnamefont {L.}~\bibnamefont {Van~Damme}},
  \bibinfo {author} {\bibfnamefont {M.}~\bibnamefont {Rossignolo}}, \bibinfo
  {author} {\bibfnamefont {L.}~\bibnamefont {Festa}}, \bibinfo {author}
  {\bibfnamefont {M.}~\bibnamefont {Melchner}}, \bibinfo {author}
  {\bibfnamefont {R.}~\bibnamefont {Eberhard}}, \bibinfo {author}
  {\bibfnamefont {D.}~\bibnamefont {Tsevas}}, \bibinfo {author} {\bibfnamefont
  {K.}~\bibnamefont {Mours}}, \bibinfo {author} {\bibfnamefont
  {E.}~\bibnamefont {Reches}}, \bibinfo {author} {\bibfnamefont
  {J.}~\bibnamefont {Zeiher}}, \bibinfo {author} {\bibfnamefont
  {S.}~\bibnamefont {Blatt}}, \bibinfo {author} {\bibfnamefont
  {I.}~\bibnamefont {Bloch}}, \bibinfo {author} {\bibfnamefont {S.~J.}\
  \bibnamefont {Glaser}},\ and\ \bibinfo {author} {\bibfnamefont
  {A.}~\bibnamefont {Alberti}},\ }\href@noop {} {\enquote {\bibinfo {title}
  {{Recoil-free Quantum Gates with Optical Qubits}},}\ } (\bibinfo {year}
  {2024}),\ \Eprint {https://arxiv.org/abs/2408.04622} {arXiv:2408.04622
  [quant-ph]} \BibitemShut {NoStop}%
\bibitem [{\citenamefont {Haeberlen}\ and\ \citenamefont
  {Waugh}(1968)}]{Haeberlen68}%
  \BibitemOpen
  \bibfield  {author} {\bibinfo {author} {\bibfnamefont {U.}~\bibnamefont
  {Haeberlen}}\ and\ \bibinfo {author} {\bibfnamefont {J.~S.}\ \bibnamefont
  {Waugh}},\ }\bibfield  {title} {\enquote {\bibinfo {title} {Coherent
  averaging effects in magnetic resonance},}\ }\href
  {https://doi.org/10.1103/PhysRev.175.453} {\bibfield  {journal} {\bibinfo
  {journal} {Phys. Rev.}\ }\textbf {\bibinfo {volume} {175}},\ \bibinfo {pages}
  {453} (\bibinfo {year} {1968})}\BibitemShut {NoStop}%
\bibitem [{\citenamefont {Brinkmann}(2016)}]{BrinkmannAVH}%
  \BibitemOpen
  \bibfield  {author} {\bibinfo {author} {\bibfnamefont {A.}~\bibnamefont
  {Brinkmann}},\ }\bibfield  {title} {\enquote {\bibinfo {title} {{Introduction
  to average Hamiltonian theory. I. Basics}},}\ }\href
  {https://doi.org/https://doi.org/10.1002/cmr.a.21414} {\bibfield  {journal}
  {\bibinfo  {journal} {Concepts Magn. Reson. Part A}\ }\textbf {\bibinfo
  {volume} {45A}},\ \bibinfo {pages} {e21414} (\bibinfo {year}
  {2016})}\BibitemShut {NoStop}%
\bibitem [{\citenamefont {Lizuain}\ \emph {et~al.}(2007)\citenamefont
  {Lizuain}, \citenamefont {Muga},\ and\ \citenamefont
  {Eschner}}]{lizuainMotionalFrequencyShifts2007}%
  \BibitemOpen
  \bibfield  {author} {\bibinfo {author} {\bibfnamefont {I.}~\bibnamefont
  {Lizuain}}, \bibinfo {author} {\bibfnamefont {J.~G.}\ \bibnamefont {Muga}},\
  and\ \bibinfo {author} {\bibfnamefont {J.}~\bibnamefont {Eschner}},\
  }\bibfield  {title} {\enquote {\bibinfo {title} {{Motional frequency shifts
  of trapped ions in the Lamb-Dicke regime}},}\ }\href
  {https://doi.org/10.1103/PhysRevA.76.033808} {\bibfield  {journal} {\bibinfo
  {journal} {Phys. Rev. A}\ }\textbf {\bibinfo {volume} {76}},\ \bibinfo
  {pages} {033808} (\bibinfo {year} {2007})}\BibitemShut {NoStop}%
\bibitem [{\citenamefont {Heinz}(2020)}]{ediss26329}%
  \BibitemOpen
  \bibfield  {author} {\bibinfo {author} {\bibfnamefont {A.}~\bibnamefont
  {Heinz}},\ }\emph {\bibinfo {title} {{Ultracold strontium in state-dependent
  optical lattices}}},\ \href@noop {} {Ph.D. thesis},\ \bibinfo  {school}
  {Ludwig-Maximilians-Universit{\"a}t M{\"u}nchen} (\bibinfo {year}
  {2020})\BibitemShut {NoStop}%
\bibitem [{\citenamefont {Legero}\ \emph {et~al.}(2007)\citenamefont {Legero},
  \citenamefont {Winfred}, \citenamefont {Riehle},\ and\ \citenamefont
  {Sterr}}]{4319045}%
  \BibitemOpen
  \bibfield  {author} {\bibinfo {author} {\bibfnamefont {T.}~\bibnamefont
  {Legero}}, \bibinfo {author} {\bibfnamefont {J.~S. R.~V.}\ \bibnamefont
  {Winfred}}, \bibinfo {author} {\bibfnamefont {F.}~\bibnamefont {Riehle}},\
  and\ \bibinfo {author} {\bibfnamefont {U.}~\bibnamefont {Sterr}},\ }\bibfield
   {title} {\enquote {\bibinfo {title} {{Ultracold ${}^{88}$Sr atoms for an
  optical lattice clock}},}\ }in\ \href
  {https://doi.org/10.1109/FREQ.2007.4319045} {\emph {\bibinfo {booktitle}
  {2007 IEEE International Frequency Control Symposium Joint with the 21st
  European Frequency and Time Forum}}}\ (\bibinfo {year} {2007})\ pp.\ \bibinfo
  {pages} {119--122}\BibitemShut {NoStop}%
\bibitem [{\citenamefont {Tao}\ \emph {et~al.}(2024)\citenamefont {Tao},
  \citenamefont {Ammenwerth}, \citenamefont {Gyger}, \citenamefont {Bloch},\
  and\ \citenamefont {Zeiher}}]{PhysRevLett.133.013401}%
  \BibitemOpen
  \bibfield  {author} {\bibinfo {author} {\bibfnamefont {R.}~\bibnamefont
  {Tao}}, \bibinfo {author} {\bibfnamefont {M.}~\bibnamefont {Ammenwerth}},
  \bibinfo {author} {\bibfnamefont {F.}~\bibnamefont {Gyger}}, \bibinfo
  {author} {\bibfnamefont {I.}~\bibnamefont {Bloch}},\ and\ \bibinfo {author}
  {\bibfnamefont {J.}~\bibnamefont {Zeiher}},\ }\bibfield  {title} {\enquote
  {\bibinfo {title} {{High-Fidelity Detection of Large-Scale Atom Arrays in an
  Optical Lattice}},}\ }\href {https://doi.org/10.1103/PhysRevLett.133.013401}
  {\bibfield  {journal} {\bibinfo  {journal} {Phys. Rev. Lett.}\ }\textbf
  {\bibinfo {volume} {133}},\ \bibinfo {pages} {013401} (\bibinfo {year}
  {2024})}\BibitemShut {NoStop}%
\bibitem [{\citenamefont {Dridi}\ \emph {et~al.}(2020)\citenamefont {Dridi},
  \citenamefont {Mejatty}, \citenamefont {Glaser},\ and\ \citenamefont
  {Sugny}}]{dridiRobustControlNot2020}%
  \BibitemOpen
  \bibfield  {author} {\bibinfo {author} {\bibfnamefont {G.}~\bibnamefont
  {Dridi}}, \bibinfo {author} {\bibfnamefont {M.}~\bibnamefont {Mejatty}},
  \bibinfo {author} {\bibfnamefont {S.~J.}\ \bibnamefont {Glaser}},\ and\
  \bibinfo {author} {\bibfnamefont {D.}~\bibnamefont {Sugny}},\ }\bibfield
  {title} {\enquote {\bibinfo {title} {Robust control of a not gate by
  composite pulses},}\ }\href {https://doi.org/10.1103/PhysRevA.101.012321}
  {\bibfield  {journal} {\bibinfo  {journal} {Phys. Rev. A}\ }\textbf {\bibinfo
  {volume} {101}},\ \bibinfo {pages} {012321} (\bibinfo {year}
  {2020})}\BibitemShut {NoStop}%
\bibitem [{\citenamefont {Torosov}\ and\ \citenamefont
  {Vitanov}(2022)}]{PhysRevApplied.18.034062}%
  \BibitemOpen
  \bibfield  {author} {\bibinfo {author} {\bibfnamefont {B.~T.}\ \bibnamefont
  {Torosov}}\ and\ \bibinfo {author} {\bibfnamefont {N.~V.}\ \bibnamefont
  {Vitanov}},\ }\bibfield  {title} {\enquote {\bibinfo {title} {Experimental
  demonstration of composite pulses on ibm's quantum computer},}\ }\href
  {https://doi.org/10.1103/PhysRevApplied.18.034062} {\bibfield  {journal}
  {\bibinfo  {journal} {Phys. Rev. Appl.}\ }\textbf {\bibinfo {volume} {18}},\
  \bibinfo {pages} {034062} (\bibinfo {year} {2022})}\BibitemShut {NoStop}%
\bibitem [{\citenamefont {Wu}\ \emph {et~al.}(2023)\citenamefont {Wu},
  \citenamefont {Zhang}, \citenamefont {Song}, \citenamefont {Xia},\ and\
  \citenamefont {Shi}}]{wuCompositePulsesOptimal2023}%
  \BibitemOpen
  \bibfield  {author} {\bibinfo {author} {\bibfnamefont {H.-N.}\ \bibnamefont
  {Wu}}, \bibinfo {author} {\bibfnamefont {C.}~\bibnamefont {Zhang}}, \bibinfo
  {author} {\bibfnamefont {J.}~\bibnamefont {Song}}, \bibinfo {author}
  {\bibfnamefont {Y.}~\bibnamefont {Xia}},\ and\ \bibinfo {author}
  {\bibfnamefont {Z.-C.}\ \bibnamefont {Shi}},\ }\bibfield  {title} {\enquote
  {\bibinfo {title} {{Composite pulses for optimal robust control in two-level
  systems}},}\ }\href {https://doi.org/10.1103/PhysRevA.107.023103} {\bibfield
  {journal} {\bibinfo  {journal} {Phys. Rev. A}\ }\textbf {\bibinfo {volume}
  {107}},\ \bibinfo {pages} {023103} (\bibinfo {year} {2023})}\BibitemShut
  {NoStop}%
\bibitem [{\citenamefont {Genov}\ \emph {et~al.}(2014)\citenamefont {Genov},
  \citenamefont {Schraft}, \citenamefont {Halfmann},\ and\ \citenamefont
  {Vitanov}}]{genovCorrectionArbitraryField2014}%
  \BibitemOpen
  \bibfield  {author} {\bibinfo {author} {\bibfnamefont {G.~T.}\ \bibnamefont
  {Genov}}, \bibinfo {author} {\bibfnamefont {D.}~\bibnamefont {Schraft}},
  \bibinfo {author} {\bibfnamefont {T.}~\bibnamefont {Halfmann}},\ and\
  \bibinfo {author} {\bibfnamefont {N.~V.}\ \bibnamefont {Vitanov}},\
  }\bibfield  {title} {\enquote {\bibinfo {title} {{Correction of Arbitrary
  Field Errors in Population Inversion of Quantum Systems by Universal
  Composite Pulses}},}\ }\href {https://doi.org/10.1103/PhysRevLett.113.043001}
  {\bibfield  {journal} {\bibinfo  {journal} {Phys. Rev. Lett.}\ }\textbf
  {\bibinfo {volume} {113}},\ \bibinfo {pages} {043001} (\bibinfo {year}
  {2014})}\BibitemShut {NoStop}%
\bibitem [{\citenamefont {Jones}(2013)}]{jonesDesigningShortRobust2013}%
  \BibitemOpen
  \bibfield  {author} {\bibinfo {author} {\bibfnamefont {J.~A.}\ \bibnamefont
  {Jones}},\ }\bibfield  {title} {\enquote {\bibinfo {title} {{Designing short
  robust not gates for quantum computation}},}\ }\href
  {https://doi.org/10.1103/PhysRevA.87.052317} {\bibfield  {journal} {\bibinfo
  {journal} {Phys. Rev. A}\ }\textbf {\bibinfo {volume} {87}},\ \bibinfo
  {pages} {052317} (\bibinfo {year} {2013})}\BibitemShut {NoStop}%
\bibitem [{\citenamefont {Levitt}(1986)}]{LEVITT198661}%
  \BibitemOpen
  \bibfield  {author} {\bibinfo {author} {\bibfnamefont {M.~H.}\ \bibnamefont
  {Levitt}},\ }\bibfield  {title} {\enquote {\bibinfo {title} {Composite
  pulses},}\ }\href
  {https://doi.org/https://doi.org/10.1016/0079-6565(86)80005-X} {\bibfield
  {journal} {\bibinfo  {journal} {Prog. Nucl. Magn.}\ }\textbf {\bibinfo
  {volume} {18}},\ \bibinfo {pages} {61} (\bibinfo {year} {1986})}\BibitemShut
  {NoStop}%
\bibitem [{\citenamefont {Ran}\ \emph {et~al.}(2020)\citenamefont {Ran},
  \citenamefont {Shan}, \citenamefont {Shi}, \citenamefont {Yang},
  \citenamefont {Song},\ and\ \citenamefont {Xia}}]{PhysRevA.101.023822}%
  \BibitemOpen
  \bibfield  {author} {\bibinfo {author} {\bibfnamefont {D.}~\bibnamefont
  {Ran}}, \bibinfo {author} {\bibfnamefont {W.-J.}\ \bibnamefont {Shan}},
  \bibinfo {author} {\bibfnamefont {Z.-C.}\ \bibnamefont {Shi}}, \bibinfo
  {author} {\bibfnamefont {Z.-B.}\ \bibnamefont {Yang}}, \bibinfo {author}
  {\bibfnamefont {J.}~\bibnamefont {Song}},\ and\ \bibinfo {author}
  {\bibfnamefont {Y.}~\bibnamefont {Xia}},\ }\bibfield  {title} {\enquote
  {\bibinfo {title} {Pulse reverse engineering for controlling two-level
  quantum systems},}\ }\href {https://doi.org/10.1103/PhysRevA.101.023822}
  {\bibfield  {journal} {\bibinfo  {journal} {Phys. Rev. A}\ }\textbf {\bibinfo
  {volume} {101}},\ \bibinfo {pages} {023822} (\bibinfo {year}
  {2020})}\BibitemShut {NoStop}%
\bibitem [{\citenamefont {Daems}\ \emph {et~al.}(2013)\citenamefont {Daems},
  \citenamefont {Ruschhaupt}, \citenamefont {Sugny},\ and\ \citenamefont
  {Gu\'erin}}]{PhysRevLett.111.050404}%
  \BibitemOpen
  \bibfield  {author} {\bibinfo {author} {\bibfnamefont {D.}~\bibnamefont
  {Daems}}, \bibinfo {author} {\bibfnamefont {A.}~\bibnamefont {Ruschhaupt}},
  \bibinfo {author} {\bibfnamefont {D.}~\bibnamefont {Sugny}},\ and\ \bibinfo
  {author} {\bibfnamefont {S.}~\bibnamefont {Gu\'erin}},\ }\bibfield  {title}
  {\enquote {\bibinfo {title} {Robust quantum control by a single-shot shaped
  pulse},}\ }\href {https://doi.org/10.1103/PhysRevLett.111.050404} {\bibfield
  {journal} {\bibinfo  {journal} {Phys. Rev. Lett.}\ }\textbf {\bibinfo
  {volume} {111}},\ \bibinfo {pages} {050404} (\bibinfo {year}
  {2013})}\BibitemShut {NoStop}%
\bibitem [{\citenamefont {Müller}\ \emph {et~al.}(2022)\citenamefont
  {Müller}, \citenamefont {Said}, \citenamefont {Jelezko}, \citenamefont
  {Calarco},\ and\ \citenamefont {Montangero}}]{Muller_2022}%
  \BibitemOpen
  \bibfield  {author} {\bibinfo {author} {\bibfnamefont {M.~M.}\ \bibnamefont
  {Müller}}, \bibinfo {author} {\bibfnamefont {R.~S.}\ \bibnamefont {Said}},
  \bibinfo {author} {\bibfnamefont {F.}~\bibnamefont {Jelezko}}, \bibinfo
  {author} {\bibfnamefont {T.}~\bibnamefont {Calarco}},\ and\ \bibinfo {author}
  {\bibfnamefont {S.}~\bibnamefont {Montangero}},\ }\bibfield  {title}
  {\enquote {\bibinfo {title} {One decade of quantum optimal control in the
  chopped random basis},}\ }\href {https://doi.org/10.1088/1361-6633/ac723c}
  {\bibfield  {journal} {\bibinfo  {journal} {Rep. Prog. Phys.}\ }\textbf
  {\bibinfo {volume} {85}},\ \bibinfo {pages} {076001} (\bibinfo {year}
  {2022})}\BibitemShut {NoStop}%
\bibitem [{\citenamefont {Khaneja}\ \emph {et~al.}(2005)\citenamefont
  {Khaneja}, \citenamefont {Reiss}, \citenamefont {Kehlet}, \citenamefont
  {Schulte-Herbr{\"u}ggen},\ and\ \citenamefont
  {Glaser}}]{khanejaOptimalControlCoupled2005}%
  \BibitemOpen
  \bibfield  {author} {\bibinfo {author} {\bibfnamefont {N.}~\bibnamefont
  {Khaneja}}, \bibinfo {author} {\bibfnamefont {T.}~\bibnamefont {Reiss}},
  \bibinfo {author} {\bibfnamefont {C.}~\bibnamefont {Kehlet}}, \bibinfo
  {author} {\bibfnamefont {T.}~\bibnamefont {Schulte-Herbr{\"u}ggen}},\ and\
  \bibinfo {author} {\bibfnamefont {S.~J.}\ \bibnamefont {Glaser}},\ }\bibfield
   {title} {\enquote {\bibinfo {title} {{Optimal control of coupled spin
  dynamics: design of NMR pulse sequences by gradient ascent algorithms}},}\
  }\href {https://doi.org/10.1016/j.jmr.2004.11.004} {\bibfield  {journal}
  {\bibinfo  {journal} {J. Magn. Reson.}\ }\textbf {\bibinfo {volume} {172}},\
  \bibinfo {pages} {296} (\bibinfo {year} {2005})}\BibitemShut {NoStop}%
\bibitem [{\citenamefont {Skinner}\ and\ \citenamefont
  {Gershenzon}(2010)}]{skinnerOptimalControlDesign2010}%
  \BibitemOpen
  \bibfield  {author} {\bibinfo {author} {\bibfnamefont {T.~E.}\ \bibnamefont
  {Skinner}}\ and\ \bibinfo {author} {\bibfnamefont {N.~I.}\ \bibnamefont
  {Gershenzon}},\ }\bibfield  {title} {\enquote {\bibinfo {title} {{Optimal
  control design of pulse shapes as analytic functions}},}\ }\href
  {https://doi.org/10.1016/j.jmr.2010.03.002} {\bibfield  {journal} {\bibinfo
  {journal} {J. Magn. Reson.}\ }\textbf {\bibinfo {volume} {204}},\ \bibinfo
  {pages} {248} (\bibinfo {year} {2010})}\BibitemShut {NoStop}%
\bibitem [{\citenamefont {Barnes}\ \emph {et~al.}(2022)\citenamefont {Barnes},
  \citenamefont {Calderon-Vargas}, \citenamefont {Dong}, \citenamefont {Li},
  \citenamefont {Zeng},\ and\ \citenamefont
  {Zhuang}}]{barnesDynamicallyCorrectedGates2022}%
  \BibitemOpen
  \bibfield  {author} {\bibinfo {author} {\bibfnamefont {E.}~\bibnamefont
  {Barnes}}, \bibinfo {author} {\bibfnamefont {F.~A.}\ \bibnamefont
  {Calderon-Vargas}}, \bibinfo {author} {\bibfnamefont {W.}~\bibnamefont
  {Dong}}, \bibinfo {author} {\bibfnamefont {B.}~\bibnamefont {Li}}, \bibinfo
  {author} {\bibfnamefont {J.}~\bibnamefont {Zeng}},\ and\ \bibinfo {author}
  {\bibfnamefont {F.}~\bibnamefont {Zhuang}},\ }\bibfield  {title} {\enquote
  {\bibinfo {title} {{Dynamically corrected gates from geometric space
  curves}},}\ }\href {https://doi.org/10.1088/2058-9565/ac4421} {\bibfield
  {journal} {\bibinfo  {journal} {Quantum Sci. Technol.}\ }\textbf {\bibinfo
  {volume} {7}},\ \bibinfo {pages} {023001} (\bibinfo {year}
  {2022})}\BibitemShut {NoStop}%
\bibitem [{\citenamefont {Nelson}\ \emph {et~al.}(2023)\citenamefont {Nelson},
  \citenamefont {Piliouras}, \citenamefont {Connelly},\ and\ \citenamefont
  {Barnes}}]{nelsonDesigningDynamicallyCorrected2023}%
  \BibitemOpen
  \bibfield  {author} {\bibinfo {author} {\bibfnamefont {H.~T.}\ \bibnamefont
  {Nelson}}, \bibinfo {author} {\bibfnamefont {E.}~\bibnamefont {Piliouras}},
  \bibinfo {author} {\bibfnamefont {K.}~\bibnamefont {Connelly}},\ and\
  \bibinfo {author} {\bibfnamefont {E.}~\bibnamefont {Barnes}},\ }\bibfield
  {title} {\enquote {\bibinfo {title} {{Designing dynamically corrected gates
  robust to multiple noise sources using geometric space curves}},}\ }\href
  {https://doi.org/10.1103/PhysRevA.108.012407} {\bibfield  {journal} {\bibinfo
   {journal} {Phys. Rev. A}\ }\textbf {\bibinfo {volume} {108}},\ \bibinfo
  {pages} {012407} (\bibinfo {year} {2023})}\BibitemShut {NoStop}%
\bibitem [{\citenamefont {Boscain}\ \emph {et~al.}(2021)\citenamefont
  {Boscain}, \citenamefont {Sigalotti},\ and\ \citenamefont
  {Sugny}}]{PRXQuantum.2.030203}%
  \BibitemOpen
  \bibfield  {author} {\bibinfo {author} {\bibfnamefont {U.}~\bibnamefont
  {Boscain}}, \bibinfo {author} {\bibfnamefont {M.}~\bibnamefont {Sigalotti}},\
  and\ \bibinfo {author} {\bibfnamefont {D.}~\bibnamefont {Sugny}},\ }\bibfield
   {title} {\enquote {\bibinfo {title} {Introduction to the pontryagin maximum
  principle for quantum optimal control},}\ }\href
  {https://doi.org/10.1103/PRXQuantum.2.030203} {\bibfield  {journal} {\bibinfo
   {journal} {PRX Quantum}\ }\textbf {\bibinfo {volume} {2}},\ \bibinfo {pages}
  {030203} (\bibinfo {year} {2021})}\BibitemShut {NoStop}%
\bibitem [{\citenamefont {Van~Damme}\ \emph {et~al.}(2017)\citenamefont
  {Van~Damme}, \citenamefont {Ansel}, \citenamefont {Glaser},\ and\
  \citenamefont {Sugny}}]{vandammeRobustOptimalControl2017}%
  \BibitemOpen
  \bibfield  {author} {\bibinfo {author} {\bibfnamefont {L.}~\bibnamefont
  {Van~Damme}}, \bibinfo {author} {\bibfnamefont {Q.}~\bibnamefont {Ansel}},
  \bibinfo {author} {\bibfnamefont {S.~J.}\ \bibnamefont {Glaser}},\ and\
  \bibinfo {author} {\bibfnamefont {D.}~\bibnamefont {Sugny}},\ }\bibfield
  {title} {\enquote {\bibinfo {title} {{Robust optimal control of two-level
  quantum systems}},}\ }\href {https://doi.org/10.1103/PhysRevA.95.063403}
  {\bibfield  {journal} {\bibinfo  {journal} {Phys. Rev. A}\ }\textbf {\bibinfo
  {volume} {95}},\ \bibinfo {pages} {063403} (\bibinfo {year}
  {2017})}\BibitemShut {NoStop}%
\bibitem [{\citenamefont {Dionis}\ and\ \citenamefont
  {Sugny}(2023)}]{dionisTimeoptimalControlTwolevel2023}%
  \BibitemOpen
  \bibfield  {author} {\bibinfo {author} {\bibfnamefont {E.}~\bibnamefont
  {Dionis}}\ and\ \bibinfo {author} {\bibfnamefont {D.}~\bibnamefont {Sugny}},\
  }\bibfield  {title} {\enquote {\bibinfo {title} {{Time-optimal control of
  two-level quantum systems by piecewise constant pulses}},}\ }\href
  {https://doi.org/10.1103/PhysRevA.107.032613} {\bibfield  {journal} {\bibinfo
   {journal} {Phys. Rev. A}\ }\textbf {\bibinfo {volume} {107}},\ \bibinfo
  {pages} {032613} (\bibinfo {year} {2023})}\BibitemShut {NoStop}%
\bibitem [{\citenamefont {Ansel}\ \emph {et~al.}(2024)\citenamefont {Ansel},
  \citenamefont {Dionis}, \citenamefont {Arrouas}, \citenamefont {Peaudecerf},
  \citenamefont {Guérin}, \citenamefont {Guéry-Odelin},\ and\ \citenamefont
  {Sugny}}]{Ansel_2024}%
  \BibitemOpen
  \bibfield  {author} {\bibinfo {author} {\bibfnamefont {Q.}~\bibnamefont
  {Ansel}}, \bibinfo {author} {\bibfnamefont {E.}~\bibnamefont {Dionis}},
  \bibinfo {author} {\bibfnamefont {F.}~\bibnamefont {Arrouas}}, \bibinfo
  {author} {\bibfnamefont {B.}~\bibnamefont {Peaudecerf}}, \bibinfo {author}
  {\bibfnamefont {S.}~\bibnamefont {Guérin}}, \bibinfo {author} {\bibfnamefont
  {D.}~\bibnamefont {Guéry-Odelin}},\ and\ \bibinfo {author} {\bibfnamefont
  {D.}~\bibnamefont {Sugny}},\ }\bibfield  {title} {\enquote {\bibinfo {title}
  {Introduction to theoretical and experimental aspects of quantum optimal
  control},}\ }\href {https://doi.org/10.1088/1361-6455/ad46a5} {\bibfield
  {journal} {\bibinfo  {journal} {J. Phys. B-At. Mol. Opt.}\ }\textbf {\bibinfo
  {volume} {57}},\ \bibinfo {pages} {133001} (\bibinfo {year}
  {2024})}\BibitemShut {NoStop}%
\bibitem [{\citenamefont {Koch}\ \emph {et~al.}(2022)\citenamefont {Koch},
  \citenamefont {Boscain}, \citenamefont {Calarco}, \citenamefont {Dirr},
  \citenamefont {Filipp}, \citenamefont {Glaser}, \citenamefont {Kosloff},
  \citenamefont {Montangero}, \citenamefont {Schulte-Herbrüggen},
  \citenamefont {Sugny},\ and\ \citenamefont {Wilhelm-Mauch}}]{Koch:909195}%
  \BibitemOpen
  \bibfield  {author} {\bibinfo {author} {\bibfnamefont {C.~P.}\ \bibnamefont
  {Koch}}, \bibinfo {author} {\bibfnamefont {U.}~\bibnamefont {Boscain}},
  \bibinfo {author} {\bibfnamefont {T.}~\bibnamefont {Calarco}}, \bibinfo
  {author} {\bibfnamefont {G.}~\bibnamefont {Dirr}}, \bibinfo {author}
  {\bibfnamefont {S.}~\bibnamefont {Filipp}}, \bibinfo {author} {\bibfnamefont
  {S.~J.}\ \bibnamefont {Glaser}}, \bibinfo {author} {\bibfnamefont
  {R.}~\bibnamefont {Kosloff}}, \bibinfo {author} {\bibfnamefont
  {S.}~\bibnamefont {Montangero}}, \bibinfo {author} {\bibfnamefont
  {T.}~\bibnamefont {Schulte-Herbrüggen}}, \bibinfo {author} {\bibfnamefont
  {D.}~\bibnamefont {Sugny}},\ and\ \bibinfo {author} {\bibfnamefont
  {F.}~\bibnamefont {Wilhelm-Mauch}},\ }\bibfield  {title} {\enquote {\bibinfo
  {title} {{Q}uantum optimal control in quantum technologies. {S}trategic
  report on current status, visions and goals for research in {E}urope},}\
  }\href {https://doi.org/10.1140/epjqt/s40507-022-00138-x} {\bibfield
  {journal} {\bibinfo  {journal} {EPJ Quantum Technol.}\ }\textbf {\bibinfo
  {volume} {9}},\ \bibinfo {pages} {19} (\bibinfo {year} {2022})}\BibitemShut
  {NoStop}%
\bibitem [{Note1()}]{Note1}%
  \BibitemOpen
  \bibinfo {note} {When $\omega /\Omega \lesssim 1$, we find solutions that are
  very similar to the bang-bang pulse of Fig.~\ref {FigSchemaTORF}, but a bit
  smooth. They even become singular when $\omega /\Omega \to 0$, in the sense
  that they contain segments where the pulse amplitude vanishes. However, these
  solutions are not relevant, as they arise in a regime where the motional
  sidebands are strongly excited, rendering the Average Hamiltonian Theory
  invalid, as briefly discussed in Appendix~\ref
  {AppPerturbTheory}.}\BibitemShut {Stop}%
\bibitem [{\citenamefont {Pontryagin}(1964)}]{PontryaginOCT}%
  \BibitemOpen
  \bibfield  {author} {\bibinfo {author} {\bibfnamefont {L.~S.}\ \bibnamefont
  {Pontryagin}},\ }\href@noop {} {\emph {\bibinfo {title} {{Mathematical theory
  of optimal processes}}}}\ (\bibinfo  {publisher} {Pergamon press},\ \bibinfo
  {address} {London},\ \bibinfo {year} {1964})\BibitemShut {NoStop}%
\bibitem [{\citenamefont {Liberzon}(2011)}]{10.5555/2209760}%
  \BibitemOpen
  \bibfield  {author} {\bibinfo {author} {\bibfnamefont {D.}~\bibnamefont
  {Liberzon}},\ }\href@noop {} {\emph {\bibinfo {title} {Calculus of Variations
  and Optimal Control Theory: A Concise Introduction}}}\ (\bibinfo  {publisher}
  {Princeton University Press},\ \bibinfo {address} {USA},\ \bibinfo {year}
  {2011})\BibitemShut {NoStop}%
\end{thebibliography}%
\end{document}